\documentclass[12pt]{aastex63}

\usepackage{float}
\usepackage{graphicx}           
\usepackage{latexsym}           
\usepackage{amsmath}            
\usepackage{nccmath}            
\usepackage{bm}                 
\usepackage[english]{babel}
\usepackage{color}              
\usepackage{booktabs}           
\usepackage{multirow}           
\usepackage{natbib}

\newcommand{\del}[1]{\relax}%
\newcommand{\DELETED}[1]{\relax}%
{\relax}%

\definecolor{violet}  {rgb}{1.0,0.0,1.0}

\definecolor{dviolet} {rgb}{0.75,0.0,1.0}

\definecolor{blue}    {rgb}{0.0,0.7,1.0}

\definecolor{lblue}   {rgb}{0.5,1,1}

\definecolor{dblue}   {rgb}{0.0,0.0,1.0}

\definecolor{blgr}    {rgb}{0.70,0.80,1.00}

\definecolor{navy}    {rgb}{0.00,0.00,0.48}

\definecolor{green}   {rgb}{0.7,1.0,0.0}

\definecolor{dgreen}  {rgb}{0.0,0.6,0.0}

\definecolor{lgreen}  {rgb}{0.0,0.8,0.0}

\definecolor{dg}      {rgb}{0.0,0.6,0.0}

\definecolor{orange}  {rgb}{1.0,0.5,0.0}

\definecolor{dorange} {rgb}{1.0,0.6,0.0}

\definecolor{brown}   {rgb}{0.1,0.1,0.0}

\definecolor{lbrown}  {rgb}{0.7,0.5,0.0}

\definecolor{red}     {rgb}{1,0.0,0.0}

\definecolor{dred}    {rgb}{0.6,0.0,0.0}

\definecolor{grey}    {rgb}{0.1,0.1,0.1}

\definecolor{lgrey}   {rgb}{0.5,0.5,0.5}

\definecolor{black}   {rgb}{0.0,0.0,0.0}

\newcommand\n            {\noindent}

\newcommand\m            {\medskip}
\newcommand\s            {\smallskip}

\newcommand\bn           {\bigskip\noindent}
\newcommand\mn           {\medskip\noindent}
\newcommand\sn           {\smallskip\noindent}
\newcommand\cl           {\centerline}
\newcommand\ve           {\vfill\eject}

\newcommand\degree       {{\ifmmode^\circ\else$^\circ$\fi}} 
\newcommand\arcm         {{\ifmmode {'\ }\else$'     $\fi}} 
\newcommand\arcs         {{\ifmmode{''\ }\else$''    $\fi}} 

\newcommand\bEcl         {{$b^{\rm Ecl}$}}
\newcommand\lEcl         {{$l^{\rm Ecl}$}}

\newcommand\bII          {{$b^{\rm II}$} }

\newcommand{\bul}        {$\bullet$\ }
\newcommand\cge          {{$\gtrsim$}}
\newcommand\cle          {{$\lesssim$}}

\newcommand\chisq        {{$\chi^{2}$} }
\newcommand\eg           {{\it e.g.},}
\newcommand\ie           {{\it i.e.},}

\newcommand\eminpixsec   {{$e^{-}$/pixel/sec}}

\newcommand\etal         {{et\thinspace al.}} 

\newcommand\Fnu          {{$F_{\nu}$}}
\newcommand\Inu          {{$I_{\nu}$}}
\newcommand\Ho           {{$H_{0}$} }

\newcommand\HAB          {{$H_{AB}$} }
\newcommand\JAB          {{$J_{AB}$} }
\newcommand{\jit}        {\emph{jit}}

\newcommand\kmsMpc       {{km\ s$^{-1}$\ Mpc$^{-1}$} }
\newcommand\leff         {{$\lambda_{eff}$} }

\newcommand\mAB          {{$m_{\rm AB}$}}

\newcommand\magarc       {{mag\ arcsec$^{-2}$}}

\newcommand\mum          {{\micron}}

\newcommand\nWsqmsr      {{nW\ m$^{-2}$\ sr$^{-1}$}}

\newcommand\WsqmHz       {{$W\ m^{-2}\ Hz^{-1}$}}

\newcommand\ProFit       {\texttt{ProFit}}
\newcommand\ProFound     {\texttt{ProFound}}

\newcommand\texp         {{$t_{\rm exp}$}}

\newcommand\Tref         {{$T_{\rm ref}$}}

\newcommand\nwat        {nW/m$^2$/sr}

\defcitealias{driver_2016}{D16}
\defcitealias{windhorst_2011}{W11}
\defcitealias{windhorst_2022}{W22}
\defcitealias{Windhorst2022}{SKYSURF-1}

\def\ltsima{$\; \buildrel < \over \sim \;$}
\def\lsim{\lower.5ex\hbox{\ltsima}}
\def\gtsima{$\; \buildrel > \over \sim \;$}
\def\gsim{\lower.5ex\hbox{\gtsima}}

\newlength{\txw}
\newlength{\txh}

\begin{document}

\vspace*{-0.50cm}
\title{SKYSURF: Constraints on Zodiacal Light and Extragalactic Background Light
through Panchromatic HST All-Sky Surface-Brightness Measurements: II. First Limits on Diffuse Light at 1.25, 1.4, and 1.6 microns} 

\author{Timothy Carleton} 
\affiliation{School of Earth and Space Exploration, Arizona State University,
	Tempe, AZ 85287-1404}

\author{Rogier A. Windhorst}
\affiliation{School of Earth and Space Exploration, Arizona State University,
Tempe, AZ 85287-1404}

\author{Rosalia O'Brien} 
\affiliation{School of Earth and Space Exploration, Arizona State University,
Tempe, AZ 85287-1404}

\author{Seth H. Cohen} 
\affiliation{School of Earth and Space Exploration, Arizona State University,
Tempe, AZ 85287-1404}

\author{Delondrae Carter} 
\affiliation{School of Earth and Space Exploration, Arizona State University,
Tempe, AZ 85287-1404}

\author{Rolf Jansen} 
\affiliation{School of Earth and Space Exploration, Arizona State University,
Tempe, AZ 85287-1404}

\author{Scott Tompkins} 
\affiliation{School of Earth and Space Exploration, Arizona State University,
Tempe, AZ 85287-1404}

\author{Richard G. Arendt} 
\affiliation{UMBC/CRESST2, NASA Goddard Space Flight Center, Greenbelt, MD 21771, USA}

\author{Sarah Caddy} 
\affiliation{Macquarie University, Sydney, NSW 2109, Australia}

\author{Norman Grogin} 
\affiliation{Space Telescope Science Institute, 3700 San Martin Drive,
Baltimore, MD 21210}

\author{Scott J. Kenyon} 
\affiliation{Smithsonian Astrophysical Observatory, 60 Garden Street,
	Cambridge, MA 02138} 

\author{Anton Koekemoer} 
\affiliation{Space Telescope Science Institute, 3700 San Martin Drive,
Baltimore, MD 21210}

\author{John MacKenty} 
\affiliation{Space Telescope Science Institute, 3700 San Martin Drive,
Baltimore, MD 21210}

\author{Stefano Casertano} 
\affiliation{Space Telescope Science Institute, 3700 San Martin Drive,
Baltimore, MD 21210}

\author{Luke J. M. Davies} 
\affiliation{The University of Western Australia, M468, 35 Stirling Highway, 
Crawley, WA 6009, Australia}

\author{Simon P. Driver} 
\affiliation{International Centre for Radio Astronomy Research (ICRAR) and the
	International Space Centre (ISC), The University of Western Australia, M468,
	35 Stirling Highway, Crawley, WA 6009, Australia}

\author{Eli Dwek} 
\affiliation{NASA Goddard Space Flight Center, Greenbelt, MD 21771}

\author{Alexander Kashlinsky} 
\affiliation{NASA Goddard Space Flight Center, Greenbelt, MD 21771}

\author{Nathan Miles} 
\affiliation{Space Telescope Science Institute, 3700 San Martin Drive,
Baltimore, MD 21210}

\author{Nor Pirzkal} 
\affiliation{Space Telescope Science Institute, 3700 San Martin Drive,
Baltimore, MD 21210}

\author{Aaron Robotham} 
\affiliation{International Centre for Radio Astronomy Research (ICRAR) and the
International Space Centre (ISC), The University of Western Australia, M468,
35 Stirling Highway, Crawley, WA 6009, Australia}

\author{Russell Ryan} 
\affiliation{Space Telescope Science Institute, 3700 San Martin Drive,
Baltimore, MD 21210}

\author{Haley Abate} 
\affiliation{School of Earth and Space Exploration, Arizona State University,
Tempe, AZ 85287-1404}

\author{Hanga Andras-Letanovszky} 
\affiliation{Steward Observatory, University of Arizona, Tucson, AZ 85721-0065}

\author{Jessica Berkheimer} 
\affiliation{School of Earth and Space Exploration, Arizona State University,
Tempe, AZ 85287-1404}

\author{Zak Goisman} 
\affiliation{School of Earth and Space Exploration, Arizona State University,
Tempe, AZ 85287-1404}

\author{Daniel Henningsen} 
\affiliation{School of Earth and Space Exploration, Arizona State University,
Tempe, AZ 85287-1404}

\author{Darby Kramer} 
\affiliation{School of Earth and Space Exploration, Arizona State University,
Tempe, AZ 85287-1404}

\author{Ci'mone Rogers} 
\affiliation{School of Earth and Space Exploration, Arizona State University,
Tempe, AZ 85287-1404}

\author{Andi Swirbul} 
\affiliation{School of Earth and Space Exploration, Arizona State University,
Tempe, AZ 85287-1404}

\email{tmcarlet@asu.edu}

\begin{abstract}
	We present the first results from the HST Archival Legacy project ``SKYSURF.” As described in \cite{Windhorst2022}, SKYSURF utilizes the large HST archive to study the diffuse UV, optical, and near-IR backgrounds and foregrounds in detail. Here we utilize SKYSURF's first sky-surface brightness measurements to constrain the level of near-IR diffuse Extragalactic Background Light (EBL) in three near-IR filters (F125W, F140W, and F160W). This is done by comparing our preliminary sky measurements of  $>30,000$ images to Zodiacal light models, carefully selecting the darkest images to avoid contamination from stray light. Our sky-surface brightness measurements have been verified to an accuracy of better than $1\%$, which when combined with systematic errors associated with HST, results in sky brightness uncertainties of $\sim2-4\% \simeq 0.005$ MJy/sr in each image. When compared to the \cite{Kelsall1998} Zodiacal model, an isotropic diffuse background of $\sim30$ \nWsqmsr\, remains, whereas using the \cite{Wright1998} Zodiacal model results in no discernible diffuse background. Based primarily on uncertainties in the foreground model subtraction, we present limits on the amount of diffuse EBL of 29 \nWsqmsr\,, 40 \nWsqmsr\,, and 29 \nWsqmsr\,, for F125W, F140W, and F160W respectively. While this light is generally isotropic, our modeling at this point does not distinguish between a cosmological origin or a Solar System origin (such as a dim, diffuse, spherical cloud of cometary dust).
\end{abstract}

\bn 

\vspace*{2em}


\n \section{Introduction} \label{sec:introduction}

\n {The cosmic optical and near-IR Extragalactic Background Light (EBL), derived from
the integrated luminosity of all extragalactic objects over all redshifts, represents a fundamental
test of our understanding of extragalactic astronomy \citep[\eg][]{McVittie1959, Partridge1967a, Partridge1967b, Hauser2001,
	Lagache2005, Kashlinsky2005a, Finke2010, Dominguez2011, Dwek2013, Khaire2015,
	Driver2016b, Koushan2021, SaldanaLopez2021}. If our census of galaxies and their luminosities are truly
complete, the total EBL level should equal that of all discrete objects. On the other hand, if the EBL is found to be in excess of predictions from galaxy counts, that suggests that galaxy surveys may be missing some discrete or diffuse sources. Despite the importance of this measurement, direct EBL measurements have yet to arrive at a value that agrees with predictions from galaxy number counts \citep[for a recent review, see][]{Cooray2016}. Project SKYSURF \citep{Windhorst2022} aims to study this discrepancy with the vast archive of HST images.}

{Because of the difficulty of characterizing the foreground signal of Earth's atmosphere, observational attempts at constraining the EBL level directly are primarily done with space missions, such as COBE \citep[\eg][]{Puget1996, Fixsen1998, Dwek1998a, Hauser1998, Finkbeiner2000, Cambresy2001, Sano2020}, {\it Spitzer}
		\citep[][]{Dole2006}, {HST} \citep[][]{Bernstein2002, Bernstein2007}, IRTS \citep[][]{Matsumoto2005,Matsumoto2011}, and AKARI \citep[][]{Matsuura2011,Tsumura2013}}. These observations have large errors and are often discrepant with each other because of the limited number of observations and the difficulty of subtracting the instrumental, Zodiacal, Galactic, and astrophysical foregrounds \citep{Cooray2016}. Regardless, these direct measurements consistently arrive at EBL levels of $\sim 20-50$ \nwat{}, significantly above the predictions from galaxy counts of $\sim 10$ \nwat{} \citep[e.g.~][]{Driver2011,Andrews2018}.  Recent advances have been made with the CIBER experiment {\citep{Matsuura2017,Korngut2022}}, and Pioneer and New Horizons missions \citep{Matsumoto2018,Lauer2021, Lauer2022} that aim to better subtract the Zodiacal foreground, using Ca absorption features and by leaving the solar system, respectively. These observations find EBL levels closer to expectations, but they still identify a significant diffuse signal and represent a relatively small number of measurements. A parallel indirect approach, using observations of attenuated $\gamma$-rays, also finds values in line with predictions from galaxy counts \citep{HESS2013}.

{While the presence of diffuse EBL may diminish as new measurements better constrain foreground levels, many astrophysical sources have been hypothesized as contributing to it. The large population of recently-identified Ultra-Diffuse Galaxies in clusters \citep{Impey1988,vanDokkum2015} and the field \citep{Dalcanton1997,Leisman2017} represents one possible source of diffuse light, although many more unidentified UDGs would have to be present to contribute significantly to the EBL \citep{Jones2018}. Diffuse light in the outskirts of galaxy halos (IGL) may contribute as well \citep{Conselice2016}, although a number of studies \citep[e.g.~][]{Ashcraft2018a,Borlaff2019,Cheng2021} find that halo light, or light in galaxy outskirts, only represents $15\%$ of the luminosity of bright galaxies. Alternatively, significant levels of difficult-to-detect diffuse intracluster \citep{Bernstein1995} or intragroup light \citep{Mihos2005} may contribute to the diffuse EBL. More exotic explanations, such as light from reionization \citep{Santos2002, Cooray2004, Kashlinsky2004} have been put forward as well.}

{The SKYSURF project, introduced in \cite{Windhorst2022}, aims to better understand the EBL level with the large volume of archival HST observations using a two-pronged approach. First, it will use HST's remarkable stability and precision as an absolute photometer to conduct precise sky brightness measurements for over $200,000$ HST images. Second, it will use the depth and large volume probed by those images to search for possible sources of diffuse EBL.}

{For the full motivation and overview of the SKYSURF project, and an overview of its methods, see \cite{Windhorst2022}; we refer to this paper as \citetalias{Windhorst2022} throughout. In this paper, we describe the first results of SKYSURF surface-brightness measurements at 1.25, 1.4, and 1.6 microns. In Section~\ref{sec:backgrounds} we further outline the diffuse foreground sources necessary to consider for SKYSURF's EBL constraints, in Section~\ref{sec:measurement}, we briefly describe our measurement procedure, Sections~\ref{sec:results} presents our results, Section~\ref{sec:discussion} includes a discussion of those results, and Section~\ref{sec:conclusions} summarizes our conclusions. 
Throughout we use Planck cosmology \citep{Ade2016}: \Ho~=~66.9 
\kmsMpc\,, matter density parameter $\Omega_{m}$=0.32 and vacuum energy
density $\Omega_{\Lambda}$=0.68. When quoting magnitudes, our fluxes are all in AB-magnitudes (hereafter
AB-mag), and our SB-values are in AB-\magarc\ \citep{OkeGunn1983} or MJy/sr, 
using flux densities $F_{\nu}$ = 10$^{-0.40({\rm AB} - 8.90\ {\rm mag})}$ in
Jy. Further details on the flux density scales used are given in
Fig.~\ref{fig:eblfig1} and the Table footnotes in \S\
\ref{sec:results}.

\n \subsection{Foregrounds}
\label{sec:backgrounds}
\n The main goal of SKYSURF is to characterize the components of sky surface brightness present in HST images, including a possible diffuse EBL component, in detail. Below, we summarize the relevant astronomical foregrounds and backgrounds that exist in the SKYSURF
images. In summary, they are the
following: Zodiacal Light (ZL), Diffuse Galactic Light (DGL), discrete stellar and extragalactic light, and diffuse EBL. The Zodiacal Light (ZL) is the main foreground in most HST images,
and SKYSURF will measure and model it as well as possible with available tools. All stars in our
galaxy (except the Sun) and all other galaxies are beyond the InterPlanetary
Dust Cloud (IPD), so the ZL is thus always referred to as a ``foreground''. Similarly, the
Diffuse Galactic Light, caused by scattered star-light in our Galaxy, can
be a background (to nearby stars), or a foreground (to more distant stars and
all external galaxies).
Most objects in an
average moderately deep (AB\cle 25--26 mag) HST image are faint galaxies close
to the peak in the cosmic star-formation history at $z$\cle $2$
\citep[e.g.~][]{Madau2014}. Most of the Extragalactic Background Light (EBL)
therefore comes from distant galaxies and AGN, and is thus referred to as a ``background''. 

Before
SKYSURF can quantify and model these astronomical foregrounds and backgrounds,
it needs to address the main contaminants, which are residual detector
systematics, orbital phase-dependent straylight from the
Earth, Sun, and/or Moon, and the WFC3/IR Thermal Dark signal.
Instrumental and stray light contaminants, as well as the contribution of discrete objects to the SKYSURF EBL constraints, are discussed in \citetalias{Windhorst2022}. Below, we discuss the diffuse Zodiacal, Galactic, and Extragalactic foregrounds in more detail. 


\n \subsubsection{Zodiacal Foreground}\label{sec:zodi}
\n By far, the brightest component of the sky brightness is Zodiacal Light from the 
InterPlanetary Dust (IPD) cloud, \ie\ from distances less than 5 AU, representing over
$95\%$ of the photons with 0.6-1.25 \mum\ wavelengths in the HST archive (see Fig.~\ref{fig:eblfig1}). Given its extremely diffuse nature, as well as its time variability, it has been
a challenge to understand in detail; observations with all-sky space missions such as COBE/DIRBE are required to fully model it. For example, the \cite{Kelsall1998} and \cite{Wright1998} Zodiacal models use the COBE/DIRBE data to model the Zodiacal emission, considering multiple dust components scattering sunlight toward Earth. The absence of an all-sky optical survey means that such modeling cannot be done in the optical to a similar extent; most authors simply assume that the Zodiacal spectrum is a Solar, or slightly reddened Solar spectrum \citep[\eg][]{Leinert1998}. Future SKYSURF studies will utilize its UV-to-optical database to improve constraints on the Zodiacal spectrum, but here we only consider observations with wavelengths similar to COBE/DIRBE wavebands for which a detailed Zodiacal model is obtainable.

\n \subsubsection{Discrete and Diffuse Light from Kuiper Belt Objects}
\label{sec:diffusezodi}

\sn The darkness of the night sky, ``Olbers' Paradox'', was one of astronomy's
oldest mysteries: an infinite and infinitely old Universe full of stars and
galaxies would have a sky as bright as the surface of an average star. The
resolution of this ``paradox'' --- an expanding Universe of finite age --- is,
of course, the central tenet of Big Bang cosmology, where the galaxy surface 
density is a finite integral over the galaxy luminosity function and the
cosmological volume element \citep{Driver1995, Metcalfe1995, Odewahn1996,
Tyson1988}. Because of their very steep observed number counts, Kuiper Belt
Objects (KBOs) can also appear to violate Olbers' Paradox, producing an
apparently \emph{diverging} sky integral when the smallest objects are taken
into account \citep[\eg][]{Kenyon2001}. To not exceed the observed the ZL sky-SB, the
counts of KBOs at distances \cge 40 AU must turn over from the non-converging
power-law slope $\gamma$$\simeq$0.6 dex/mag observed for R\cle 27 mag
\citep{Fraser2014} to a converging slope flatter than $\gamma$=0.4 dex/mag at
R-band fluxes of AB\cge 45--55 mag, in combination with a limited volume over
which KBOs occur \citep{Kenyon2001}. Assuming albedos of a few percent
\citep[\eg][]{Kenyon1999} and a physical size distribution of $N(r) \propto\
r^{-\beta}$, such a slope change in the KBO number counts implies that the 
size-slope of unresolved Solar System debris at $\sim$40 AU must flatten from
$\beta$$\simeq$4 for larger objects to $\beta$\cle 3.25--3.5 for objects with
sizes $r$$\sim$0.05--5 m. A flattening of the size distribution of the
planetesimal population with radii $r$\cge 10 km from $\beta$$\simeq$4 to
$\beta$\cle 3.5 is consistent with simulations for the debris population with
$r$\cle 1 km, which suggest that collisions with $r$\cle 100 m objects tend to
produce debris rather than larger planetesimals \citep{Kenyon1999, Kenyon2004,
Kenyon2020}. It is also consistent with ground-based observations of KBOs with
$r$\cle 50 km \citep[\eg][]{Fuentes2009, Shankman2013}, and with New Horizons
(NH) crater counts on Pluto and Charon, which suggest a flattening of the KBO
count slope for $r$\cle 1 km \citep[\eg][]{Singer2019}.

To refine these constraints across the Kuiper Belt, SKYSURF will measure the
panchromatic Zodiacal foreground in the ecliptic plane in places where other
foregrounds are small. Better {\it SB-limits} on the small KBO population may
constrain the slope of the KBO counts, and hence the total Kuiper Belt mass at
35--50 AU. Time-tagged monitoring of the sky-SB in the Ecliptic may also yield
constraints to the integral of Plutinos in Neptune's L4 and L5 Lagrange points,
which have moved significantly in Ecliptic Longitude (\lEcl) during the 32-year
HST mission. \citet{Kelsall1998} fit the data from the Cosmic Background
Explorer/Diffuse InfraRed Background Experiment (COBE/DIRBE) as a family of 3-D
(flaring) disk models of decreasing density with increasing radius and distance
from the ecliptic plane. This model accounts for the variation with Solar phase
angle for realistic properties of dust grains. Other ZL models and refinements
were presented by, \eg\ \citet{Reach1997}, \citet{Leinert1998},
\citet{Wright1998}, \citet{Wright2001}, \citet{Jorgensen2020},
\citet{Arendt2014}, and \citet{Arendt2016}. \citet{Kelsall1998} adopt an albedo
at 1.25 \mum\ wavelength for their Zodiacal ``Smooth Cloud'', ``Dust Bands'',
and ``Ring+Blob'' components of $a$=0.204$\pm$0.0013. Recent thermal IR
observations of Trans-Neptunian Objects (TNOs) with typical sizes of
$\sim$20--400 km imply geometric albedo values of \cle 20--30\%, whereas TNOs have
albedos as large as $\sim$60\% \citep[\eg][]{Duffard2014, Kovalenko2017,
Vilenius2012, Vilenius2014, Vilenius2018}, possibly indicating a more icy
surface for some TNOs. Possible variation in Solar System objects, and the impact that they may
have on our results is discussed further in \S\ \ref{sec:discussion}. 

\n \subsubsection{Diffuse Galactic Light} \label{sec:diffusegal} 

\sn Diffuse Galactic Light (DGL) in the UV--optical is mainly caused by
scattered light or reflection nebulae from early-type (O and B) stars,
scattered by dust and gas in the Interstellar Medium (ISM). The DGL is thus a
strong function of Galactic coordinates ($l^{II}$, $b^{II}$). SKYSURF's
SB-measurements may thus also constrain the DGL at low Galactic latitudes
(\bII\cle 20--30\degree), although these fields are very likely not useful for
background galaxy counts. The All-sky Infrared
Astronomical Satellite \citep[``IRAS''][]{Soifer1984, Helou1985}, COBE/DIRBE
\citep{Kelsall1998, Schlegel1998}, {\it Planck} \citep{Ade2016}, Wide-field
Infrared Survey Explorer \citep[WISE;][]{Wright2010}, and AKARI
\citep{Tsumura2013} maps in the near to far-IR help identify Galactic infrared
cirrus and regions of likely enhanced Galactic scattered light. Possible high
spatial frequency structures in the DGL appear in deep ground-based images of
low Galactic latitude at SB-levels of B$\simeq$26--27 mag/arcsec$^2$, and at
much fainter levels sometimes also at high Galactic latitudes
\citep[\eg][]{Szomoru1998, Guhathakurta1989}. While not a main goal of SKYSURF,
the DGL needs to be estimated and subtracted in order to better estimate the
levels of the ZL and the EBL at higher Galactic latitudes, as discussed in \S\
\ref{sec:results}. Panchromatic HST constraints on the DGL in the Galactic plane
($\vert$$b^{II}$$\vert$\cle 20\degree) are interesting in their own right and
are a byproduct of SKYSURF. We refer to \S\ \ref{sec:results_dgl} for the DGL levels we
subtract from any diffuse light levels implied by the comparison between our
HST sky-SB measurements and the ZL models. 

\n \section{Measurements}
\label{sec:measurement}
\n An overview of the SKYSURF database and our sky measurement procedure can be found in \citetalias{Windhorst2022}. Further details on the multiple sky measurements procedures, as well as the full results of the sky surface brightness measurements across the entire SKYSURF database will come in O'Brien et. al. (2022, in preparation). For context, we give a brief overview of the database and methods here.

First, the HST archive was searched for images taken with its wide-band filters, excluding grism images, quad/linear ramp or polarizing filters, subarray images, time-series, moving targets, or spatial scans. This resulted in $249,861$ images that made up the initial database. Further cuts on target selection, HST orbital phase, and exposure time will be conducted to avoid possible contamination and minimize measurement errors.

To measure the sky background of these images, the SKYSURF team tested multiple sky-measurement algorithms on realistic simulated images to identify the most robust method of estimating the uncontaminated sky background. All algorithms that were tested had an accuracy of better than $0.2\%$ for flat images, and slightly worse for images with gradients \citepalias[Fig.~8 in][]{Windhorst2022}. At this point, it is worth identifying the general philosophy of the SKYSURF program as to identify the \emph{Lowest Estimated Sky} (LES) value --- {defined as the lowest sky-SB in an image} --- as the fiducial sky measurement. While electronic errors within the cameras can introduce either positive or negative errors in sky estimation, errors deriving from contamination (i.e. stray light from nearby bright sources like the Earth and the Sun or thermal emission from the telescope) are more common and more significant. To make full use of our large dataset, we aimed to develop and use algorithms that are the \emph{most robust} across our database, which contains a wide variety of images. The full results with the most robust algorithms will be presented in O'Brien et. al. (2022, in preparation); here we present the first results using an initial estimation done by fitting a Gaussian to the sigma-clipped image (described as method 2 in O'Brien et. al. (2022, in preparation) and \citetalias{Windhorst2022}).

Combining the sky measurement uncertainties with the systematic uncertainties associated with HST's detectors, the overall absolute uncertainty on the sky measurements is $\sim2.7\%$ for the F125W, $\sim2.8\%$F140W, and $\sim3.8\%$ for the F160W filter. The systematic uncertainties come from Bias/Darkframe subtraction ($\lesssim1.0\%$), the global flat field correction ($0.5-2\%$), zeropoint accuracy ($\sim1.5\%$), and thermal dark subtraction ($\sim0.2\%$ for F125W, $0.5\%$ for F140W, and $\sim3.8\%$ for F160W).

\n \section{First SKYSURF Results on Diffuse Near-IR Sky-SB Estimates at
1.25--1.6 \mum} \label{sec:results}

\mn For the final analysis of 249,861 SKYSURF images, we expect \cle 50\% to be
usable for sky-SB measurements. Although these images are not completely randomly
distributed on the sky, they {\it on average} provide
$\sim$4400 sky-SB measurements in each of the 28 broad-band SKYSURF filters. In this section, we will use two
complementary analyses of the HST sky-SB estimates to make our first assessment
of available Zodiacal Light models, identify any diffuse light that may be present,
and check on the consistency of our methods.

The results from both methods will be compared to the \citet{Kelsall1998} and 
\citet{Wright1998} models, which predict the ZL brightness as a function of
sky position and time of the year. Both \cite{Kelsall1998} and \cite{Wright1998} models are fit to COBE/DIRBE measurements at $1.25-2.2~\mum$. \cite{Kelsall1998} is a physical model that contains multiple dust components, whereas the \cite{Wright1998} model is a more parametric model normalized at $25$ \mum\ to ensure $0$ residual diffuse light at the ecliptic poles. Because
their ZL model predictions are anchored to the COBE/DIRBE 1.25--2.2 \mum\ data,
we will limit our analysis in this paper to the SKYSURF WFC3/IR filters F125W,
F140W, and F160W. We will deal with the uneven sky-sampling of the HST data by
comparing the HST sky-SB data with the
corresponding ZL model predictions. Again, our premise throughout is that the lowest estimated sky-SB values
measured amongst the HST images in each direction will be the least affected by
HST systematics or discrete foreground objects, and therefore be closest to the
true sky-SB in that direction. 

\begin{figure*}[!hptb]
	\begin{tabular}{cc}
		\includegraphics[width=0.48\txw]{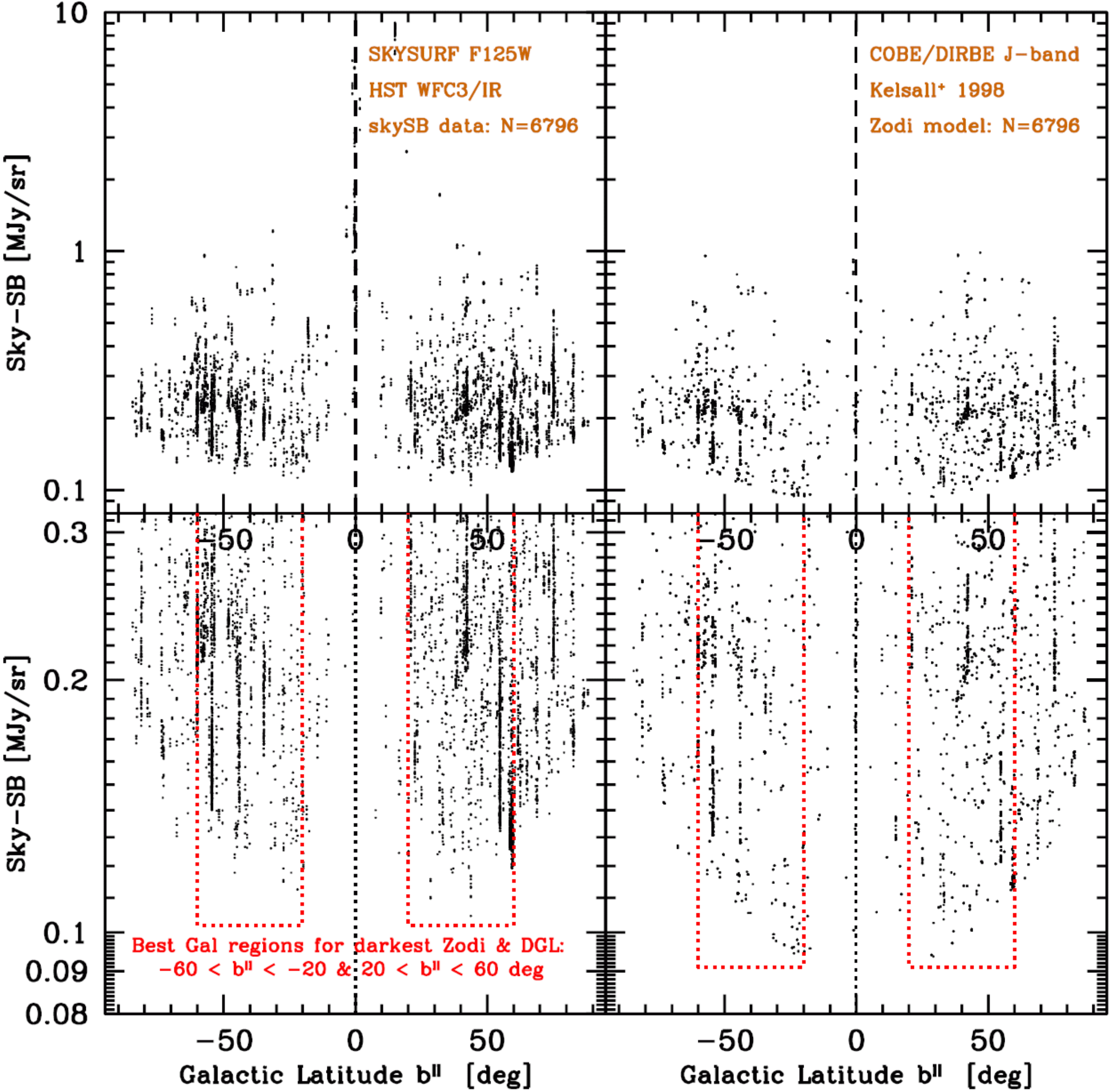} & \includegraphics[width=0.48\txw]{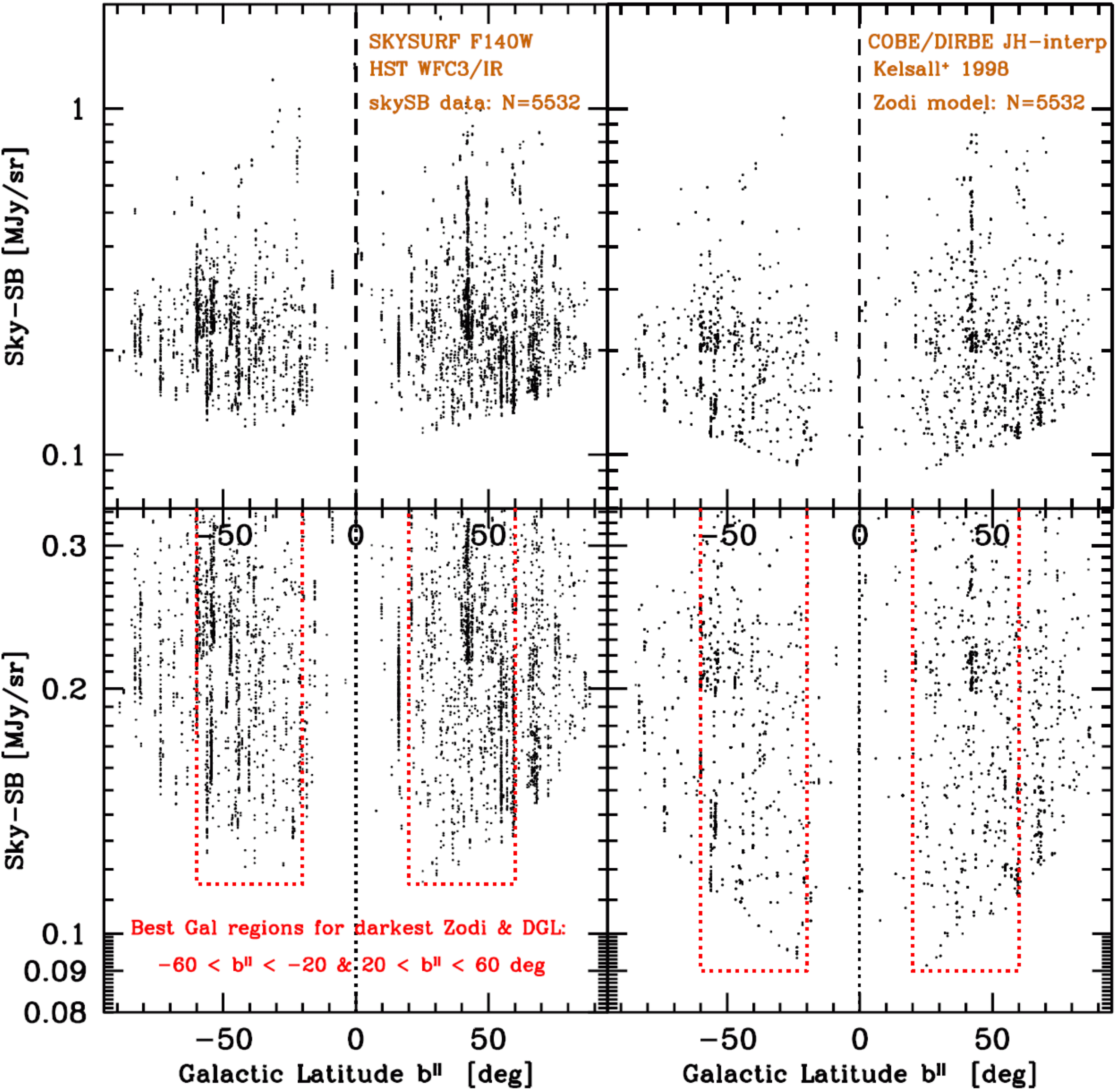}\\
		\includegraphics[width=0.48\txw]{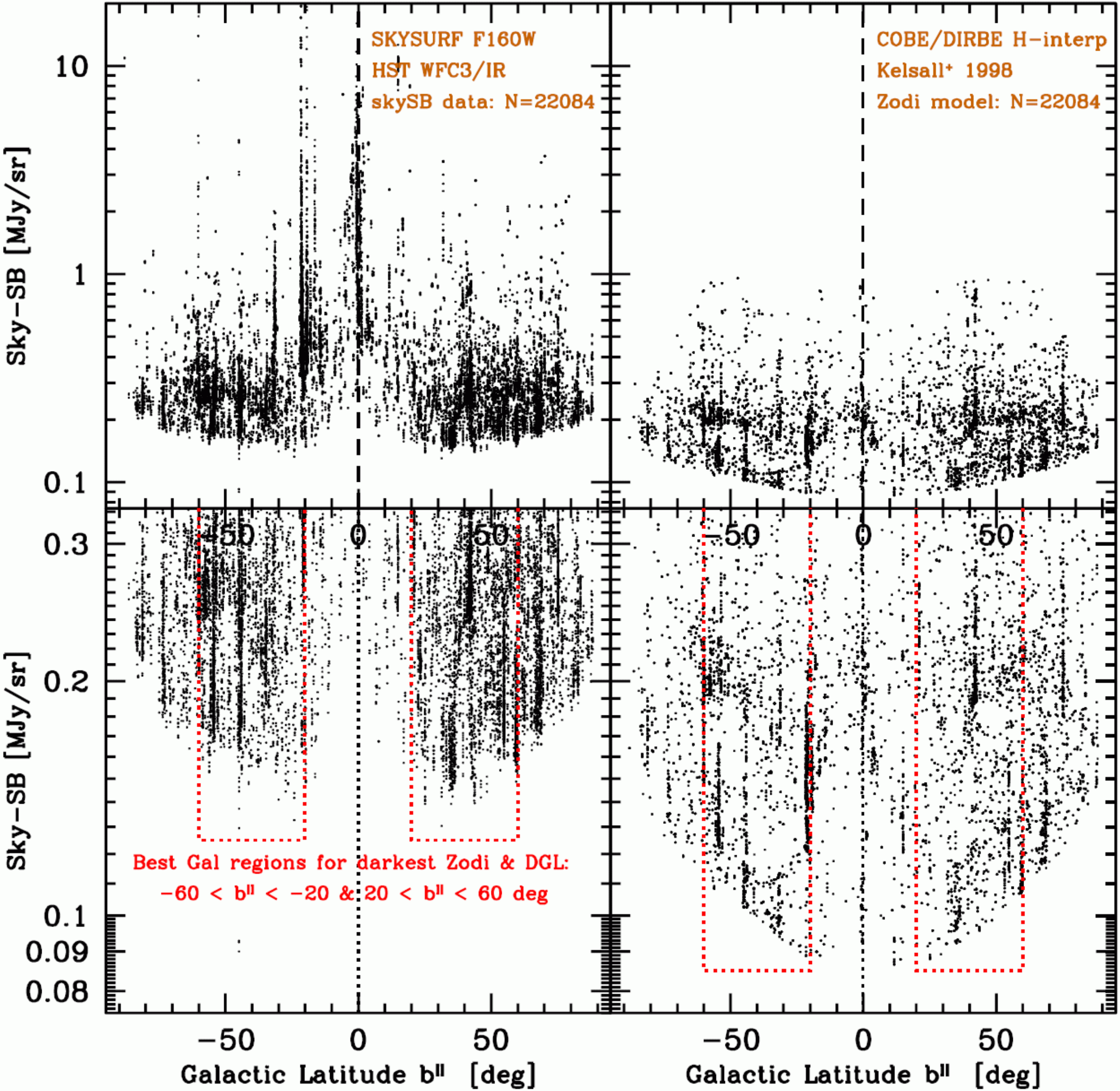} &
	\end{tabular}
		\vspace*{+0.20cm}
\n \caption{
	 All SKYSURF F125W, F140W, and F160W sky-SB measurements vs. Galactic Latitude for
	the HST data [left sub-panels] and the \citet{Kelsall1998} COBE/DIRBE models [right
	sub-panels]. The upper left plot is for F125W, the upper right plot is for F140W, and the lower left plot is for F160W. All bottom sub-panels are enlargements of top sub-panels. Because of the $\sim$60\degree\ tilt of the Galactic plane with
	respect to the Ecliptic, the darkest sky-SB occurs for 20\degree\cle
	$\vert$$b^{II}$$\vert$\cle 60\degree, highlighted by the red columns in all bottom sub-panels. Fields with $\vert$$b^{II}$$\vert$\cle
	20\degree\ have significant DGL, and are ignored
	in the final sky-SB analysis.}
\label{fig:f125gal}
\end{figure*}

The first approach uses the Lowest Estimated Sky-SB values from the HST 
images. Both the HST LES-data
and the \citet{Kelsall1998} model predictions are fit with analytic
functions as a function of Ecliptic
Latitude ($b^{Ecl}$) in the darkest parts of the Galactic sky. These fits will be
referred to as the Lowest Fitted Sky-SB (``LFS'') method. To avoid regions with significant DGL, the LFS
method will first select the LES-data and model predictions as a function of
Galactic Latitude ($b^{II}$), to identify the darkest regions of the
Galactic sky.

Next, the LFS method will identify the lowest sky-SB as a function of Ecliptic
Latitude ($b^{Ecl}$) to constrain the ZL+EBL sky-SB in each direction (see Fig.~\ref{fig:f125ecl_all}). For
$\vert$$b^{II}$$\vert$\cge 20\degree, where the DGL contribution is lower, the
LFS fits provide analytical functions describing the lowest sky-SB as a function of Ecliptic
Latitude for both the HST data and the model predictions in the same directions
of the sky. The limitation of the LFS method is that not all sky-SB
measurements are done at {\it constant} Sun Angles (SA; defined as the Sun-HST-target angle), which ranges from
SA$\simeq$85--180\degree\ at the Ecliptic to SA=90\degree\ at the
Ecliptic poles. Although many HST observations are scheduled around SA$\sim90$\degree, many others are done with higher solar elongations for which the Zodiacal sky-SB is lower (the Zodiacal sky-SB reaches a minimum in the Ecliptic at Solar
Elongations of 120--150\degree\ \citep{Leinert1998}). This method will thus focus on observations with SA$\sim$150\degree\ in the Ecliptic Plane and SA$\sim$90\degree\ at the Ecliptic Poles. However, because the analysis is conducted on the \emph{Zodiacal models in parallel}, this is not expected to bias our results.
In particular, this method aligns with the SKYSURF philosophy that most sources of error are positive, and thus the lowest sky values are likely the most accurate. 

The second method more closely follows the actual selection of the COBE/DIRBE
data, on which both the \citet{Kelsall1998} and \citet{Wright1998} models were
based. The COBE/DIRBE data were measured at Sun Angles
SA$\simeq$94$\pm$30\degree\ \citep[\eg][]{Leinert1998}. The HST data are
observed over a range of Sun Angles, but a significant fraction is also
observed at SA$\simeq$90$\pm$10\degree, \ie\ over
a Sun Angle range similar to, but somewhat narrower than that of the COBE/DIRBE
data. Hence, our second method will only select the HST
LES-data and COBE/DIRBE-based model predictions in the Sun Angle range of
SA$\simeq$90$\pm$10\degree. This ``SA90 method'' has the advantage of the
selected HST data being more directly comparable to the COBE/DIRBE based models, but
because of their SA-selection, it may also have somewhat higher levels of
(unrecognized) Earthshine. The HST {\it data} from the SA90 method may thus be
systematically somewhat higher than the minimum Zodiacal sky-SB level that is 
traced with the LFS-method.

Stated differently, the LFS method fits a ($sech$)
function to the {\it lowest} sky-SB levels observed at each Ecliptic latitude,
and is thus based on fewer data points. The LFS method is therefore more
reliable, but statistically less precise than the SA90 method. The SA90 method
fits regions with sky-SB more comparable to the COBE/DIRBE SA-range, and thus
has better statistics in this SA-range, but also subject to higher straylight
levels. A comparison between the two methods will
then give us an assessment of the uncertainties in any remaining diffuse light. In this initial analysis, as we are simply looking for a possible diffuse excess above the \cite{Kelsall1998} and \cite{Wright1998} models, these approaches work well. Future SKYSURF analysis will investigate stray-light contamination, as well as the structure of offsets between SKYSURF sky values and model predictions, in more detail.

\n \subsection{HST 1.25--1.6 \mum\ Sky-SB Measurements Compared to COBE/DIRBE
Predictions} \label{sec:hstkelsallresults}

\begin{figure*}[!hptb]
	\begin{tabular}{cc}
		\includegraphics[width=0.48\txw]{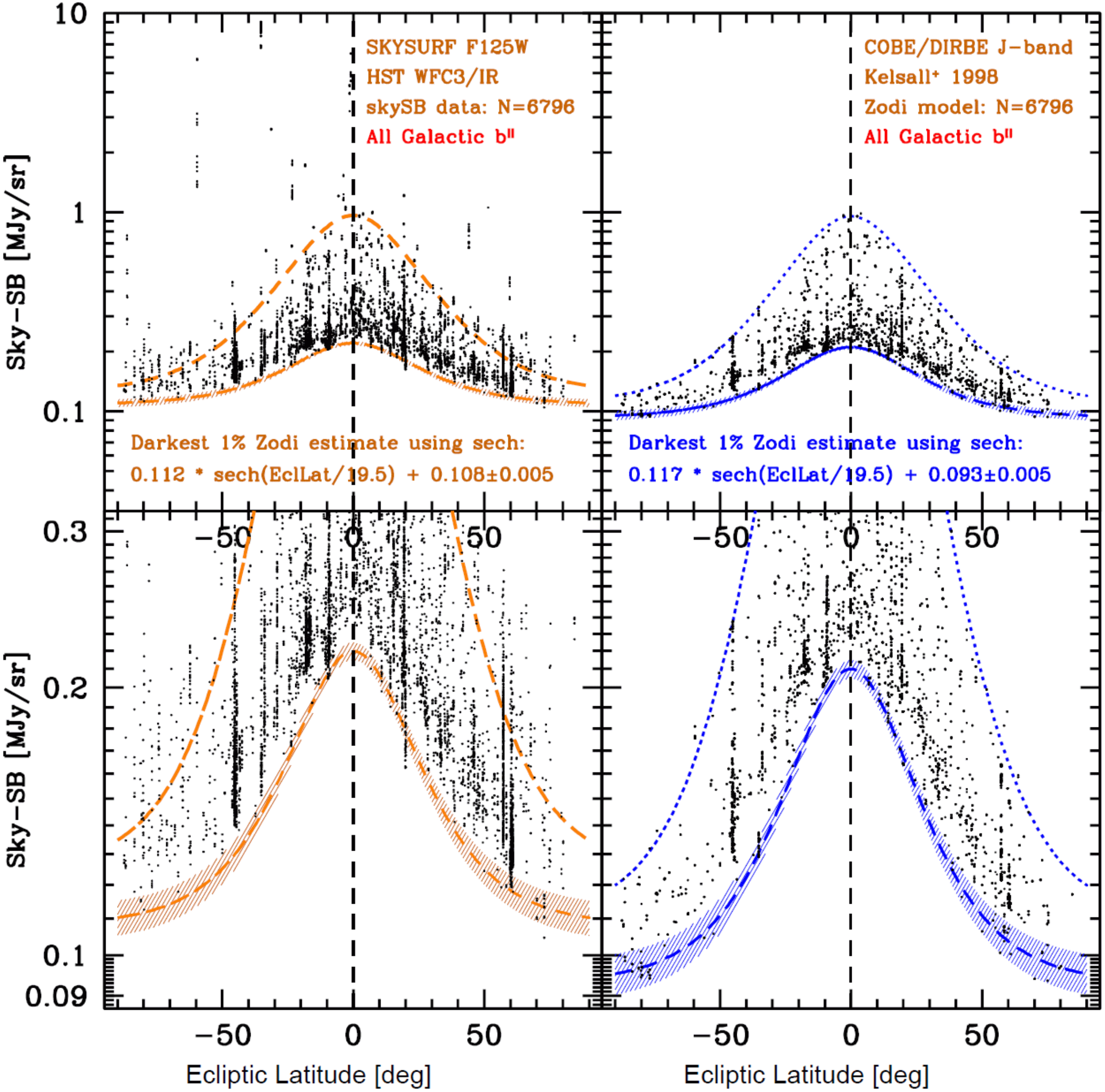} & \includegraphics[width=0.48\txw]{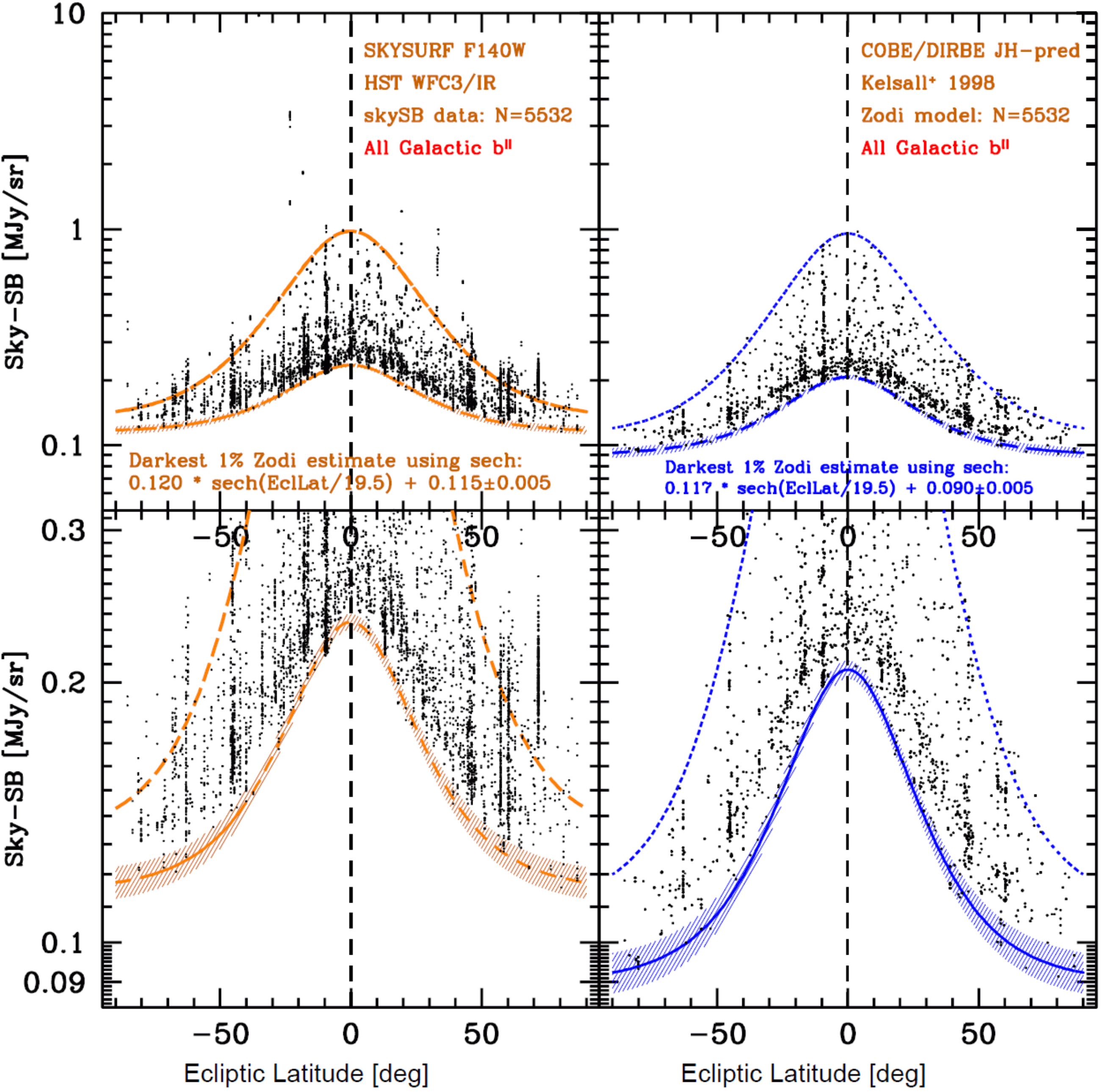}\\
		\includegraphics[width=0.48\txw]{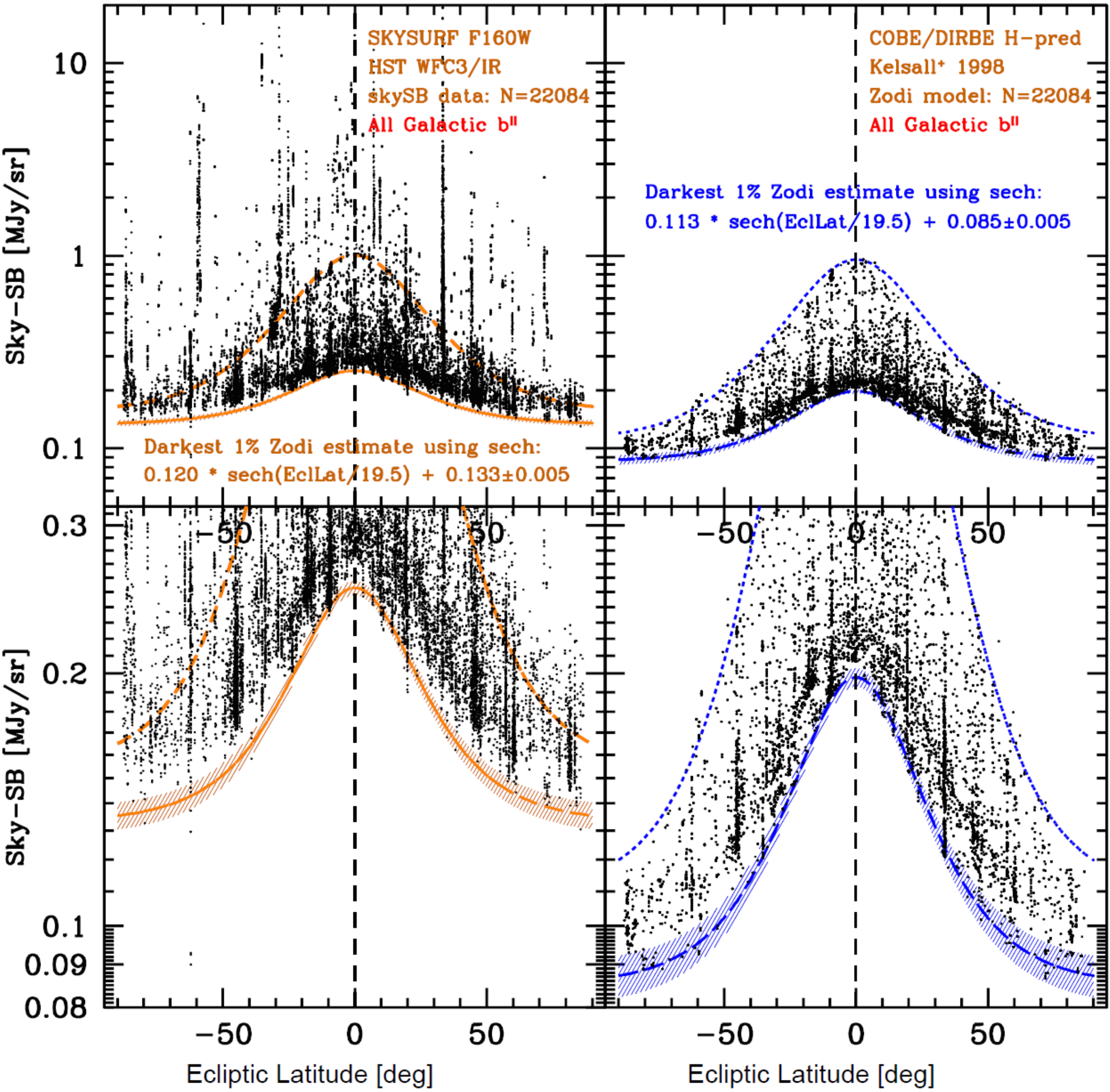} & 
	\end{tabular}
	\vspace*{+0.20cm}
	\n \caption{
		All SKYSURF F125W, F140W, and F160W sky-SB measurements vs. Ecliptic Latitude for
		HST data (left sub-panels) and the Kelsall COBE/DIRBE models (right sub-panels). The
		orange and blue $sech$ functions and error wedges outline the darkest 
		$\sim$1\% of the sky-SB measurements (magnify the PDF figure as needed to see this). Bottom sub-panels give enlargements of the
		top sub-panels. As in Figure~\ref{fig:f125gal}, the upper left plot is for F125W, the upper right plot is for F140W, and the lower left plot is for F160W. The short-dashed blue line represents the upper envelope to the
		\citet{Kelsall1998} model predictions, and the long-dashed orange line the
		correspondingly scaled {\it upper} envelope to the HST data that does not
		suffer excessive DGL or straylight, as described in \S\ \ref{sec:hstkelsallresults} and
		Table~\ref{tab:sech_vals}.}
	\label{fig:f125ecl_all}
\end{figure*}

\begin{figure*}[!hptb]
	\begin{tabular}{cc}
		\includegraphics[width=0.48\txw]{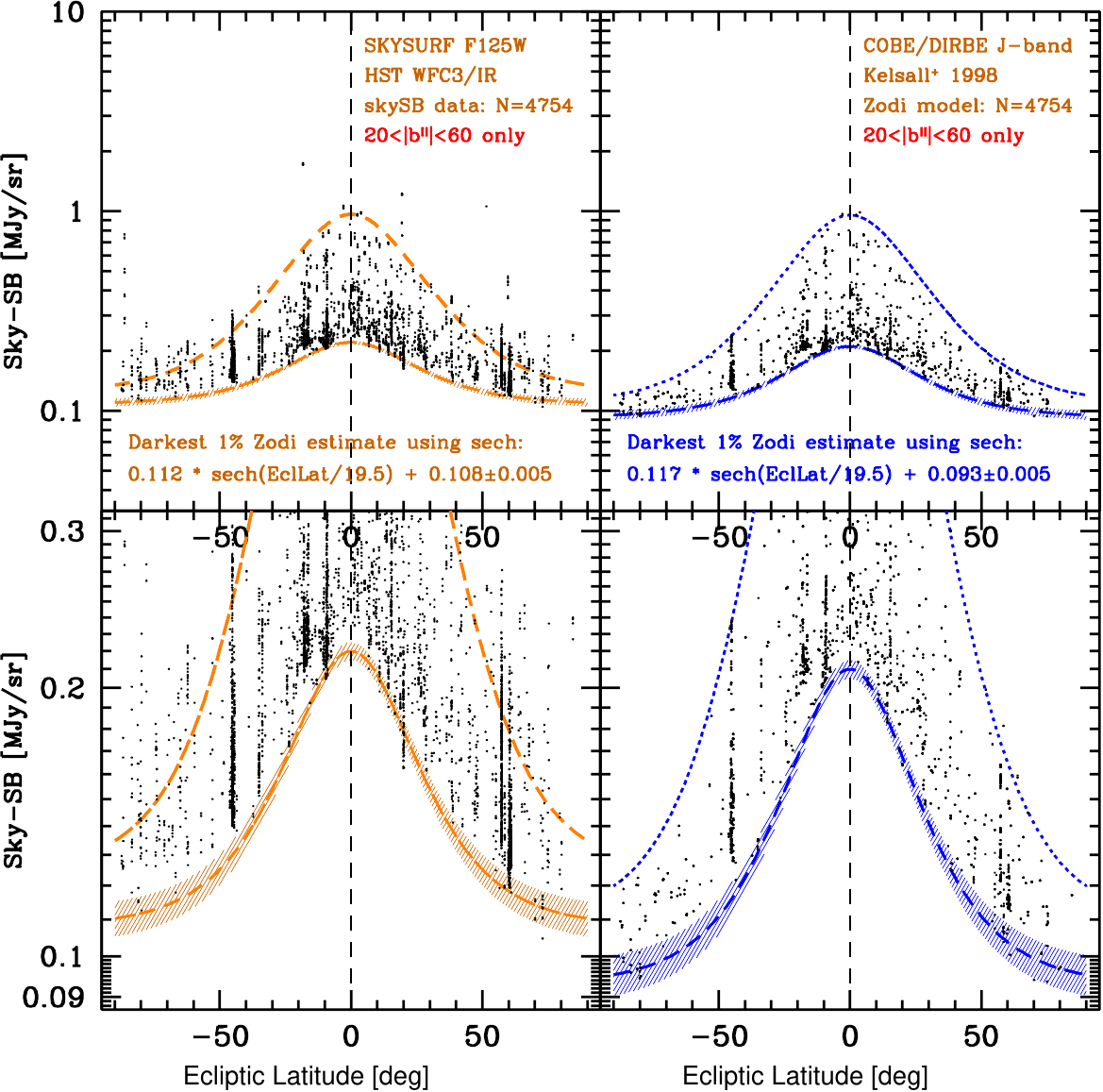} & \includegraphics[width=0.48\txw]{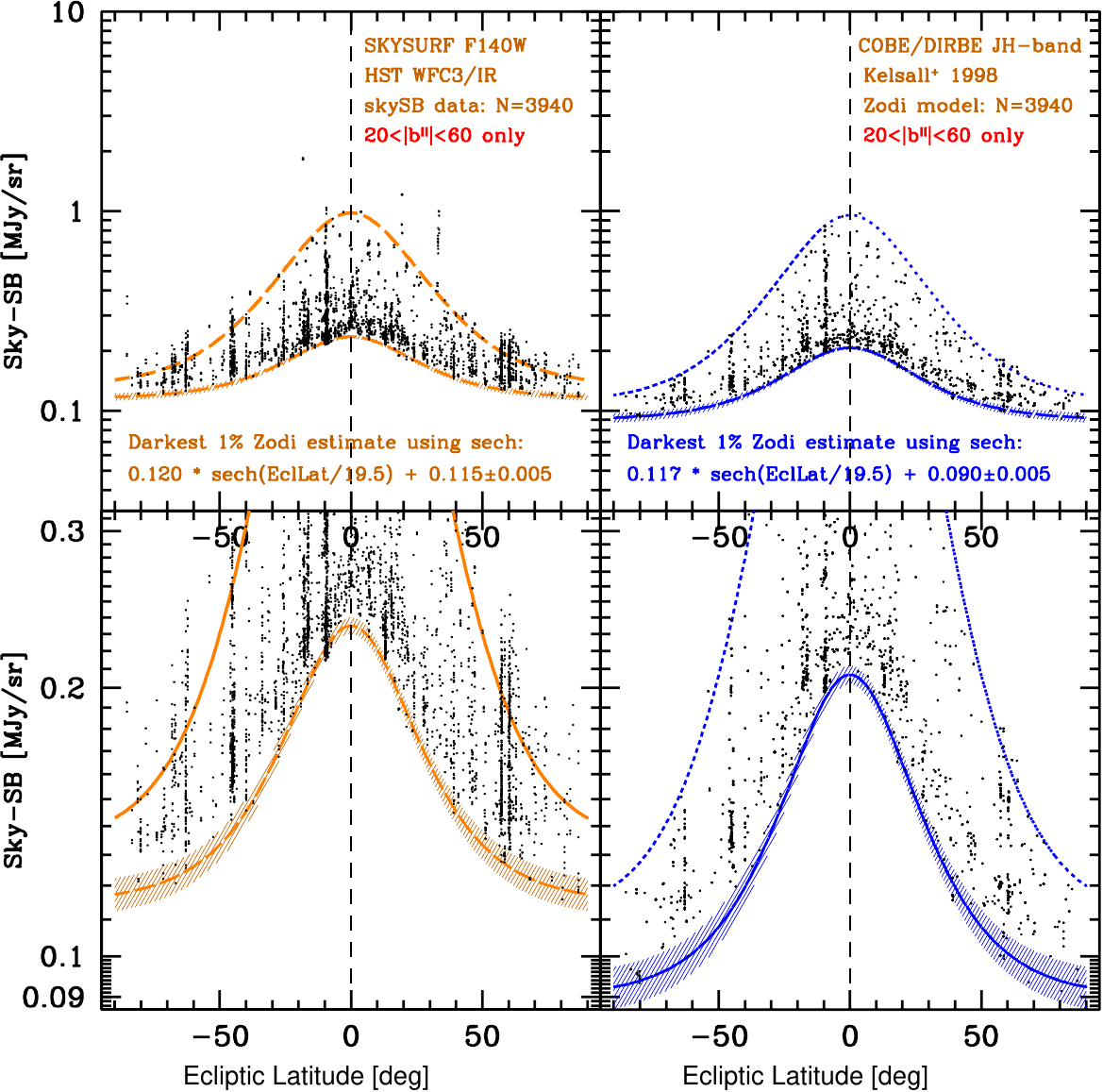}\\
		\includegraphics[width=0.48\txw]{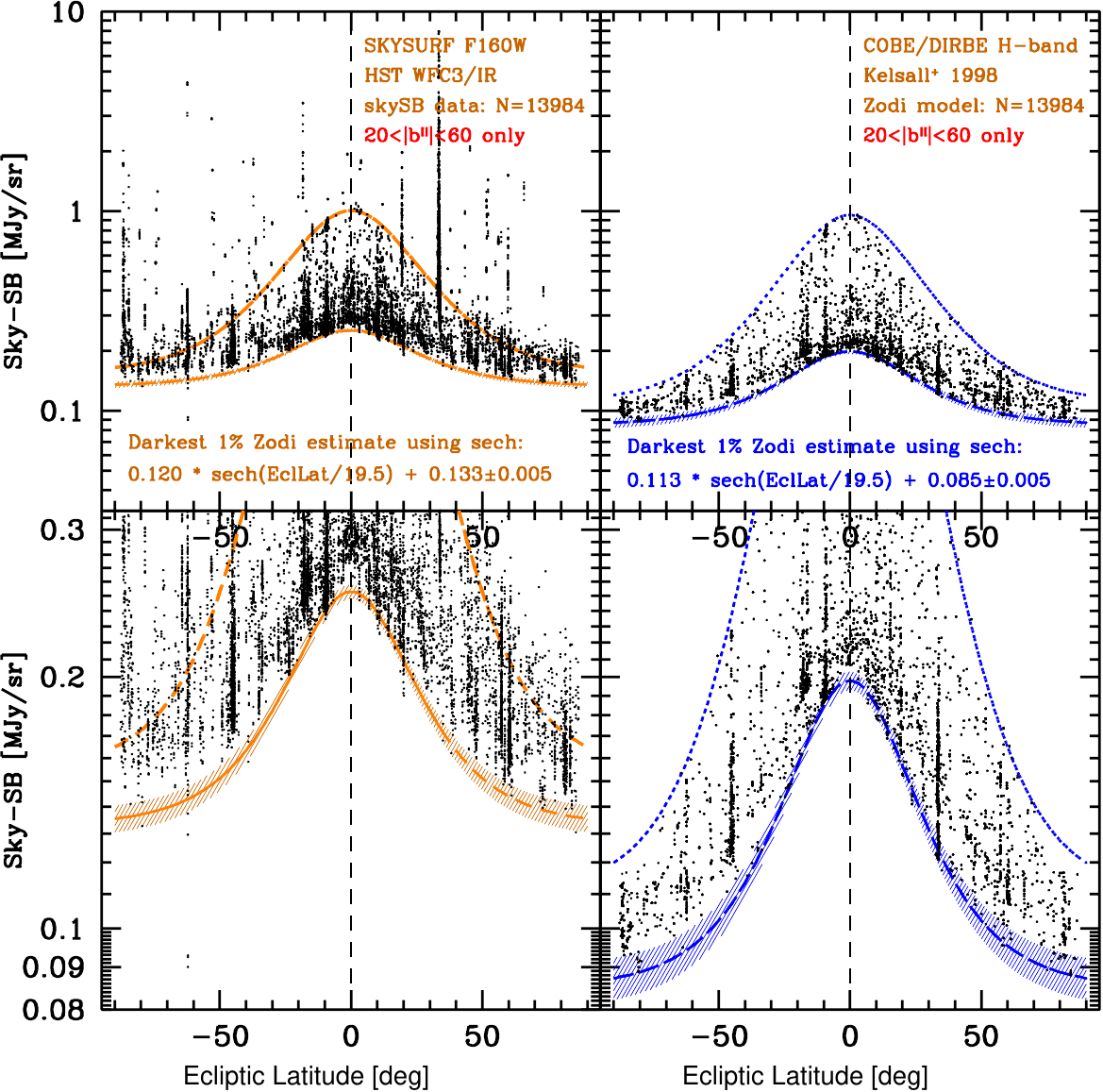} & 
	\end{tabular}
\caption{
	All SKYSURF sky-SB measurements vs. Ecliptic Latitude for
	HST data (left sub-panels) and the Kelsall COBE/DIRBE models (right sub-panels), but
	only for the darkest Galactic regions with 20\degree\cle
	$\vert$$b^{II}$$\vert$\cle 60\degree\ (see Fig.~\ref{fig:f125gal}). As in Figure~\ref{fig:f125gal}, the upper left plot is for F125W, the upper right plot is for F140W, and the lower left plot is for F160W. The orange
	and blue $sech$ functions and error wedges outline the darkest $\sim$1\% of
	the sky-SB measurements (magnify the PDF figure as needed to see this). Bottom panels give enlargements of the top panels.}
\label{fig:f125ecl_sub}
\end{figure*}

\sn In this section, we present our first SKYSURF results from 34,412 images
observed in the WFC3/IR filters F125W, F140W, and F160W. 
Figs.~\ref{fig:f125gal} and \ref{fig:f125ecl_all} show the sky-SB in F125W, F140W, and
F160W as a function of Galactic Latitude and Ecliptic Latitude. In these
Figures, we simply attempt to find the minimum sky-SB signal in the darkest
parts of the sky.

For example, in Fig.~\ref{fig:f125gal}a the sample of WFC3/IR sky-SB
measurements is first plotted vs. Galactic Latitude to find and exclude the
regions with significant DGL. Fig.~\ref{fig:f125ecl_all} and Fig.~\ref{fig:f125ecl_sub}
then plot the sky-SB vs. Ecliptic Latitude to find in this subset the
regions with the lowest LES values of all images in each \bEcl-bin.
Next, Fig.~\ref{fig:f125gal}b
plots the predictions of the 1.25 \mum\ sky-SB for all HST locations in
the sky {\it and} at the same Sun Angles at the time of the HST observations as
provided by the Zodiacal COBE/DIRBE model of \citet{Kelsall1998}. \del{To bring HST and COBE onto exactly the same 1.25 \mum\
flux scale, we multiply the predicted COBE 1.25 \mum\ sky-SB values by 1.0056.}
Given the large range in sky-SB values, and the fact that most of the relevant
information is at the low-end of the SB-range in all these Figures, the bottom
panels in Fig.~\ref{fig:f125gal}cd provide enlargements of the
top panels in Fig.~\ref{fig:f125gal}ab. 

The WFC3/IR ZPs used in the F125W, F140W, and F160W filters are 26.232, 26.450,
25.936 AB-mag, respectively, for an object with 1.000 \eminpixsec.
Fig.~\ref{fig:filtcomparison} shows the HST WFC3/IR F125W~\footnote{\footnotesize{
		\url{https://www.stsci.edu/hst/instrumentation/wfc3/data-analysis/photometric-calibration},\\
		\url{https://www.stsci.edu/hst/instrumentation/wfc3/data-analysis/photometric-calibration/ir-photometric-calibration},
		see also \url{http://svo2.cab.inta-csic.es/svo/theory/fps3/index.php}}} and the
COBE/DIRBE J-band total system 
responses~\footnote{\url{https://lambda.gsfc.nasa.gov/product/cobe/dirbe_ancil_sr_get.cfm},\
	\ \url{https://lambda.gsfc.nasa.gov/product/cobe/c_spectral_res.cfm}\\ and
	Section 2.2.2.3 and Fig. 2.2-2 of 
	\url{https://lambda.gsfc.nasa.gov/product/cobe/dirbe_exsup.cfm}} compared to
the Solar spectrum in \Fnu\ \citep[\eg][]{Arvesen1969}~\footnote{see also
	\url{https://www.nrel.gov/grid/solar-resource/spectra-astm-e490.html}}, which
is fairly flat across both these filters. From this, we calculate that for a
Solar type spectrum like the ZL that the $\Delta$(HST
data--Kelsall COBE/DIRBE model) flux is --0.0061 AB-mag due to the small J-band
filter differences. This was calculated three independent ways: using 
integration in $F_{\lambda}$, {\it pysynphot}, and black body interpolation
between the two very similar filters, resulting in a scaling factor of
HST/Kelsall = 1.00557$\pm$0.0008. That is, for an SED with a Zodiacal spectrum,
the HST 1.25 \mum\ fluxes will be $\sim$0.56\% brighter than in the COBE/DIRBE
J-band filter. Hence, we will multiply the 
\citet{Kelsall1998} model predictions, which are based on COBE/DIRBE
observations, by 1.00557 to bring them onto exactly the same J-band flux scale
as the HST WFC3/IR F125W filter for a Solar type spectrum. ZL model predictions
for the HST WFC3/IR F140W and F160W filters were derived by interpolation
between the \citet{Kelsall1998} COBE/DIRBE J-band and K-band predictions using
the slope of the slightly reddened near-IR Zodiacal spectrum of
\citet{Aldering2001}, with uncertainties that include the errors in the
\citet{Kelsall1998} model. While HST and COBE are at different orbits, MSISE-90 Upper
Atmospheric models of the
Earth~\footnote{\url{http://www.braeunig.us/space/atmos.htm}} list that
the mean atmospheric pressure is 2.27$\times$10$^{-7}$ Pa at 540 km and
1.04$\times$10$^{-8}$ Pa at 885 km, so it is unlikely that the differences in altitudes between HST and COBE
contribute significantly to systematic differences in sky-SB levels between the two missions.

\n\begin{figure*}[!hptb]
	\n \cl{
		\includegraphics[width=0.500\txw]{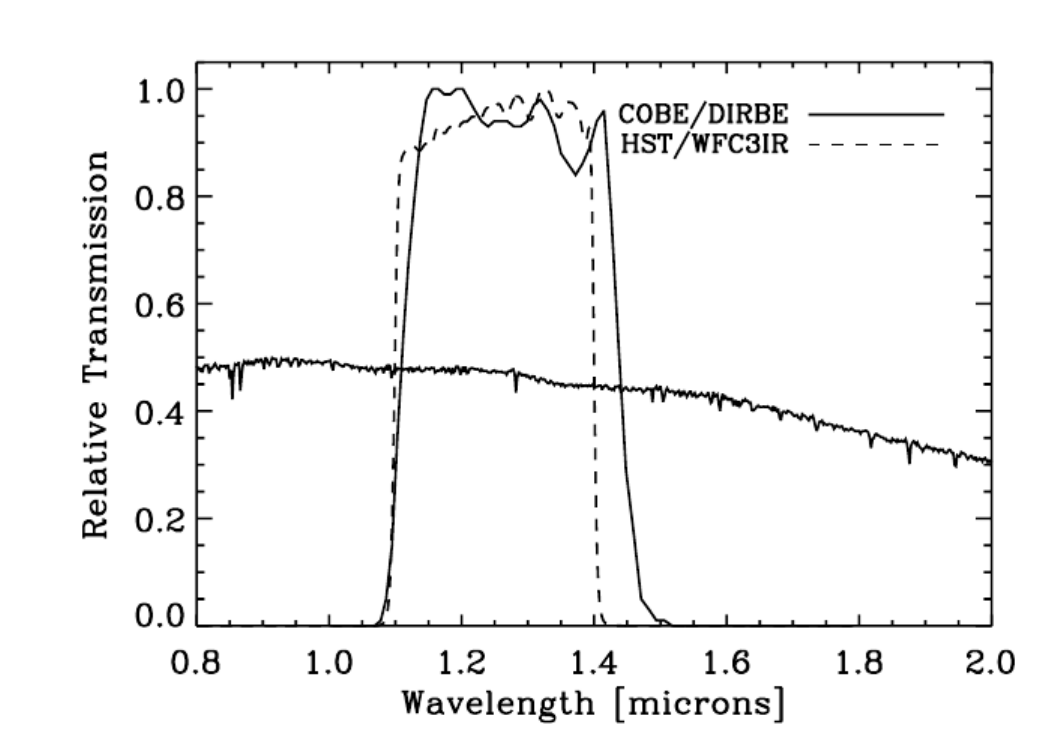}
	}
	
	\n \caption{
		A comparison of the HST F125W and COBE/DIRBE J-band filters, and the to Solar spectrum in \Fnu,
		which is fairly flat across these filters. For a Zodiacal spectrum, the
		expected J-filter flux difference between HST WFC3/IR measurements and the
		COBE/DIRBE models due to small differences in their respective filter 
		efficiency curves corresponds to --0.0061 AB-mag. We apply this ZP difference
		to our diffuse light estimates in the Tables of \S\ \ref{sec:hstkelsallresults}. Note that the
		Geocoronal 1.083 \mum\ He II line is essentially suppressed in the wing of
		both filters.}
	\label{fig:filtcomparison}
\end{figure*}

 Because of the 
$\sim$60\degree\ inclination of the Galactic plane with respect to the
Ecliptic, the darkest sky-SB occurs for 20\degree\cle $\vert$$b^{II}$$\vert$\cle
60\degree\ and {\it not} at the Galactic poles. Fields with
$\vert$$b^{II}$$\vert$\cle 20\degree\ have significant DGL, and are ignored in the analysis of \S\ 
\ref{sec:hstkelsallresults}--\ref{sec:results_diffuselim}.
Fig.~\ref{fig:f125ecl_all} shows all HST WFC3/IR F125W, F140W, and
F160W sky-SB measurements as in Fig.~\ref{fig:f125gal}, but now
plotted vs. {\it Ecliptic Latitude}. The orange and blue {\it sech} functions
and their error wedges outline the dimmest 1\% of the sky-SB measurements as
described below. Fig.~\ref{fig:f125ecl_sub} show the SKYSURF F125W,
F140W, and F160W sky-SB values vs. {\it Ecliptic Latitude} as in
Fig.~\ref{fig:f125ecl_all}, but {\it only for the darkest Galactic
regions with 20\degree\cle $\vert$$b^{II}$$\vert$\cle 60\degree} as selected
from Fig.~\ref{fig:f125gal}. 

Natural fits to galaxy disks seen edge-on are $sech$($z$) functions
\citep[\eg][]{vanderKruit1988, deGrijs1997}, written as SB in AB-mag vs.
vertical distance $z$ from the edge-on disk's central plane: 

\begin{equation} 
SB\ =\ a_4\ -2.5\ log\ [a_1\ sech\ (z/a_2)\ +\ a_3].
\label{eqn:sech} 
\end{equation}

\n According to these authors, the $sech$ model provides a better fit to the 
{\it vertical or} $z$-direction SB-distribution of flattened or ellipsoidal 
light-distributions seen edge-on than cosine, Gaussian, exponential, single, or
squared hyperbolic secant functions. The IPD cloud has a number of modeled
components that \cite{Kelsall1998} identify as ``Cloud'', ``Bands'', and
``Ring'' around the Sun, within which the Earth orbits. These Zodiacal
components have a ratio of their size in the Ecliptic plane to their vertical
Ecliptic Height of approximately 4:1, \ie\ a rather flattened or ``edge-on''
distribution as viewed from the Earth. As we will see, $sech$-functions
describe the vertical ZL distribution as a function of Ecliptic Latitude {\it 
as observed from the Earth} remarkably well. 

Inspired by the work that resulted in Eq.~\ref{eqn:sech}, we will use $sech$-type
functions to describe the LFS as a function of Ecliptic latitude \bEcl. While the actual dependence of ZL brightness with \bEcl may be more complicated than Eq.~\ref{eqn:sech} in reality (notably having a significant Sun Angle dependence as discussed below), we find that Eq.~\ref{eqn:sech} is a good description of the dimmest 1\% of the sky-SB 
values for {\it both} the HST sky-SB measurements {\it and} the
\cite{Kelsall1998} model predictions. Furthermore, this fitting procedure allows us to focus on the lower envelope of measurements, which we assume are the least affected by
straylight. By repeating the same fitting procedure on the lowest 1\% of the
\cite{Kelsall1998} model predictions, which predict the ZL brightness for 
{\it the same direction and at the same time of the year} as the
HST sky-SB measurements, we can search for any systematic \emph{offset} between HST measurements and the \cite{Kelsall1998} predictions. This offset could be, an additional unrecognized
Thermal Dark component (\S\ \ref{sec:thermal}), a dim spherical or mostly spheroidal
Zodiacal component not present in the model, or a dim spherical diffuse EBL
component, or some combination of these possibilities. 

In the case of HST F125W, F140W, and F160W sky-SB measurements, we use the
following $sech$ functions that are simpler than Eq.~\ref{eqn:sech} and {\it
linear} in flux density to represent the {\it lowest} 1\% envelope of {
both} the HST data {\it and} the \cite{Kelsall1998} models in
Fig.~\ref{fig:f125ecl_all} \& \ref{fig:f125ecl_sub}. The LFS of the HST data is best 
represented by: 

\begin{equation} 
LFS\ (HST)\ = a_1(HST)\ sech\ (b_{Ecl}\ /\ a_2(HST))\ +\ a_3(HST)\ \ \ 
[MJy/sr], 
\label{eqn:sech_hst} 
\end{equation}

\n while the lowest 1\% envelope of the COBE model predictions by
\citet{Kelsall1998} is best represented by:

\begin{equation} 
LFS\ (Kelsall)\ = a_1(Kel)\ sech\ (b_{Ecl}\ /\ a_2(Kel))\ +\ a_3(Kel)\ \
\ [MJy/sr].
\label{eqn:sech_kel} 
\end{equation}

\n Here, $a_3$ is the plateau value that the $sech$ function attains when
\bEcl\ reaches $\pm$$\infty$. Next, $a_1$ is a constant that captures the
maximum vertical amplitude that the $sech$ function reaches at \bEcl=0\degree\ 
{\it above} this plateau. Last, $a_2$$\simeq$19.5\degree\ measures the
effective thickness of the Zodiacal disk (or ``vertical scale
height'') as seen edge-on from HST. Coefficient $a_4$ in Eq.~\ref{eqn:sech} is
a constant that converts the SB in MJy/sr to AB \magarc, and is not used in the
linear flux density representation of Eq.~\ref{eqn:sech_hst}--\ref{eqn:sech_kel}. The best estimate parameters of the
$sech$ constants $a_1$, $a_2$, and $a_3$ are given in Table~\ref{tab:sech_vals} for
both the {\it lower} envelope to the HST data and the Kelsall models at
1.25--1.6 \mum. The upper and lower $sech$ envelope $a_2$ values are best determined from F160W measurements, which
have the best statistics, so we adopt the same $a_2$ values and their
errors for the F125W and F140W filters in Table~\ref{tab:sech_vals}, which seem to
bound the \citet{Kelsall1998} model predictions well for the F125W and F140W measurements. These $sech$ functions are indicated by the bottom orange and
blue lines plus their uncertainty wedges in
Fig.~\ref{fig:f125ecl_all}-\ref{fig:f125ecl_sub}, respectively. The main result we are
after in Table~\ref{tab:sech_vals} is the (boldfaced) difference in the bottom
envelopes (or $a_3$-values) between the HST data and the ZL models\footnote{The restriction of our data to $\pm90\degree$ means that the derivative of the model is not continuous at the ecliptic poles. However, the difference between the value at $90\degree$ and $\infty$ is $<2\%$ for our fits, and this detail does not affect our fitting procedure regardless.}. Because the best fit $a_1$
and $a_2$ values turn out to be very similar in Table~\ref{tab:sech_vals} for both
the HST data and the ZL models, we adopt the differences in $a_3$-values as a direct measure of the HST-ZL model
differences. 

The first four lines of Table~\ref{tab:sech_vals} also list the same $a_1$--$a_3$
parameters (and their estimated uncertainties) for the {\it upper} envelope to
the Kelsall models in the right-most panels, and for the HST data in the
left-most sub-panels of Fig.~\ref{fig:f125ecl_all}-\ref{fig:f125ecl_sub} (upper blue and
orange dashed lines, respectively). The $sech$ upper envelope to the 
\citet{Kelsall1998} models was directly estimated from the predictions in
Fig.~\ref{fig:f125ecl_all}-\ref{fig:f125ecl_sub}, which show a very good empirical 
$sech$-type fit to the upper envelope of the \citet{Kelsall1998} model values.

The amplitude of the {\it upper} envelope to the HST data was scaled upward
using the (HST--Kelsall) difference from the {\it lower} envelopes in
Fig.~\ref{fig:f125ecl_all} and Table~\ref{tab:sech_vals}. The orange
dashed lines indicating the {\it upper} envelopes to the HST data in
Fig.~\ref{fig:f125ecl_all}a thus provide another way to identify
HST exposures with excessive sky-SB, which could be due to several reasons: (a)
targets with higher DGL; (b) large nearby galaxy
targets, such as the LMC or M31; or (c) exposures with higher straylight 
levels, including those that got too close to the Earth's limb.
The presence of such images is most noticeable in the F160W filter.


Fig.~\ref{fig:f125_skydiff} shows a comparison of SKYSURF's F125W,
F140W, and F160W sky-SB measurements from the HST data to the Kelsall
COBE/DIRBE models as a function of Ecliptic Latitude (the top sub-panels show all data, and bottom sub-panels only for the
darkest Galactic regions at 20\degree\cle $\vert$$b^{II}$$\vert$\cle 60\degree\
as selected from Fig.~\ref{fig:f125gal}). The left sub-panels give
the HST/Kelsall model flux density {\it ratio}, while the right sub-panels give the
linear flux density {\it difference} between the HST data and the Kelsall
COBE/DIRBE models for the same subsample. In the top sub-panels of
Fig.~\ref{fig:f125_skydiff}, the orange {\it sech} functions in
Eq.~\ref{eqn:sech_hst}--\ref{eqn:sech_kel}, and their error wedges outline the darkest 
sky-SB measurements from Fig.~\ref{fig:f125ecl_all}. The bottom
sub-panels of Fig.~\ref{fig:f125_skydiff} give enlargements of the top
sub-panels, and show a significant Ecliptic Latitude dependence of the HST/Kelsall
model {\it flux ratios}, suggesting that the difference between the bottom
envelopes of the HST data and the Kelsall models are {\it not} due to a flux
density scale issue. 

The green wedges in the bottom right panels of
Fig.~\ref{fig:f125_skydiff} indicate our best estimate of the
$\Delta$(HST--Kelsall) offsets. For each filter, these {\it linear} flux density
differences between the bottom envelopes of the HST data and the Kelsall models
are fairly constant for $\vert$$b^{II}$$\vert$\cge 20\degree\ and well above
zero, suggesting {\it a somewhat wavelength-dependent constant linear offset}
between the bottom envelopes of the HST data and the Kelsall models.
For $\vert$$b^{II}$$\vert$\cle 20\degree, the differences between the data and model
have more scatter, suggesting that complex and subtle adjustments to the Kelsall model in the 
ecliptic plane may be required.
We thus
discard all data with $\vert$$b^{II}$$\vert$\cle 20\degree\ to estimate the LFS
difference between the HST data and Kelsall models. 

The LFS values from Fig.~\ref{fig:f125gal} are summarized
in Table~\ref{tab:sech_vals}. For example, Table \ref{tab:sech_vals} shows that the {\it 
plateau value $a_3$} of the $sech$ function in Eq.~\ref{eqn:sech_hst}--\ref{eqn:sech_kel}
that best captures the LFS values at high Ecliptic Latitudes in the F125W
filter amounts to $a_3$(HST) = +0.108$\pm$0.005 MJy/sr, which best fits the
lowest $\sim$1\% of the sky-SB values, while for COBE/DIRBE model predictions for 
the {\it same} sky pointings and filters, observing day of the year, {\it and}
Sun Angles, the \citet{Kelsall1998} model predicts a lowest $\sim$1\% envelope
with $sech$ parameter $a_3$(COBE) = +0.093$\pm$0.006 MJy/sr. {\it The most
likely HST--Kelsall difference from Fig.~\ref{fig:f125_skydiff}d is thus
$\sim$(0.108--0.093*1.0056)$\simeq$+0.0145$\pm$0.008 MJy/sr,} which includes
the correction for the --0.0061 mag ZP difference between the HST F125W and
COBE/DIRBE J-band flux scales. Similar but somewhat larger
values are listed in Table~\ref{tab:sech_vals} for the F140W and F160W filters,
where the \citet{Kelsall1998} models were interpolated between the COBE/DIRBE
predictions at 1.25 and 2.2 \mum\ following the discussion in \S\ \ref{sec:interpolate}.
This interpolation also results in somewhat larger $a_3$ errors for the lower
envelope to the \citet{Kelsall1998} model predictions in the F140W and F160W
filters in Table~\ref{tab:sech_vals} (see \S \ref{sec:interpolate}), and in somewhat larger
errors of $\sim$0.009 MJy/sr in the F140W and F160W HST--Kelsall difference
signal listed in Table~\ref{tab:sech_vals}. 


\begin{figure*}
	\begin{tabular}{cc}
		\includegraphics[width=0.48\txw]{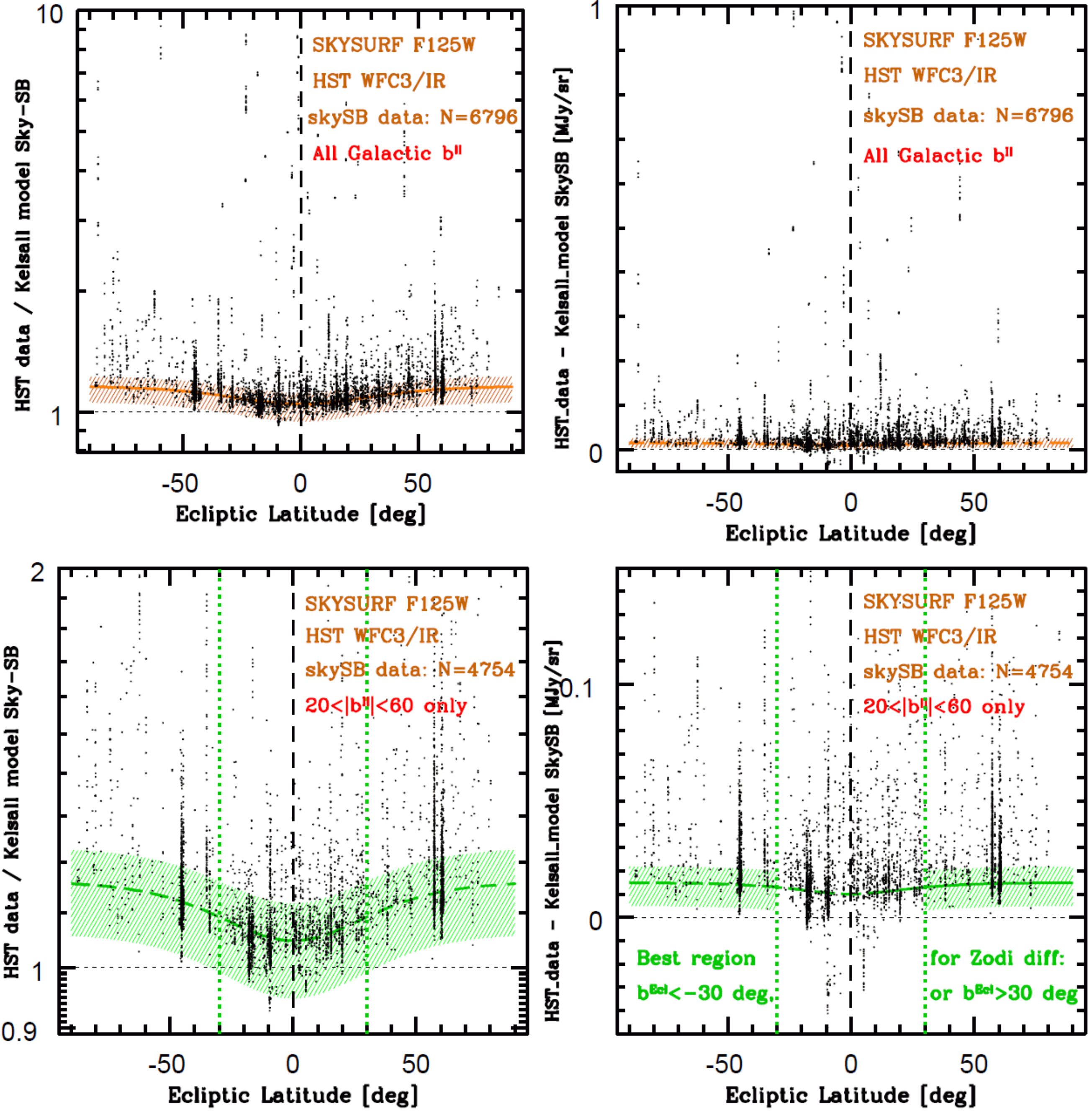} & \includegraphics[width=0.48\txw]{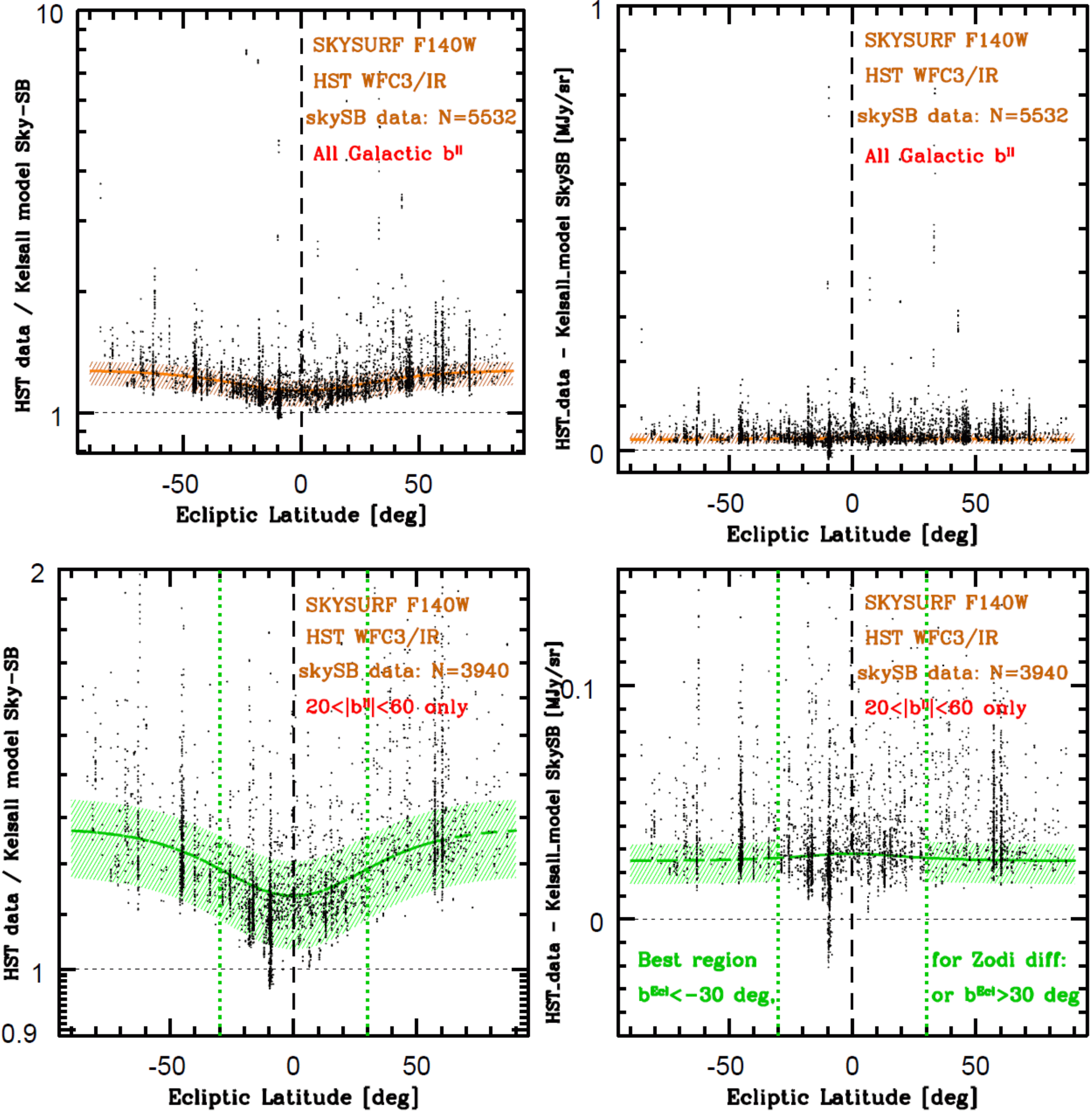}\\
		\includegraphics[width=0.48\txw]{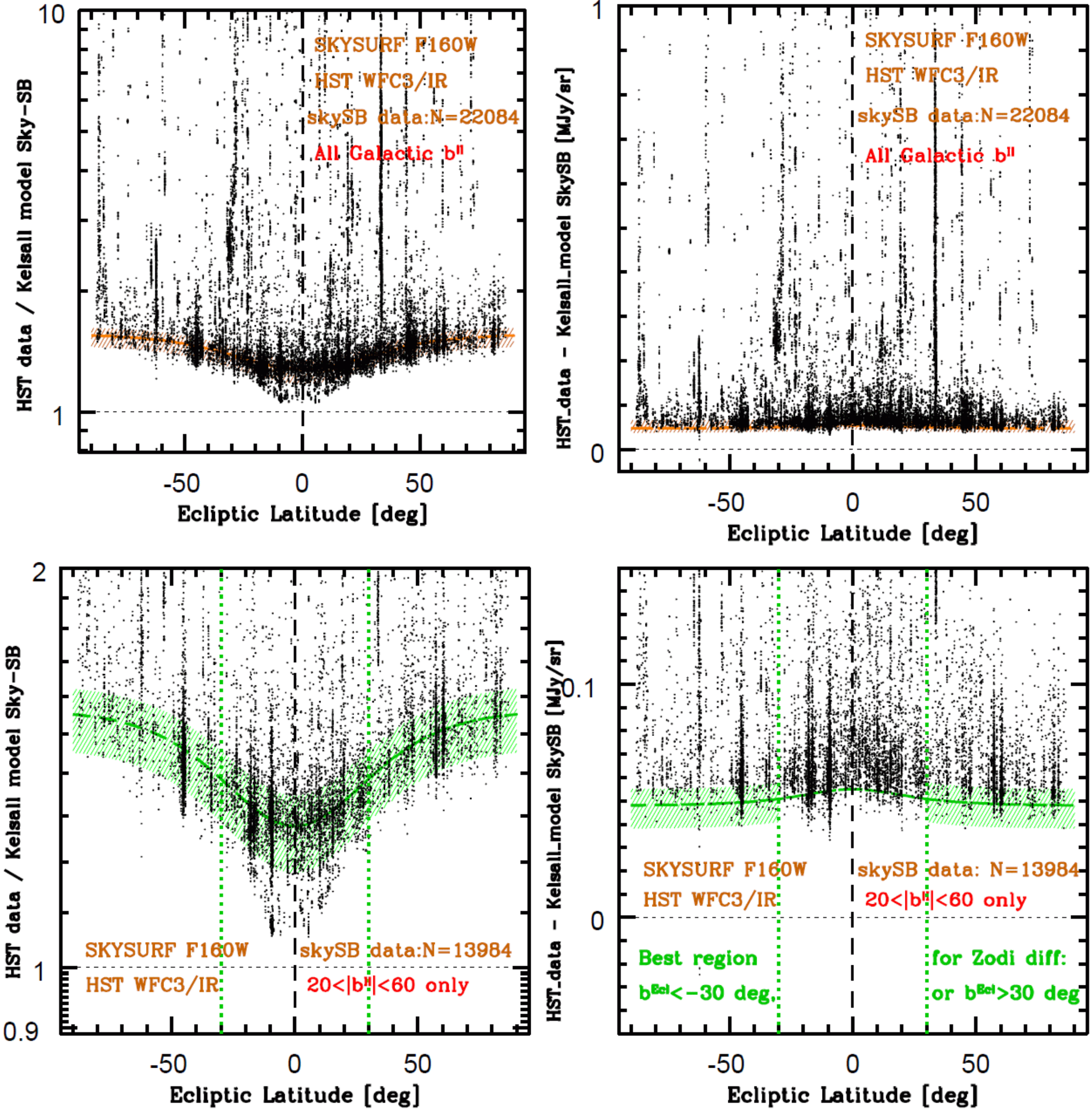}
	\end{tabular}
	\vspace*{+0.20cm}
	\n \caption{
		SKYSURF F125W, F140W, and F160W sky-SB measurements from the HST data compared to 
		the Kelsall COBE/DIRBE models [in Myr/sr] vs. Ecliptic Latitude \bEcl. Data in the bottom sub-panels is restricted to the darkest Galactic regions with 20\degree\cle
		$\vert$$b^{II}$$\vert$\cle 60\degree\ (see
		Fig.~\ref{fig:f125gal}) and zoomed into the most relevant range. Left sub-panels give the HST F125W/Kelsall
		1.25 \mum\ model flux density {\it ratio}, while the right sub-panels give the
		linear flux {\it difference} between the HST data and the Kelsall COBE/DIRBE
		models for the same subsample. The orange $sech$ functions and error wedges
		outline the darkest sky-SB measurements from
		Fig.~\ref{fig:f125ecl_all}. The green wedges in the bottom sub-panels illustrate the HST-Kelsall difference between sech fits from Figure~\ref{fig:f125ecl_sub}, highlighting our best estimate of
		the $\Delta$(HST data -- Kelsall COBE/DIRBE model) of 
		0.0145 $\pm$0.007 MJy/sr for F125W, 0.025 $\pm$0.010 MJy/sr for F140W, and  0.048 $\pm$0.010 for F160W (Table~\ref{tab:sech_vals}). The bottom right sub-panels best capture the
		{\it linear} SB-difference between the HST
		sky-SB observations and the COBE/DIRBE Kelsall 1.25 \mum\ model (magnify the PDF figure as needed to see the difference).} 
	\label{fig:f125_skydiff}
\end{figure*}

\n \subsection{Interpolating the Zodiacal Spectrum for F140W and F160W Observations}
\label{sec:interpolate}
\n To estimate Kelsall model predictions for F140W and F160W observations, as
well as the thermal modeling described in \S \ref{sec:results_thermal}, a model of how 
the Zodiacal spectrum behaves in the near-IR is necessary.
For short-wavelength IR images (up to 2.2 \mum), the sky-SB SED
closely resembles a power law in the form of:

\begin{equation} 
log(F_\lambda)\ =\ -17.755 - \alpha\ (\lambda - 0.61 \mum)\ \ \ \
[erg/cm^2/s/\AA/arcsec^2] 
\label{eqn:aldering} 
\end{equation}

\n following \citet{Aldering2001}, who adopted a power-law slope $\alpha$=0.730
for wavelengths 0.61\cle $\lambda$\cle 2.20 \mum. Hence, in our
analysis, we will use Eq.~\ref{eqn:aldering} to represent the Zodiacal spectrum for
0.61\cle $\lambda$\cle 2.20 \mum. Fig.~\ref{fig:thermfig}a shows the spectral
index distribution N($\alpha$) when interpolating the \citet{Kelsall1998}
Zodiacal sky-SB prediction in the COBE/DIRBE J and K-band filters for all HST pointings in
the F160W filter (which is very similar to the distribution of slopes for all HST pointings in the F140W filter).
The resulting median spectral index and its 1$\sigma$ range is
$\alpha$=0.713$\pm$0.023, consistent with the value adopted by
\citet{Aldering2001}'s power-law approximation of Eq.~\ref{eqn:aldering} to within
the error. We verified through numerical integration that the power-law
interpolation in Eq.~\ref{eqn:aldering} produces a \cle 2\% error in the prediction
of the reddened Zodiacal spectrum at 1.4--1.6 \mum\ wavelengths, compared to
the \citet{Kelsall1998} model that was fit to the COBE/DIRBE 1.25 and 2.2 \mum\
data and interpolated to 1.4--1.6 \mum. This is folded into the error budget of
Table \ref{tab:sech_vals}, resulting in somewhat larger $a_3$ errors for the lower
envelope to the \citet{Kelsall1998} model predictions in the F140W and F160W
filters. 

\n \subsection{Assessment of the WFC3/IR Thermal Dark Signal Levels} 
\label{sec:thermal}

\sn Possibly the most significant source of uncertainty regarding our
measurement of the near-IR diffuse light is the level of WFC3 Thermal Dark signal.
Based on onboard temperature measurements and emissivity calculations, the
WFC3 IHB lists the IR Thermal Dark signal levels as 0.052 \eminpixsec, 0.070
\eminpixsec, and 0.134 \eminpixsec\ for the F125W, F140W, and F160W filters,
respectively \citep{Dressel2021}. However, modest changes in HST component 
temperatures ($\pm$2.5 K) can impact the TD signal at a level comparable to
the diffuse signal. For example, Fig.~\ref{fig:thermfig}b shows how much changing the
overall telescope temperature can affect the TD signal. A sequel paper (Carleton \etal 2022b, in preparation), will
explore the TD signal as a function of orbital phase and HST component
temperatures in more detail. Here, we show a preliminary analysis constraining
the TD signal in SKYSURF data by fitting the spectral energy distribution (SED)
of the near-IR sky with a Zodiacal component and a temperature-dependent
thermal signal. 

We queried the HST archive for IR images that were taken of the {\it same target
within two days of each other}, such that the overall Zodiacal sky-SB level does
not change substantially. We further identified image sets where at least one
image was in the WFC3/IR F125W filter, and another in either the F098M, F105W,
F110W, F125W, F127M, F139M, F140W, F159M, and/or the F160W filter.\deleted{The
images were corrected for dark current variations described in
\citet{Sunnquist17b} by subtracting the difference between the predicted dark
current level (from their model) and the dark level used in the pipeline.} We
then ran the adjusted calibration program for the individual WFC3/IR ramps, as
described in \citetalias{Windhorst2022}, and measured the
minimum sky-SB levels in these images. Based on the orbital phase-dependent 
straylight constraints in Fig.~10 of \citetalias{Windhorst2022}, we only
selected those WFC3/IR exposures in the above filters that have minimal stray
light in order to better estimate the most likely TD levels. This resulted in a
sample of over 500 useful images in these filter pairs, predominantly from the
BORG pure-parallel program PID 12572 (PI: M. Trenti). By dividing the sky
value in each filter's image by the sky in the associated F125W filter taken in
that same direction, we construct a spectral energy distribution of the
Zodiacal sky. 

The sky-SB levels in the F140W and F160W filters can be significantly elevated
due to the foreground Thermal Dark signal. To model this thermal signal, we
use the $pysynphot$\footnote{\url{https://pysynphot.readthedocs.io/en/latest/index.html}} package, modeling each component in the
optical path as a blackbody with an effective temperature and emissivity. The
fiducial temperatures and emissivities are taken from the HST 
database~\footnote{\url{https://hst-crds.stsci.edu/}}. Using these fiducial
temperatures and emissivities, the $synphot$ model recovers the published TD
values. Subtracting this TD signal from the F140W and F160W sky values makes
them match the power-law in Eq. \ref{eqn:aldering} better. However, it is unclear if
the fiducial temperatures are the ones that best fit all available HST data. To
identify the HST temperatures that best fit the data --- which we take as more
accurately reflecting the real HST temperatures producing the Thermal Dark
signal --- we take the given effective temperatures as free parameters and
allow them to vary as:

\begin{equation} 
T\ =\ T_{ref}\ +\ \Delta T\ 
\label{eqn:deltat} 
\end{equation}

\n where \Tref\ is the ambient temperature of components listed in the HST
references files, and $\Delta$T is a parameter describing the average change 
in temperature (compared to \Tref) of the HST components that is most consistent
with the data below. Note that small values of $\Delta$T consistent with onboard
measurements can alter the TD signal significantly, especially in the F160W
filter, and thereby affect the values of any inferred diffuse light levels:
e.g., a $\pm$1 K change in temperature corresponds to a $\pm$0.04 MJy/sr change
in the Thermal Dark signal level in F160W. For the above WFC3/IR filter pairs,
we define the goodness of fit as:

\begin{equation}
\chi^2=\frac{[{\rm Sky_i}/{\rm Sky_j}{\rm (obs)}-{\rm Sky_i}/{\rm Sky_j}{\rm
(model)}]^2}{[\sigma_{\rm i}^2/{\rm Sky_j}^2+\sigma_{\rm j}^2{\rm Sky_i^2}/{\rm
Sky_j^4}]},
\label{eqn:thermalchi2}
\end{equation}

\n where $\sigma$ is the error in the sky-SB measurements, index $j$ indicates
the F125W filter and $i$ any of the other available WFC3/IR filters that
paired-up with a given F125W observation within two days. Next, we find the
best fit model by minimizing $\chi^2$.  The F105W and F110W
exposures with F105W/F125W and F110W/F125W ratios \cge 1.20 in
Fig.~\ref{fig:thermfig}cd were not used, since they may have significant
Geocoronal He II line emission at 1.083 \mum\ that can elevate their sky-SB. Using the data described above, we
obtain a formal best fit of $\Delta$T$\simeq$+1.52 K for the
\citet{Aldering2001} Zodiacal power-law slope of $\alpha$=0.73
(Fig.~\ref{fig:thermfig}c). If the slope $\alpha$ is allowed to vary as well, we
can obtain temperatures as low as $\Delta$T=--1.62 K for a slope of
$\alpha$=0.65 (Fig.~\ref{fig:thermfig}d). Hence, the best fit $\Delta$T and
$\alpha$ are correlated such that somewhat larger $\Delta$T values imply a
warmer telescope and therefore larger $synphot$ TD-values primarily in the
longer wavelength WFC3/IR filters, which --- when subtracted from the above
data --- imply a Zodiacal spectrum with a somewhat steeper power-law slope in
Fig.~\ref{fig:thermfig}a--\ref{fig:thermfig}d. The best $\chi^2$ fit occurs for
$\alpha$=0.66 and $\Delta$T = --1.15 K, which we adopt in the Tables of \S\
\ref{sec:results_diffuselim} as our nominal TD case. {Non-linearities in the 
Zodiacal spectrum have a relatively small impact on the implied thermal background.
For example, adding a $\sim7\%$ bump in the spectrum from $1.4-1.6$ microns, 
similar to what is seen in \cite{Matsuura2017}, changes the best fit slope to $0.69$, 
and the $\Delta$T to --2.72K (which is consistent with our estimated uncertainties of $\sim2K$).}

The results are shown in Fig.~\ref{fig:thermfig}c--\ref{fig:thermfig}d for this
range of $\Delta$T and $\alpha$-values, with their associated range in Thermal
Dark signal values given in Fig.~\ref{fig:thermfig}b.
The cases shown in Fig.~\ref{fig:thermfig}b--\ref{fig:thermfig}d
bracket the likely range in telescope ambient temperature values (Appendix
\ref{sec:thermalappendix}). This results in a plausible range of F125W--F160W Thermal Dark
signal values, with the most plausible ones subtracted from any diffuse sky-SB
signal in \S\ \ref{sec:results_diffuselim}. The error range resulting from the TD signal
predictions is summarized in Fig.~\ref{fig:eblfig_zoom} and bracket the range of
$\Delta$T temperature variations that the above analysis implies (see \S\
\ref{sec:results_diffuselim}). 


\n \subsection{Implications for Limits on Diffuse Light at 1.25--1.6 \mum}
\label{sec:results_diffuselim}

\sn In Fig.~\ref{fig:eblfig1} and Fig.~\ref{fig:eblfig_zoom}, we compute and plot our
limits to any diffuse light at 1.25-1.6 \mum\ as follows. Summarizing
Fig.~\ref{fig:f125_skydiff}, Table~\ref{tab:sech_vals} suggests average
offsets of the HST LFS-values minus the \citet{Kelsall1998} COBE/DIRBE model
predictions of 0.0145, 0.025, and 0.048 MJy/sr at the effective
wavelengths of the F125W, F140W, and F160W filters, respectively. Below we
will convert these differences to our limits on diffuse light. 

\n \subsubsection{The HST WFC3/IR Sky-SB Corrected for Thermal Dark Signal}
\label{sec:results_thermal}

\sn First, we need to subtract the true WFC3/IR Thermal Dark signal, which has
not yet been subtracted in any of the processing. Here, we cannot simply use
the F125W thermal foreground of 0.052 \eminpixsec\ from Table 7.11 in the WFC3
IHB \citep{Dressel2021}, as it is larger than our 1.25 \mum\ SB upper limit. 
The reason that the IHB thermal foreground is higher is that it includes a
modeled Thermal Dark signal from the instrument housing, which is subtracted
during dark-frame removal. All SKYSURF's WFC3/IR images have been dark-frame
subtracted, and so our modeled Thermal Dark signal values do not contain the
instrument housing contribution. The Thermal Dark signals predicted with
{\it synphot} (in units of \eminpixsec) for the plausible range in the temperatures of the
HST optical and instrument components across a typical orbit are listed in the first set of three columns of
Table~\ref{tab:thermal} for the F125W, F140W, and F160W filters, respectively.
With the WFC3/IR pixel scale and zeropoints of Sec. 4 of \citetalias{Windhorst2022}, these are
converted to equivalent sky-SB values in units of MJy/sr and \nWsqmsr. The
conversion factors needed for these calculations are also given in the footnotes of
Table~\ref{tab:sech_vals}--Table~\ref{tab:sa90_vals}. These TD values are subtracted from
the net HST data--Kelsall model differences listed in boldface on the bottom
line of Table~\ref{tab:sech_vals}, which are repeated on the top line of
Table~\ref{tab:thermal}. 

To give a specific example, for the {\it nominal} temperature difference of
$\Delta$T = (T--\Tref) = --1.15 K (\S\ \ref{sec:thermal}), the Thermal Dark value in
the F125W filter is predicted to be 0.00399 \eminpixsec, which corresponds to
0.00123 MJy/sr. This value is subtracted from the HST--Kelsall difference of
0.0145 MJy/sr in F125W listed in Table~\ref{tab:sech_vals} to arrive at the net
signal of 0.0133 MJy/sr listed in Table~\ref{tab:thermal} (2nd column for F125W)
or 32.1 \nWsqmsr\ (3rd column for F125W). To be conservative, we quote the
values derived in the 3rd column for each filter in Table~\ref{tab:thermal} (in
\nWsqmsr) as {\it upper limits}, given the uncertainties in the $\Delta$T to be
used for the TD subtraction, the absolute errors in the \citet{Kelsall1998} 
model (footnote $e$ of Table~\ref{tab:sech_vals}), as well as the uncertainties in
the {\it discrete} eEBL (\S\ \ref{sec:results_iebl}--\ref{sec:results_eebl}) and the DGL (\S\
\ref{sec:results_dgl}), which still need to be subtracted. 

\n \subsubsection{The iEBL Component Already Subtracted from the Diffuse Light
Limits} \label{sec:results_iebl}

\sn One of the strengths of the SKYSURF experiment is that it is very effective at removing discrete object light from our \emph{diffuse} EBL constraint. As discussed in \citetalias{Windhorst2022}, the median SKYSURF exposure is complete down to a limit of $\sim26$~\mAB, whereas most discrete extragalactic light comes from galaxies between $17-22$~\mAB. Here we describe the magnitude of this discrete object light for context with other diffuse EBL measurements. 

\sn The J-band sky-SB integral of detected objects over 40 flux bins from
AB=10 mag to AB=30 mag amounts to 1.396
$\times$ 10$^{-26}$ W/Hz/m$^2$/deg$^2$ when extrapolating the converging
integral to AB=$\infty$ following \citet{Driver2016b} (see also Fig. 2 of \citetalias{Windhorst2022}). Because the
sky-integral converges strongly for AB$>$22 mag, integrating to AB=$\infty$
only increases this sum by $\sim$0.7\% compared to when integrating to AB=30
mag. In the units of \nWsqmsr\ used in \citet{Driver2016b} and
Fig.~\ref{fig:eblfig1} here, this integral corresponds to a total sky-SB in the
F125W filter of:

\begin{equation} 
\nu . I_{\nu} \simeq 1.396 \times 10^{-26}\ W/Hz/m^{2}/deg^{2} \times 
10{^9}\ nW/W \times 2.4246 \times 10^{14}\ Hz \times\ 3282.8\ deg^{2}/sr
\simeq 11.11\ nW/m^{2}/sr
\label{eqn:iebl}
\end{equation}

\n Similarly, in \citetalias{Windhorst2022} we find that the F160W sky-SB integral of
objects detected to AB\cle 30 mag amounts to a total sky-SB of 1.813 $\times$
10$^{-26}$ W/Hz/m$^2$/deg$^2$ or 11.68 \nWsqmsr. 
 The fraction of these integrals
that comes from discrete objects detected to AB\cle 26.5 mag is 10.74 \nWsqmsr\
in the F125W filter, and 11.31 \nWsqmsr\ in the F160W filter, respectively.
Hence, to AB\cle 26.5 mag even the average shallow single HST/WFC3 exposures in
the F125W and F160W filters already resolve and detect \cge 96.6--96.8\% of the
total {\it discrete} EBL, respectively.

Many published {\it direct} EBL measurements --- or upper limits --- {\it do}
include the full {\it discrete} iEBL+eEBL signal above, since these methods
traditionally measure the total {\it diffuse+discrete} galaxy light. By the
nature of our SKYSURF methods, we have already removed {\it
almost all} of the {\it discrete} iEBL signal, except for the last 
$\sim$0.4--0.6 \nWsqmsr\ that comes from unresolved objects with AB\cge 26.5
mag (see \S\ \ref{sec:results_eebl}). Other {\it direct} EBL limits should
appear higher than our diffuse light limits in part because their values
include the {\it discrete} EBL signal of 11.11--11.68 \nWsqmsr\ at 1.25--1.6
\mum, while our SKYSURF method already has subtracted \cge
96.7\% of the {\it discrete} EBL signal from the typical 500 sec HST WFC3/IR
exposures.


\vspace{-0.00cm}
\begin{deluxetable*}{| l | ccc | ccc | ccc |}
\tablecolumns{10}
\tablewidth{1.0\linewidth}
\tablecaption{HST Data \& Kelsall Model: Zodiacal $Sech$ Parameters using the
LFS Method, and Sky-SB Differences
\label{tab:sech_vals}}
\tablehead{
\colhead{$\vert$\ Filter\ \ \ \ \ \ \ }        $\vert$ & 
\multicolumn{3}{c}{--- F125W / J-band$^a$ ---} $\vert$ & 
\multicolumn{3}{c}{--- F140W / JH-band ---}    $\vert$ & 
\multicolumn{3}{c}{--- F160W / H-band  ---}    $\vert$ \\[-4pt]
\colhead{$\vert$\ $Sech$-\ \ \ \ \ \ }         $\vert$ & 
\colhead{$a_1$}                                        &
\colhead{$a_2$}                                        &
\colhead{$a_3$}                                $\vert$ &
\colhead{$a_1$}                                        &
\colhead{$a_2$}                                        &
\colhead{$a_3$}                                $\vert$ &
\colhead{$a_1$}                                        &
\colhead{$a_2$}                                        &
\colhead{$a_3$}                                $\vert$ \\[-4pt]
\colhead{$\vert$\ parameter\ }                 $\vert$ & 
\colhead{(MJy/sr)}                                     &
\colhead{(\degree)}                                    &
\colhead{(MJy/sr)}                             $\vert$ &
\colhead{(MJy/sr)}                                     &
\colhead{(\degree)}                                    &
\colhead{(MJy/sr)}                             $\vert$ &
\colhead{(MJy/sr)}                                     &
\colhead{(\degree)}                                    &
\colhead{(MJy/sr)}                             $\vert$ 
}
\startdata
              &         &         &             &       &       &             &       &       &            \\
HST upper     &0.838    &17.5$^c$ &0.125        &0.848  &17.5   &0.133        &0.853  &17.5   &0.155       \\
              &($^b$)   &    &(0.005)      &       &  &(0.005)      &       &  &(0.005)     \\
Kelsall upper &0.846    &17.5     &0.110        &0.846  &17.5   &0.110        &0.846  &17.5   &0.110       \\
              &         &   &(0.007)      &       &  &(0.007)      &       &  &(0.007)     \\
Fig.$^d$     &         &         &[Fig.~\ref{fig:f125ecl_all}]    &       &       &[Fig.~\ref{fig:f125ecl_all}]    &       &       &[Fig.~\ref{fig:f125ecl_all}]   \\
              &         &         &             &       &       &             &       &       &            \\
HST lowest    &0.112    &19.5     &{\bf 0.108}  &0.120  &19.5   &{\bf 0.115}  &0.120  &19.5   &{\bf 0.133} \\
              &         &    &(0.005)      &       & &(0.005)      &       &  &(0.005)     \\
Kelsall lowest$^e$&0.117&19.5     &{\bf 0.093}  &0.117  &19.5   &{\bf 0.090}  &0.113  &19.5   &{\bf 0.085} \\
              &         &    &(0.006)      &       &  &(0.007)      &       &  &(0.007)     \\
Figs.$^d$     &         &         &[Fig.~\ref{fig:f125ecl_sub}]    &       &       &[Fig.~\ref{fig:f125ecl_sub}]    &       &       &[Fig.~\ref{fig:f125ecl_sub}]   \\
              &         &         &             &       &       &             &       &       &            \\
HST--Kelsall  &         &     &{\bf 0.0145}$^e$ &       &       &{\bf 0.025}  &       &       &{\bf 0.048} \\
LFS (MJy/sr)  &         &         &(0.008)      &       &       &(0.009)      &       &       &(0.009)     \\
Figs.$^d$     &         &         &[Fig.~\ref{fig:f125_skydiff}]   &       &       &[Fig.~\ref{fig:f125_skydiff}]   &       &       &[Fig.~\ref{fig:f125_skydiff}]  \\
              &         &         &             &       &       &             &       &       &            \\
HST--Kelsall  &         &         &{\bf 35.2}$^f$&      &       &{\bf 54.6}   &       &       &{\bf 94.2}  \\
(\nWsqmsr)    &         &         &(19)         &       &       &(19)         &       &       &(17)        \\
              &         &         &             &       &       &             &       &       &            \\
\enddata
\mn
\tablenotetext{a}{The effective central wavelengths used for the WFC3/IR F125W,
F140W, and F160W filters are $\lambda_c$ = 1.2364 \mum, 1.3735 \mum, and 1.5278
\mum, or central frequencies of 2.4248 $\times$ 10$^{14}$ Hz, 2.1827 $\times$ 
10$^{14}$ Hz, and 1.9622 $\times$ 10$^{14}$ Hz.}

\tablenotetext{b}{The second row of the $a_3$ parameter gives its estimated
errors in parentheses. The estimated errors in $a_1$ and $a_2$ from 
Eq.~\ref{eqn:sech_hst}--\ref{eqn:sech_kel} are not independent from the error in $a_3$, and
is of the same order. Hence, only the error in $a_3$ is listed, which is most
relevant for estimating the resulting diffuse sky-SB limits in the bottom five
rows.}

\tablenotetext{c}{The estimated values of $a_2$ are approximately the same for
all three filters F125W, F140W, and F160W for both the HST data and the Kelsall
models to within the errors (approximately $1$\degree), so the same value is adopted for all filters. The $a_2$ values are slightly narrower for the
{\it upper} envelope to the Kelsall models compared to the {\it lower}-bound 
$a_2$ values and were assumed to be equally narrow for the upper envelopes of
those HST data where the sky-SB was not enhanced by the Earth's limb.}

\tablenotetext{d}{[Between square brackets we list the Figure numbers, from 
which the $sech$ coefficients on the lines directly above were determined].}

\tablenotetext{e}{The \citet{Kelsall1998} COBE/DIRBE J-band model prediction
has been corrected for the --0.0061 mag ZP difference between the HST F125W and
COBE/DIRBE J-band flux scales. The ZL model predictions for the HST WFC3/IR
F140W and F160W filters were derived by interpolation between the
\citet{Kelsall1998} J-band and K-band predictions. The
errors in the (HST--Kelsall) differences in MJy/sr are derived in quadrature
from the $a_3$ fitting errors in the previous rows. \citet{Kelsall1998}
reported errors in their ZL model of 15 \nWsqmsr\ at 1.25 \mum\ and 6 \nWsqmsr\
at 2.2 \mum, respectively (see their Table 7). We propagate these also into the
errors of our adopted HST--Kelsall differences at 1.25--1.6 \mum\ in \nWsqmsr\
(bottom row; see also Table~\ref{tab:thermal}), which correspond to
$\sim$47--18\% errors in these differences at 1.25--1.6 \mum, respectively.}

\tablenotetext{f}{The units in these last two rows were converted from MJy/sr 
to \nWsqmsr, using multipliers of 2425, 2183, and 1962 (=10$^{-11}$ 
c/$\lambda_{c}$), respectively, yielding the upper limit to the total diffuse
light in \nWsqmsr.}

\end{deluxetable*}



\vspace{-0.00cm}
\begin{deluxetable*}{| l | l | ccc | ccc | ccc |}
\tablecolumns{11}
\tablewidth{1.0\linewidth}
\tablecaption{WFC3 Thermal Dark Signal, HST Data--Kelsall Model LFS Summary, 
and Diffuse Sky-SB Limits
\label{tab:thermal}}
\tablehead{
\colhead{$\vert$\ \ \ \ \ \ \ \ }               $\vert$ & 
\colhead{ }                                     $\vert$ & 
\multicolumn{3}{c}{--- F125W / J-band  ---}     $\vert$ & 
\multicolumn{3}{c}{--- F140W / JH-band ---}     $\vert$ & 
\multicolumn{3}{c}{--- F160W / H-band  ---}     $\vert$ \\[-4pt]
\colhead{$\vert$\ $\Delta$T$^b$\ }              $\vert$ & 
\colhead{$\alpha$$^c$ }                         $\vert$ & 
\colhead{TD$^d$ }                                       &
\multicolumn{2}{c}{[ (HST--TD)--Kelsall ]}      $\vert$ &
\colhead{TD}                                            &
\multicolumn{2}{c}{[ (HST--TD)--Kelsall ]}      $\vert$ &
\colhead{TD}                                            &
\multicolumn{2}{c}{[ (HST--TD)--Kelsall ]}      $\vert$ \\[-4pt]
\colhead{$\vert$ (K)\ \ \ }                     $\vert$ & 
\colhead{F$_{\lambda}$}                         $\vert$ & 
\colhead{e$^-$/pix/s}                                   &
\colhead{MJy/sr}                                        &
\colhead{nW/m$^2$/sr}                           $\vert$ &
\colhead{e$^-$/pix/s}                                   &
\colhead{MJy/sr}                                        &
\colhead{nW/m$^2$/sr}                           $\vert$ &
\colhead{e$^-$/pix/s}                                   &
\colhead{MJy/sr}                                        &
\colhead{nW/m$^2$/sr}                           $\vert$ 
}
\startdata
{\bf Raw$^a$} & &   &{\bf 0.0145}     &{\bf 35.2}  &    &{\bf 0.0250}    &{\bf 54.6}   &   &{\bf 0.0480}   &{\bf 94.2}   \\
        &       &        &(0.008)     &    (19)    &         &(0.009)    &    (19)     &        &(0.009)   &    (17)     \\
+2.44   &0.76   &0.00678 &0.0124      &     30.1   &0.0308   &0.0173     &     37.7    &0.1138  &0.00254   &     4.99    \\
        &       &        &            &            &         &           &             &        &          &             \\
+2.0    &0.75   &0.00636 &0.0125      &     30.4   &0.0293   &0.0177     &     38.5    &0.1086  &0.00464   &     9.10    \\
        &       &        &            &            &         &           &             &        &          &             \\
+1.84   &0.74   &0.00621 &0.0126      &     30.5   &0.0287   &0.0178     &     38.9    &0.1067  &0.00538   &     10.6    \\
        &       &        &            &            &         &           &             &        &          &             \\
+1.19   &0.72   &0.00564 &0.0127      &     30.9   &0.0266   &0.0183     &     40.0    &0.0995  &0.00826   &     16.2    \\
        &       &        &            &            &         &           &             &        &          &             \\
+0.90   &0.71   &0.00541 &0.0128      &     31.1   &0.0257   &0.0186     &     40.5    &0.0964  &0.00949   &     18.6    \\
        &       &        &            &            &         &           &             &        &          &             \\
+0.48   &0.70   &0.00509 &0.0129      &     31.3   &0.0245   &0.0189     &     41.2    &0.0921  &0.0112    &     22.0    \\
        &       &        &            &            &         &           &             &        &          &             \\
+0.0    &0.69   &0.00474 &0.0130      &     31.6   &0.0231   &0.0192     &     41.9    &0.0875  &0.0131    &     25.6    \\
        &       &        &            &            &         &           &             &        &          &             \\
--0.30  &0.68   &0.00453 &0.0131      &     31.7   &0.0223   &0.0194     &     42.4    &0.0847  &0.0142    &     27.8    \\
        &       &        &            &            &         &           &             &        &          &             \\
--1.15  &0.66   &0.00399 &0.0133      &     32.1   &0.0201   &0.0200     &     43.6    &0.0772  &0.0172    &     33.7    \\
        &       &        &            &            &         &           &             &        &          &             \\
--2.0   &0.64   &0.00351 &0.0134      &     32.5   &0.0182   &0.0204     &     44.6    &0.0703  &0.0199    &     39.1    \\
        &       &        &            &            &         &           &             &        &          &             \\
--3.19  &0.62   &0.00293 &0.0136      &     32.9   &0.0157   &0.0211     &     46.0    &0.0617  &0.0234    &     45.8    \\
        &       &        &            &            &         &           &             &        &          &             \\
{\bf Adopt$^e$}&&   &{\bf 0.0133}     &{\bf 32.1}  &   &{\bf 0.0200}     &{\bf 43.6}   &   &{\bf 0.0172}   &{\bf 33.7}   \\
        &       &        &            &            &         &           &             &        &          &             \\
DGL$^f$ &       &   &\cge 0.0009      &\cge  2.1   &   &\cge 0.0015      &\cge  3.2    &   &\cge 0.0021    &\cge  4.1    \\
        &       &        &            &            &         &           &             &        &          &             \\
eEBL$^g$&       &  &$\sim$0.0002      &$\sim$0.6   &  &$\sim$0.0003      &$\sim$0.6    &  &$\sim$0.0003    &$\sim$0.6    \\
(AB\cge 26)&    &        &            &            &         &           &             &        &          &             \\
{\bf Diff.Lim}$^h$&&&{\bf\cle 0.0122}&{\bf\cle 29} & &{\bf\cle 0.0182}   &{\bf\cle 40} & &{\bf\cle 0.0148} &{\bf\cle 29} \\
\enddata
\sn
\tablenotetext{a}{The raw HST--Kelsall LFS $sech$ differences from
Table~\ref{tab:sech_vals} are repeated in MJy/sr and \nWsqmsr\ {\bf before} TD
subtraction.}

\tablenotetext{b}{$\Delta$T = Reference Temperature
minus the $synphot$ Temperature for --3.2\cle $\Delta$T\cle +2.4 K and the
assumed $\alpha$ value.}

\tablenotetext{c}{Assumed power-law spectral index in F$_{\lambda}$ of the
Zodiacal spectrum for that model. Changing this from the fiducial value of
$\alpha$=0.66 results in different best-fit $\Delta$T values and predicted TD 
signal levels.}

\tablenotetext{d}{The first column for each filter lists the predicted
$pysynphot$ Thermal Dark signal (TD) in e$^-$/pix/s at the quoted $\Delta$T (\S\
\ref{sec:thermal}). Dividing by 3.25, 3.99 and 2.50 to fold in the F125W, F140W and
F160W filter ZP's, respectively, converts this TD to MJy/sr. In each filter's
second column, this TD is subtracted from the raw lower-envelope HST
$sech$-values in MJy/sr at the {\bf top}. Each filter's third column converts 
the (HST--TD)--Kelsall difference to from MJy/sr to \nWsqmsr\ using footnote
$f$ of Table~\ref{tab:sech_vals}.}

\tablenotetext{e}{The (HST--TD)--Kelsall differences adopted for the TD values
as predicted for the best fit $\Delta$T = --1.15 K. }

\tablenotetext{f}{Estimated lower limit to the Diffuse Galactic Light for the
	used SKYSURF regions also subtracted from the adopted values, using the IPAC
	IRSA DGL estimator as in \S\ \ref{sec:results_dgl}.}

\tablenotetext{g}{Our HST SKYSURF analysis
already automatically subtracted from the diffuse signal most of the {\it discrete}
EBL integral from discrete objects with AB\cle 26.5 mag, but the undetected
eEBL integral for AB\cge 26.5 mag, which amounts to 0.56 \nWsqmsr\ (see \S\
\ref{sec:results_eebl}) which is also subtracted here, resulting in the boldface numbers on the bottom row.}

\tablenotetext{h}{The last row lists our resulting estimated limits to {\it any
remaining diffuse light} (boldface in MJy/sr and \nWsqmsr).}

\end{deluxetable*}



\vspace{-0.00cm}
\begin{deluxetable*}{| cc | cc | cc | cc | cc | cc | }
\tablecolumns{12}
\tablewidth{1.0\linewidth}
\tablecaption{Net HST--Zodiacal Light Model Summary and 1.25--1.6 \mum\ 
Diffuse Sky-SB Limits using the SA90 Method
\label{tab:sa90_vals}}
\tablehead{
\multicolumn{4}{c}{$\vert$\ \ \ \ \ \ \ \ 
--- F125W or J-band$^a$ (per sr) ---}                    $\vert$ & 
\multicolumn{4}{c}{--- F140W or JH-band (per sr) ---}    $\vert$ & 
\multicolumn{4}{c}{--- F160W or H-band (per sr) ---}     $\vert$ \\[-0pt]
\multicolumn{2}{c}{$\vert$\ \ --- HST--Kelsall ---}      $\vert$ &
\multicolumn{2}{c}{--- HST--Wright ---}                  $\vert$ &
\multicolumn{2}{c}{--- HST--Kelsall ---}                 $\vert$ &
\multicolumn{2}{c}{--- HST--Wright ---}                  $\vert$ &
\multicolumn{2}{c}{--- HST--Kelsall ---}                 $\vert$ & 
\multicolumn{2}{c}{--- HST--Wright ---}                  $\vert$ \\[-0pt]
\multicolumn{2}{c}{$\vert$\ MJy\ \ \ \ \ \ \ \ nW/m$^2$} $\vert$ &
\multicolumn{2}{c}{MJy\ \ \ \ nW/m$^2$}                  $\vert$ &
\multicolumn{2}{c}{MJy\ \ \ \ \ \ \ nW/m$^2$}            $\vert$ &
\multicolumn{2}{c}{MJy\ \ \ \ nW/m$^2$}                  $\vert$ &
\multicolumn{2}{c}{MJy\ \ \ \ \ \ \ nW/m$^2$}            $\vert$ &
\multicolumn{2}{c}{MJy\ \ \ \ \ \ nW/m$^2$}              $\vert$ 
}
\startdata
                &            &         &          &            &          &         &          &            &          &           &          \\
\cle 0.0148$^b$ &{\bf \cle 35}$^c$ &\cle 0 &{\bf \cle 0}     &\cle 0.0205&{\bf \cle 44}&\cle 0 &{\bf \cle 0}&\cle 0.0296 &{\bf \cle 58} &\cle 0.0077 &{\bf \cle 15} \\
(0.0059)        &            &         &          &(0.0060)    &          &         &          &(0.0133)    &          &           &          \\
(N=589)         &            &         &          &(N=400)     &          &         &          &(N=2171)    &          &           &          \\
                &            &         &          &            &          &         &          &            &          &           &          \\
                &            &         &          &            &          &         &          &            &          &           &          \\
\enddata
\sn
\tablenotetext{a}{For the HST WFC3/IR F125W, F140W, and F160W filters, each set
of 4 columns lists the two-sided 1-sigma clipped median differences\\ (green
dashed lines in Figs.~\ref{fig:f125_sa90}--\ref{fig:f160_sa90}) {\it between} the HST
sky-SB values using the SA90 method (\S\ \ref{sec:results_final}) --- from which the best
Thermal Dark signal (Table~\ref{tab:thermal}) and DGL estimates (\S\ \ref{sec:results_dgl})
have been subtracted --- {\it and} the \citet{Kelsall1998} or
\citet{Wright1998} ZL model prediction for each HST field with
SA=90$\pm$10\degree\ and $\vert$\bEcl$\vert$\cge 30\degree, respectively. The
second row lists the rms value of the clipped median sky-SB, and the third row
lists the number of points used in this clipped median. The quantity pairs are
listed in units of MJy/sr and \nWsqmsr, respectively, following the footnotes
in Table~\ref{tab:sech_vals}--\ref{tab:thermal}.}

\tablenotetext{b}{The best fit WFC3/IR Thermal Dark signal values that have
been subtracted here (for $\Delta$T = --1.15 K in Table~\ref{tab:thermal})
correspond to 0.00123 MJy/sr in the F125W filter, 0.00504 MJy/sr in F140W, and
0.03088 MJy/sr in the F160W filter, respectively.}

\deleted{\tablenotetext{c}{The WFC3/IR Thermal Dark signal values subtracted are
now those for the expected warmer temperatures at SA$\simeq$90$\pm$10\degree\ 
\\ ($\Delta$T=+2.4 K in Table~\ref{tab:thermal}). At these higher temperature,
these TD values now correspond to 0.00209 MJy/sr in the F125W filter, \\
0.00772 MJy/sr in F140W, and 0.04553 MJy/sr in the F160W filter, respectively.}}

\tablenotetext{c}{Each second column includes the correction for the
undetected discrete EBL integral extrapolated for AB\cge 26.5 mag, which
amounts\\ to 0.56 \nWsqmsr\ (\S\ \ref{sec:results_eebl}).}

\end{deluxetable*}


\n\begin{figure*}[!hptb]
	\includegraphics[width=1.000\txw]{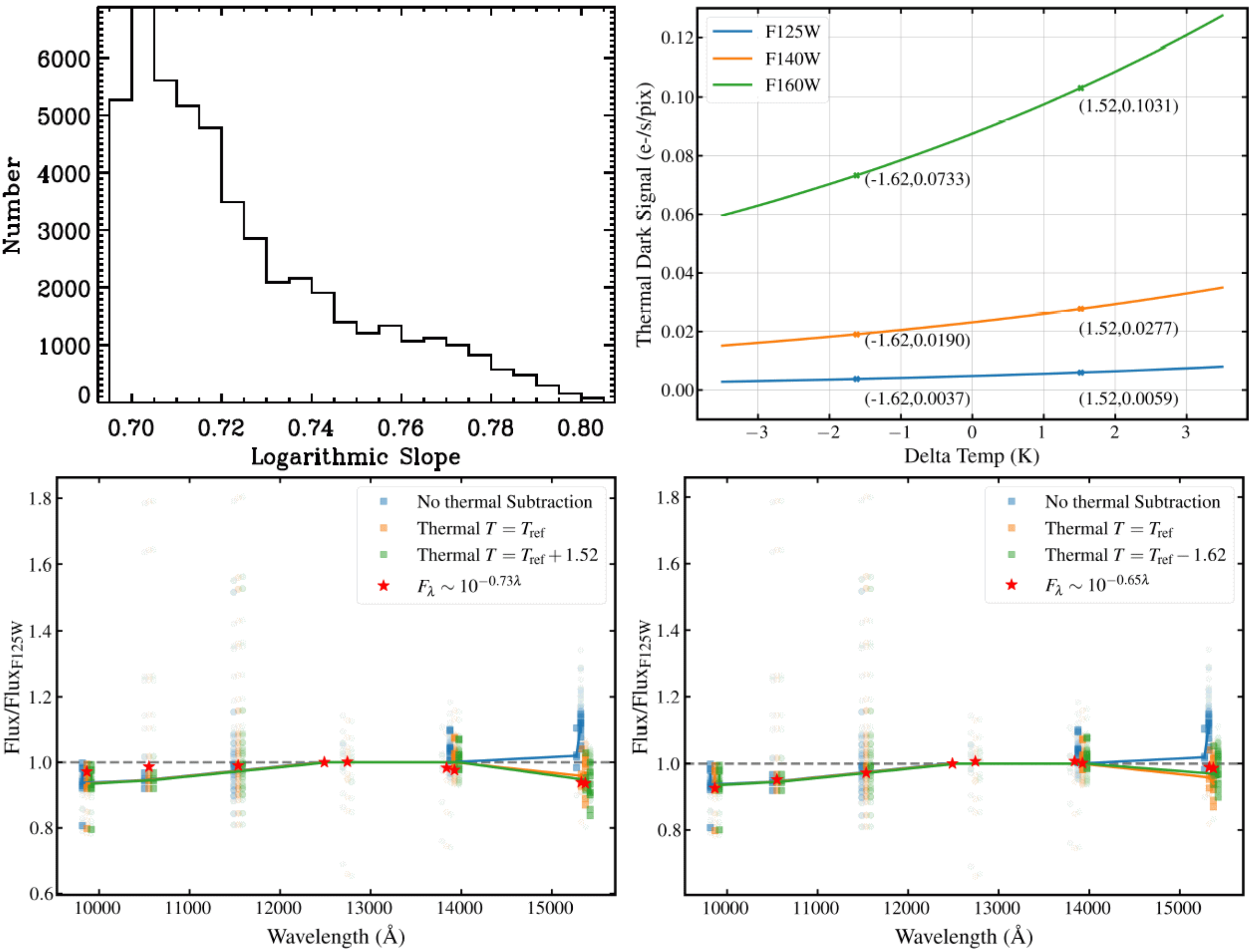}
	\vspace*{+0.00cm}
	\n \caption{
		{\bf (a) [Top Left]:}\ Spectral index distribution N($\alpha$) when
		interpolating the \citet{Kelsall1998} Zodiacal sky-SB prediction of SKYSURF F160W pointings in the
		COBE/DIRBE J and K-band filters. The median power-law spectral index and its 1$\sigma$
		range is $\alpha$=0.713$\pm$0.023. 
		{\bf (b) [Top Right]:}\ Thermal Dark signal corrections predicted by $synphot$ 
		in the WFC3 F125W, F140W, and F160W filters as a function of the change in
		telescope ambient temperature $\Delta$T required to provide a best fit to the
		observed Thermal Dark signals shown in panels (c) and (d). The ($\Delta$T, TD
		signal) numbers listed bracket the range of temperature and TD signal
		considered here (see Table~\ref{tab:thermal}). 
		{\bf (c) [Bottom Left]:}\ Mean Sky-SB values averaged over 500 WFC3/IR
		exposures in the filters F098M, F105W, F110W, F125W, F127M, F139M, F140W,
		F159M, and F160W taken in distinct fields, subtracting various different
		Thermal Dark signal scenarios as a function of ambient temperature \Tref. The
		lightly shaded points were affected by the Geocoronal He II 1.083 \mum-line (in
		F105W and F110W), or got too close to the Earth's limb. The red asterisks show
		the Zodiacal spectrum of \citet{Aldering2001} at each filter's effective
		central wavelength using Eq.~\ref{eqn:aldering}. All $synphot$ Thermal Dark
		subtracted scenarios are normalized to the ratio at 1.25 \mum, which is by
		definition unity. Blue squares and their median fit (blue solid line) assume
		that no Thermal Dark signal needs to be subtracted. The orange squares and
		green line assume T=\Tref\ as the best fit, which is taken from the
		reference files. The green squares and green line assume the
		\citet{Aldering2001} near-IR Zodiacal spectral slope of $\alpha$=0.73 and an
		HST temperature T=\Tref\ + 1.52 K. 
		{\bf (d) [Bottom Right]:}\ As in panel (c), but with the green squares and 
		green line assuming a near-IR Zodiacal spectral slope $\alpha$=0.65 and an HST
		temperature of T=\Tref\ --1.62 K. The two $\Delta$T cases in panels (b)--(d)
		bracket the likely range in HST telescope ambient temperatures, which are used
		to estimate the possible range in F125W--F160W Thermal Dark signal that needs
		to be subtracted from any diffuse sky-SB signal (\S\ \ref{sec:results_diffuselim}).}
	\label{fig:thermfig}
\end{figure*}

\n \subsubsection{The eEBL Component Yet to be Subtracted from the Diffuse Light
Limits} \label{sec:results_eebl}

\sn While the discrete EBL down to $\sim26$~\mAB\ is already automatically excluded from the diffuse EBL limits, we do need to subtract from the upper limits in Tables
\ref{tab:thermal}--\ref{tab:sa90_vals} the expected eEBL sky-integral of galaxies {\it
beyond the detection limits} of the typical short F125W, F140W and F160W
exposures in which the HST sky-SB measurements were made. In \citetalias{Windhorst2022}, we showed that for typical exposure times of
\texp$\simeq$500 sec the WFC3/IR detection limit is AB \cle 26.5 mag for
compact objects in the F125W filter. For similar median exposure times, this detection limit is about 0.3 mag shallower in the
F160W filter \citep[see Table 1 and Fig. 10 of][]{Windhorst2011}. Hence, we
assume that all objects with \JAB\cge 26.5 mag or \HAB\cge 26.2 mag have been
{\it undetected} in SKYSURF's individual $\sim$500 sec WFC3/IR F125W
or F160W exposures, respectively, and so their
sky-integral is {\it still included} in the diffuse sky-SB measurement. We will therefore estimate and subtract it here.

First, we need to correct the total sky-integral values of all objects ---
including low-SB objects --- discussed in \S\
\ref{sec:results_iebl} for the SB-incompleteness that sets in at AB\cge 22 mag due to the
galaxy size distribution. This correction is identified in \citetalias{Windhorst2022} for the F125W filter and repeated below as Eqn.~\ref{eqn:sbcomplete}:
\begin{equation}
Incompleteness\ Correction\ =\ 1.0\ +\ [1.00\ +\ 6.184\ (J_{AB} - 22.0\ 
mag)]/100\%.
\label{eqn:sbcomplete}
\end{equation}
 This incompleteness
correction was also applied to the F160W counts, accounting for the fact that
the F160W catalogs have $\sim$0.3 mag lower sensitivity per unit time. This is
justified by the similarity of the J- and H-band versions of
Fig.~11 of \citetalias{Windhorst2022}, and as shown in Fig. 10 of \citet{Windhorst2011}.
Fig.~2 of \citetalias{Windhorst2022} showed that 75\% of the discrete EBL is already reached
for objects with AB\cle 22.0 mag in the F125W filter, so in essence, this
procedure corrects the faintest 25\% of the EBL integral for SB-incompleteness
of objects known to exist in deeper HST images. The potential impact of very
low-SB discrete objects that are beyond the SB-limits of {\it all} HST images
including the HUDF --- and thus not captured by Eq.~\ref{eqn:sbcomplete} --- will be
discussed in \S\ \ref{sec:discussion}. 

As yet uncorrected for SB-incompleteness, the fraction of the discrete EBL
detected to AB\cle 26.5 mag is $\sim$96.8\%. When we fold
in the SB-incompleteness correction of
Eq.~\ref{eqn:sbcomplete}, this number increases to 99.1\%. Hence, while the SB-incompleteness
correction to the iEBL for discrete sources missed at AB\cge 26.5 mag is
substantial (\cge 26\% at AB\cge 26.5 mag; Eq.~\ref{eqn:sbcomplete}), the actual
correction to the iEBL value due to SB-incompleteness from objects {\it known to
exist in deeper HST images} is small (\cle 3\%), since objects at AB\cge 26.5
mag contribute such a small fraction to the iEBL to begin with. 

Corrected for SB-incompleteness, the above discrete EBL integral to 
AB$\simeq$26.5 mag increases to 1.381$\times$$10^{-26}$ W/Hz/m$^2$/deg$^2$ or
10.99 \nWsqmsr\ in the F125W filter, and to 1.793$\times$$10^{-26}$
W/Hz/m$^2$/deg$^2$ or 11.55 \nWsqmsr\ in the F160W filter, respectively.
Extrapolating Eq.~\ref{eqn:sbcomplete} to AB\cge 30 mag, the converging discrete and
extrapolated EBL integral (iEBL+eEBL) --- corrected for SB-incompleteness ---
amounts to 1.451$\times$$10^{-26}$ W/Hz/m$^2$/deg$^2$ or 11.55 \nWsqmsr\ in the
F125W filter, and to 1.880$\times$$10^{-26}$ W/Hz/m$^2$/deg$^2$ or 12.11
\nWsqmsr\ in the F160W filter, respectively. After correction for
SB-incompleteness, the total sky-integral of objects
detected in typical short HST exposures at AB\cle 26.5 mag is thus still \cge
95\% of the total discrete EBL integral in the F125W and F160W filters,
respectively.

We can now estimate the integrated and extrapolated EBL integral for undetected
sources with AB\cge 26.5 mag that is also corrected for missing low-SB sources that
we know to exist in deeper HST images. Taking the difference between the above
incompleteness-corrected sky-integrals to AB\cle 26.5 mag and AB\cge 30 mag, we
find that the sky-integral for discrete objects with AB\cge 26.5 mag amounts to
$\sim$0.56 \nWsqmsr\ in both the F125W and F160W filters. The amounts are very
similar in both filters, simply because the galaxy counts are very similar in
the F125W and F160W filters \citep{Windhorst2011}, since both filters sample
redwards of the redshifted Balmer or 4000\AA\ breaks for most objects. Hence,
the integrated and extrapolated EBL values in J- and H-band filters are also
very similar \citep[][Fig.~\ref{fig:eblfig1} here]{Driver2016b, Koushan2021}. 

We cannot make an estimate of the sky-integral values in the F140W filter from
existing data, because this filter is not available in ground-based surveys due
to atmospheric water absorption. When project SKYSURF
is completed, it will also provide discrete F140W object counts for 17\cle
AB\cle 28 mag (see Appendix C of \citetalias{Windhorst2022} and Tompkins \etal, 2022, in
preparation). Given the similarity of the above sky-integral values in both the
F125W and F160W filters, we will thus assume that the sky-integral for discrete
objects undetected at AB\cge 26.5 mag in the F140W filter is also $\sim$0.56
\nWsqmsr. 

In conclusion, we subtract $\sim$0.56 \nWsqmsr\ to obtain the diffuse light
limits in Table~\ref{tab:thermal}--\ref{tab:sa90_vals} to account for the sky integral
of discrete objects that remain undetected in typical SKYSURF exposures at
AB\cge 26.5 mag in both the F125W, F140W, and F160W filters. Our diffuse light
limits have thus the {\it discrete} integrated and extrapolated EBL (iEBL+eEBL),
and the Zodiacal model prediction, fully removed from the HST sky-SB data. 


\n \subsection{Corrections for Diffuse Galactic Light} \label{sec:results_dgl}

\mn The DGL is subtracted using the IPAC IRSA 
model~\footnote{\url{https://irsa.ipac.caltech.edu/applications/BackgroundModel/}},
as shown in Table~\ref{tab:thermal}. The IRSA tool presents a model for the
emission from the diffuse interstellar medium of our Galaxy, which uses a
combination of the \citet{Arendt1998} Galactic emission and
\citet{Schlegel1998} dust maps. These models are anchored to the COBE/DIRBE
data at 100~\mum\ wavelength, where the ZL is minimal. {This DGL model relies on accurate 100~\mum\ maps and a dust emission model describing the ratio of NIR-to-100~\mum\ emission. COBE/DIRBE galactic maps have zero-point uncertainties of $\sim3$~nW/m$^2$/sr \citep{Schlegel1998}. Systematic uncertainties related to converting 100 \mum\ emission to our near-IR wavelengths may be up to a factor of $2$ \citep{Onishi2018} and have a complex Galactic Latitude dependence due to differing amounts of thermal emission and scattered light \citep{Sano2017}. However, this uncertainty typically corresponds to $\sim0.002$ MJy/sr, much less than other systematic uncertainties in our analysis.} } The IRSA tool also includes an estimate of diffuse
scattered starlight down to 0.5 \mum\ wavelength based on the \citet{Zubko2004}
model integrated with observations of \citet{Brandt2012}. The DGL correction
to our HST--Kelsall differences in Fig.~\ref{fig:f125_skydiff} is small (typically $<0.003$ MJy/sr) since the {\it darkest Galactic and Ecliptic} regions have
already been sub-selected. {Furthermore, there is no discernible trend between HST-Kelsall and Galactic Latitude, suggesting that our measurements are not sensitive to the uncertainties in DGL described above.}

From the HST--Kelsall differences, corrected for the most plausible TD values
in Table~\ref{tab:thermal}, we plot the resulting upper limits to the amount of 
diffuse light at 1.25, 1.37, and 1.53 \mum\ as the brown downward arrows in
Fig.~\ref{fig:eblfig1} and Fig.~\ref{fig:eblfig_zoom}. This includes an orange shaded
uncertainty wedge in Fig.~\ref{fig:eblfig_zoom} that captures the TD values predicted
for --3.2\cle $\Delta$T\cle +2.4 K. Given the uncertainty in the Thermal Dark
Signal subtraction (\S\ \ref{sec:thermal} and Table~\ref{tab:thermal}, as well as uncertainties
in the ZL models (\S \ref{sec:hstkelsallresults}), we will quote these values as upper limits, even though in
the nominal range of HST component temperatures ($\Delta$T\cle 2K), the
remaining TD-subtracted diffuse light signal in the F125W and F140W filters
remains significant (Fig.~\ref{fig:eblfig_zoom}).

\n \subsection{Comparison of the (HST--TD--DGL) Estimates vs. the Kelsall and
Wright Zodiacal Light models} \label{sec:results_final}

\begin{figure*}[!hptb]
	\n\cl{
		\includegraphics[width=1.050\txw]{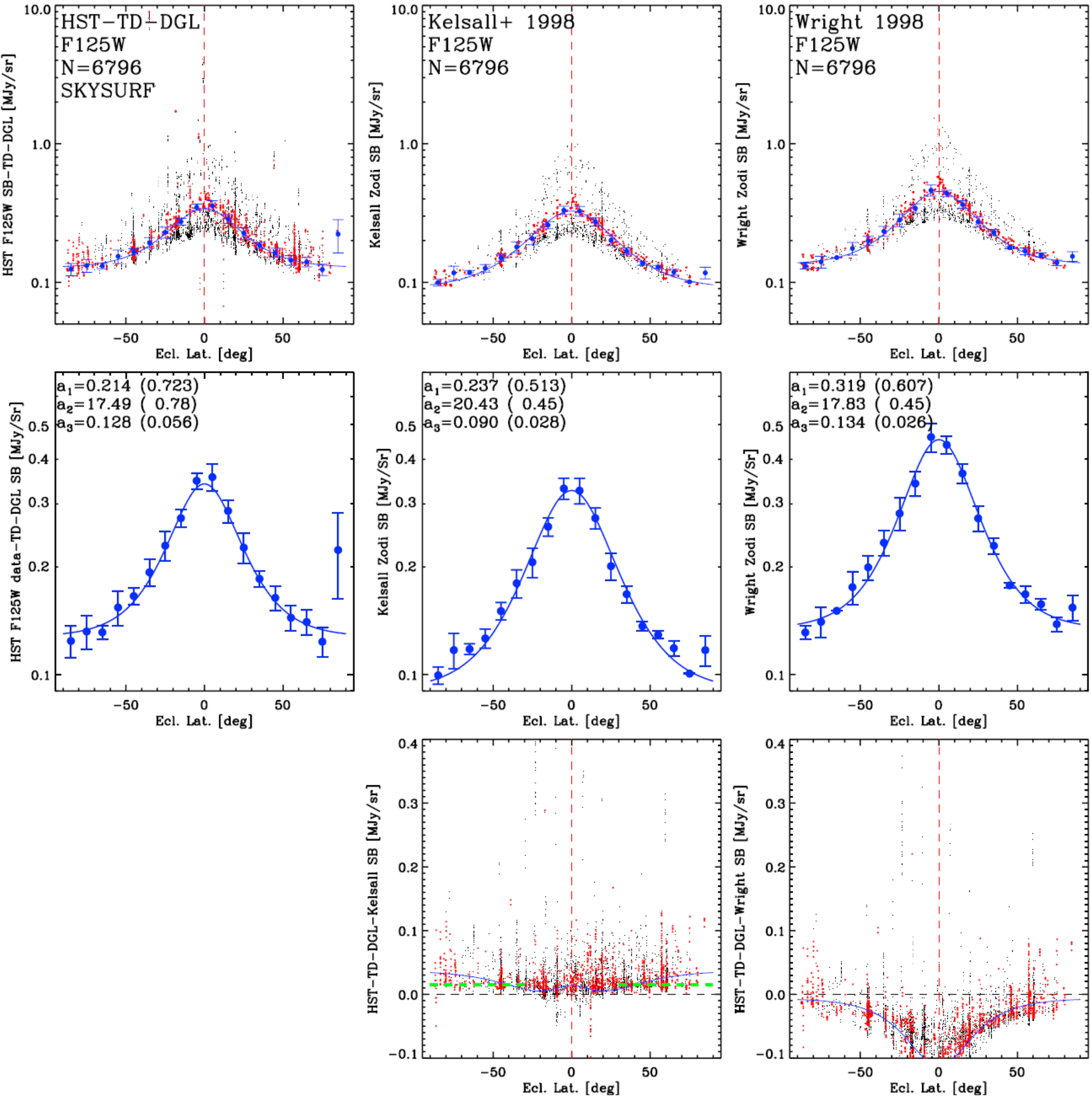}
	}
	
	\vspace*{+0.10cm}
	\n \caption{
		{\bf [(a) Top Left]:}\ HST WFC3/IR sky-SB measurements in the F125W filter 
		vs. Ecliptic Latitude after subtracting the best WFC3/IR Thermal Dark signal
		estimate for each exposure from \S\ \ref{sec:thermal} and the Diffuse Galactic Light
		(DGL) signal from \S\ \ref{sec:results_dgl}. Black dots indicate all observations from
		Fig.~\ref{fig:f125ecl_all}, and red dots only those with Sun Angle
		SA=90$\pm$10\degree. The COBE/DIRBE data, to which the models were fit, were
		taken at a comparable SA-range. The blue points indicate the one-sided clipped
		medians of the SA=90$\pm$10\degree\ points in each 10\degree\ \bEcl-bin, and
		the blue line indicates the best $sech$-fit to these medians. 
		{\bf [(b) Top Middle]:}\ \citet{Kelsall1998} ZL model prediction for the {same} 
		sky pointings and filters, observing day of the year, {\it and} Sun
		Angles as the HST data in (a) with their medians and best fit $sech$-model. 
		{\bf [(c) Top Right]:}\ As in (b), but for the \citet{Wright1998} Zodiacal 
		model prediction. 
		{\bf [(d), (e), (f) Middle Row]:}\ The clipped medians for each \bEcl-bin from
		panels (a), (b), (c) are reproduced for clarity, respectively, with their errors
		plus the best $sech$-fits and their coefficients. 
		{\bf [(g) Bottom Middle]:}\ {\it Difference} between each (HST--TD--DGL) data
		point from (a) and its \citet{Kelsall1998} ZL model prediction from (b), with
		the difference in their $sech$-fits (blue line). The green dashed line
		indicates the two-sided 1-sigma clipped median for SA=90$\pm$10\degree\ and
		$\vert$\bEcl$\vert$\cge 30\degree\ (Table~\ref{tab:sa90_vals}). 
		{\bf [(h) Bottom Right]:}\ As in (g), but for the \citet{Wright1998} Zodiacal
		model prediction from (c).}
	\label{fig:f125_sa90}
\end{figure*}

\begin{figure*}[!hptb]
	\n\cl{
		\includegraphics[width=1.050\txw]{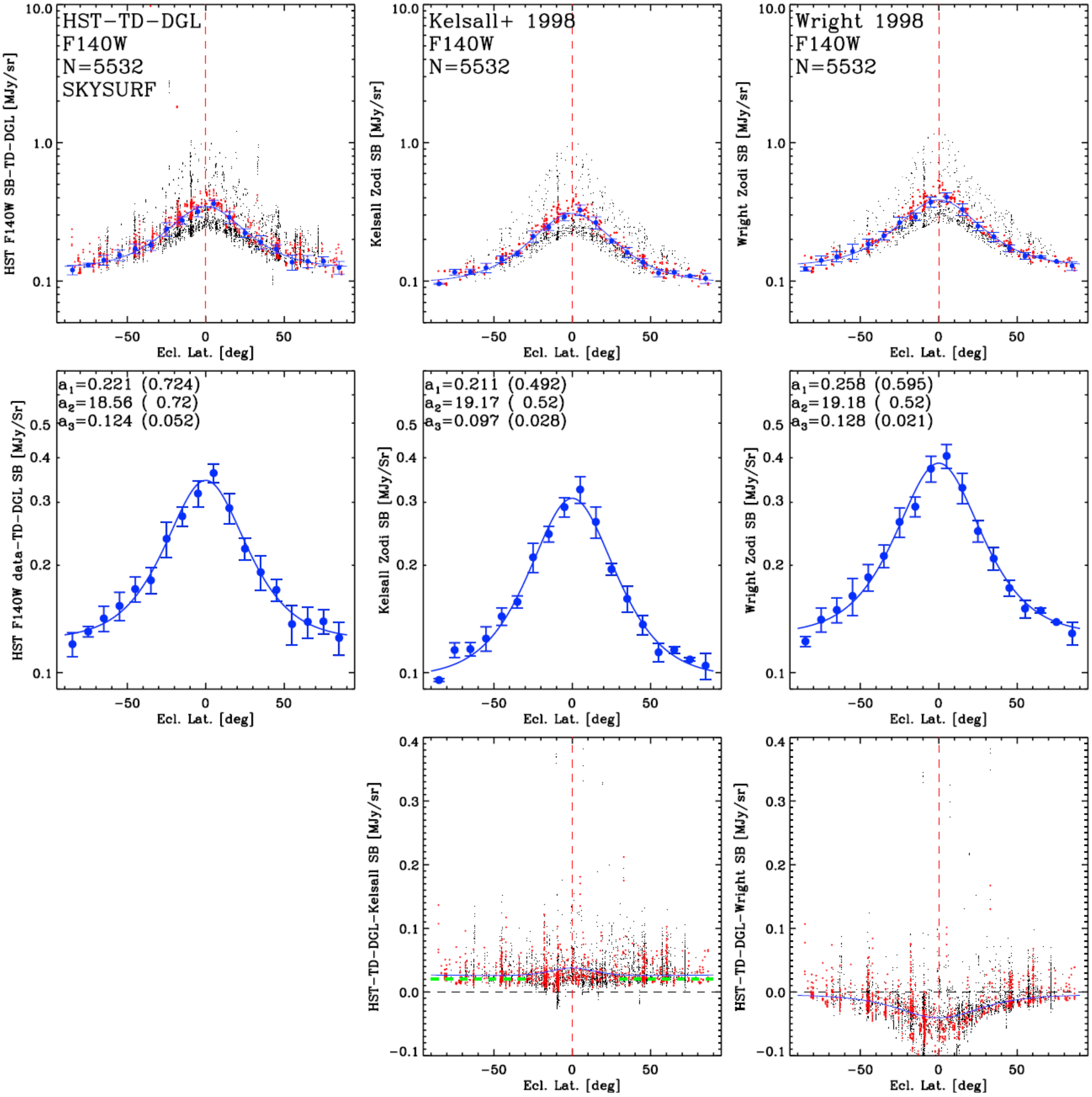}
	}
	
	\vspace*{+0.10cm}
	\n \caption{
		{\bf [(a) Top Left]:}\ HST WFC3/IR sky-SB measurements in the F140W filter 
		vs. Ecliptic Latitude after subtracting the best WFC3/IR Thermal Dark signal
		estimate for each exposure from \S\ \ref{sec:thermal} and the Diffuse Galactic Light
		(DGL) signal from \S\ \ref{sec:results_dgl}. Black dots indicate all observations from
		Fig.~\ref{fig:f125ecl_all}, and red dots only those with Sun Angle
		SA=90$\pm$10\degree. The COBE/DIRBE data, to which the models were fit, were
		taken at a comparable SA-range. The blue points indicate the one-sided clipped
		medians of the SA=90$\pm$10\degree\ points in each 10\degree\ \bEcl-bin, and
		the blue line indicates the best $sech$-fit to these medians. 
		{\bf [(b) Top Middle]:}\ \citet{Kelsall1998} ZL model prediction for the {same}
		sky pointings and filters, observing day of the year, {\it and} Sun
		Angles as the HST data in (a) with their medians and best fit $sech$-model. 
		{\bf [(c) Top Right]:}\ As in (b), but for the \citet{Wright1998} Zodiacal 
		model prediction. 
		{\bf [(d), (e), (f) Middle Row]:}\ The clipped medians for each \bEcl-bin from
		panels (a), (b), (c) are reproduced for clarity, respectively, with their errors
		plus the best $sech$-fits and their coefficients. 
		{\bf [(g) Bottom Middle]:}\ {\it Difference} between each (HST--TD--DGL) data
		point from (a) and its \citet{Kelsall1998} ZL model prediction from (b), with
		the difference in their $sech$-fits (blue line). The green dashed line
		indicates the two-sided 1-sigma clipped median for SA=90$\pm$10\degree\ and
		$\vert$\bEcl$\vert$\cge 30\degree\ (Table~\ref{tab:sa90_vals}). 
		{\bf [(h) Bottom Right]:}\ As in (g), but for the \citet{Wright1998} Zodiacal
		model prediction from (c).}
	\label{fig:f140_sa90}
\end{figure*}

\begin{figure*}[!hptb]
	\n\cl{
		\includegraphics[width=1.050\txw]{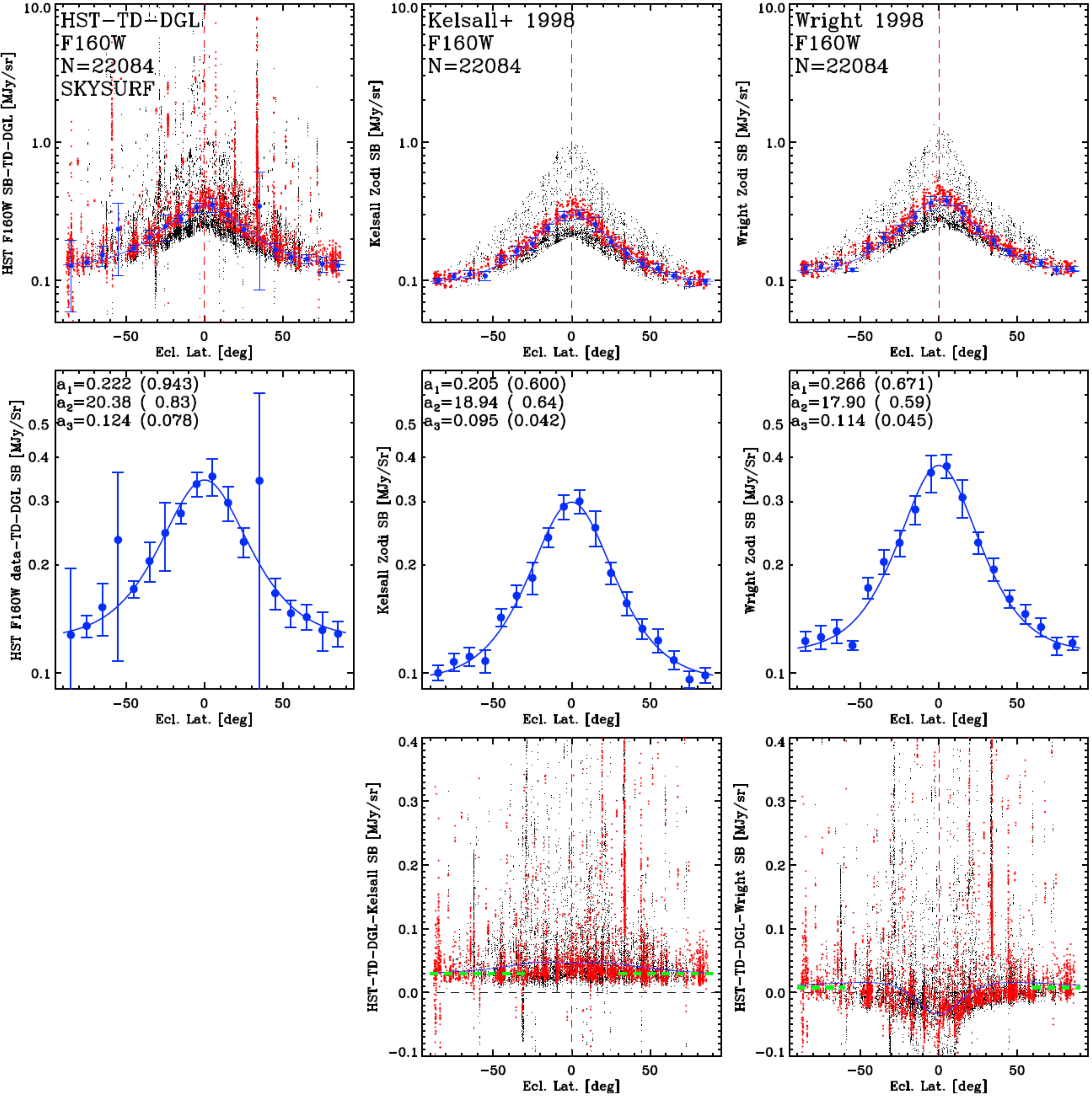}
	}
	
	\vspace*{+0.10cm}
	\n \caption{
		{\bf [(a) Top Left]:}\ HST WFC3/IR sky-SB measurements in the F160W filter 
		vs. Ecliptic Latitude after subtracting the best WFC3/IR Thermal Dark signal
		estimate for each exposure from \S\ \ref{sec:thermal} and the Diffuse Galactic Light
		(DGL) signal from \S\ \ref{sec:results_dgl}. Black dots indicate all observations from
		Fig.~\ref{fig:f125ecl_all}, and red dots only those with Sun Angle
		SA=90$\pm$10\degree. The COBE/DIRBE data, to which the models were fit, were
		taken at a comparable SA-range. The blue points indicate the one-sided clipped
		medians of the SA=90$\pm$10\degree\ points in each 10\degree\ \bEcl-bin, and
		the blue line indicates the best $sech$-fit to these medians. 
		{\bf [(b) Top Middle]:}\ \citet{Kelsall1998} ZL model prediction for the {same}
		sky pointings and filters, observing day of the year, {\it and} Sun
		Angles as the HST data in (a) with their medians and best fit $sech$-model. 
		{\bf [(c) Top Right]:}\ As in (b), but for the \citet{Wright1998} Zodiacal 
		model prediction. 
		{\bf [(d), (e), (f) Middle Row]:}\ The clipped medians for each \bEcl-bin from
		panels (a), (b), (c) are reproduced for clarity, respectively, with their errors
		plus the best $sech$-fits and their coefficients. 
		{\bf [(g) Bottom Middle]:}\ {\it Difference} between each (HST--TD--DGL) data
		point from (a) and its \citet{Kelsall1998} ZL model prediction from (b), with
		the difference in their $sech$-fits (blue line). The green dashed line
		indicates the two-sided 1-sigma clipped median for SA=90$\pm$10\degree\ and
		$\vert$\bEcl$\vert$\cge 30\degree\ (Table~\ref{tab:sa90_vals}). 
		{\bf [(h) Bottom Right]:}\ As in (g), but for the \citet{Wright1998} Zodiacal
		model prediction from (c).}
	\label{fig:f160_sa90}
\end{figure*}

\sn For the WFC3/IR F125W, F140W, and F160W filters, respectively, the top three
panels of Fig.~\ref{fig:f125_sa90}--\ref{fig:f160_sa90} show the following comparison. 
The top left panels show the HST WFC3/IR sky-SB measurements vs. Ecliptic
Latitude after subtracting the best WFC3/IR Thermal Dark signal estimate for
each exposure from \S\ \ref{sec:thermal}, {and} the Diffuse Galactic Light signal
from \S\ \ref{sec:results_dgl}. The top middle panels of
Fig.~\ref{fig:f125_sa90}--\ref{fig:f160_sa90} show the \citet{Kelsall1998} Zodiacal
model prediction for the {same} observation date and SA as the HST data.
The top right panels similarly show the \citet{Wright1998} ZL model prediction
with parameters that were updated by \citet{Gorjian2000}, as provided by the
IRSA tool. 

In Fig.~\ref{fig:f125_sa90}--\ref{fig:f160_sa90}, black dots indicate {all}
observations from Fig.~\ref{fig:f125ecl_all}, and the red dots are
only those with Sun Angle SA=90$\pm$10\degree. Both the \citet{Kelsall1998} and
the \citet{Wright1998} ZL models were fit to the COBE/DIRBE data that were
taken at a comparable but somewhat wider SA-range (SA=94$\pm$30\degree). The
blue-filled circles indicate one-sided clipped medians of the
SA=90$\pm$10\degree\ points in each 10\degree\ \bEcl-bin, and the blue line
indicates the best $sech$-fit to these medians following
Eq.~\ref{eqn:sech_hst}--\ref{eqn:sech_kel}. 

The middle row of panels in Fig.~\ref{fig:f125_sa90}--\ref{fig:f160_sa90} shows the
clipped medians for each \bEcl-bin from the top panels separately for clarity,
together with their best-fit $sech$ and its coefficients. The bottom two panels
show the {\it difference} between each (HST--TD--DGL) data point from the top
left panel after subtracting either the \citet{Kelsall1998} ZL model prediction (bottom middle) or the \citet{Wright1998} prediction (bottom right). The
difference in the (HST-TD-DGL)--Kelsall or (HST-TD-DGL)--Wright $sech$-fits is
indicated by the thin full-drawn blue lines. 

The HST--Kelsall sky-SB differences clearly show positive offsets similar to
those in Fig.~\ref{fig:f125_skydiff}, where the best-fit TD and
DGL were not yet subtracted. For $\vert$\bEcl$\vert$\cle 30\degree, the
HST--Kelsall differences show a somewhat stronger dependence on Ecliptic
Latitude than at higher $\vert$\bEcl$\vert$-values. Specifically, the
HST--Kelsall differences for $\vert$\bEcl$\vert$\cle 30\degree\ are either
slightly smaller (in F125W) or slightly larger (in F140W and F160W) than at
higher Ecliptic Latitudes. For Ecliptic Latitudes $\vert$\bEcl$\vert$\cge
30\degree, these difference plots have an almost straight bottom envelope that
is above zero. We therefore quantified these {\it positive net HST--Kelsall
offsets} for $\vert$\bEcl$\vert$\cge 30\degree\ as a {\it single constant}
using a two-sided 1-sigma clipped median for the HST--Kelsall differences at
SA=90$\pm$10\degree. These numbers are given in Table~\ref{tab:sa90_vals} and
indicated by the thick green dashed lines in the lower-left panels of
Fig.~\ref{fig:f125_sa90}--\ref{fig:f160_sa90}. These offsets are our best estimate for
any difference in diffuse light that may remain between the (HST--TD--DGL)
sky-SB values and the \citet{Kelsall1998} ZL model predictions using the
SA90-method. 

Formally, these average (HST--TD--DGL)--Kelsall differences for 
$\vert$\bEcl$\vert$\cge 30\degree\ in Table~\ref{tab:sa90_vals} indicate a detection
of a positive signal within the quoted errors using the SA90 method. However,
the bottom middle panels Fig.~\ref{fig:f125_sa90}--\ref{fig:f160_sa90} show some
Ecliptic Latitude and wavelength dependence of these differences across all
\bEcl-values, more so than in Fig.~\ref{fig:f125_skydiff} using
the LFS method. Since the precise cause of this Ecliptic Latitude or wavelength
dependence is not known, we quote the (HST--Kelsall) differences from the SA90
method as upper limits in Table~\ref{tab:sa90_vals}. Since the upper limit values
from the SA90 method are somewhat larger than those from the LFS method in
Table~\ref{tab:thermal} (due to the nature of both methods discussed at the start
of \S\ \ref{sec:results}), we plot the latter as the {\it upper limits} in
Fig.~\ref{fig:eblfig1} and Fig.~\ref{fig:eblfig_zoom}, and the larger values of the
former as the {\it upper envelope} of the allowed range, which is indicated by
the orange wedge in Fig.~\ref{fig:eblfig_zoom}. 

The HST--Wright sky-SB differences are mostly negative in both the F125W,
F140W, and F160W filters, especially for Ecliptic Latitudes
$\vert$\bEcl$\vert$\cle 30\degree. The {\it ratios} of the \citet{Wright1998}
and \citet{Kelsall1998} ZL model predictions for all HST observations at their
respective observing dates and Sun Angles are as follows: Wright/Kelsall
1.346$\pm$0.05 in F125W, 1.268$\pm$0.05 in F140W,
1.223$\pm$0.05 in F160W, respectively. These ratios are not quite uniform
with \bEcl, which may suggest some remaining Ecliptic Latitude dependence in the
\citet{Wright1998} model, and also some wavelength dependence at 1.25--1.6
\mum. In Fig.~\ref{fig:f125_sa90}--\ref{fig:f160_sa90}, some Latitude dependence
remains visible in the HST--Wright sky-SB differences even at high Ecliptic
Latitudes of 40\degree\cle $\vert$\bEcl$\vert$\cle 90\degree. As noted on the 
IRSA tool, \citet{Wright1998} and \citet{Gorjian2000} adopted a ``strong
no-Zodiacal'' condition at 25 \mum\ wavelength, which requires that the
minimum 25 \mum\ residual at high Galactic latitude after subtraction of a ZL
model from the COBE/DIRBE observations has to be zero. At this wavelength, the
thermal Zodiacal dust contribution is indeed approximately maximal compared to
the Zodiacal scattered Sun-light contribution (brown dot-dashed and green
dotted lines in Fig.~\ref{fig:eblfig1}, respectively). \citet{Kelsall1998} do not
enforce this condition, and thereby obtain lower values for the ZL intensity,
also at shorter wavelengths. In conclusion, the net HST--Wright differences are
consistent with being \cle 0 in the F125W and F140W filters, and for
$\vert$\bEcl$\vert$\cle 30\degree, they are at most \cle 0.0077 MJy/sr (or \cle 
15 \nWsqmsr) in the F160W filter. These numbers are also given in
Table~\ref{tab:sa90_vals}. Based on our preliminary results, these offsets are thus
our best current limits to any difference in diffuse light that may exist
between the (HST--TD--DGL) sky-SB values and the \citet{Wright1998} ZL model
predictions.

We end with a cautionary note that our current near-IR diffuse light limits in
Fig.~\ref{fig:eblfig_zoom} may still contain some residual {\it time-varying} WFC3/IR
Thermal Dark component as a function of HST temperature and orbital phase. All
our diffuse light values in Tables~\ref{tab:thermal}--\ref{tab:sa90_vals} and 
Fig.~\ref{fig:eblfig_zoom} are derived using {\it average} orbital component
temperatures, and for this reason (in addition to the uncertainties in ZL model subtraction) our near-IR diffuse light values are listed
as upper limits. 




\n \section{Discussion of SKYSURF's First Results} \label{sec:discussion}

\mn In conclusion, the HST data--\citet{Kelsall1998} model allows for a
diffuse light component of \cle 29--40 \nWsqmsr\ at 1.25--1.6 \mum\ 
wavelength (Table~\ref{tab:thermal}). Given the relatively constant values of these
HST--Kelsall offsets at most Ecliptic Latitudes
(Fig.~\ref{fig:f125_sa90}--\ref{fig:f160_sa90}), these values may indicate a very dim,
possibly spherical or ellipsoidal component of diffuse light in the net HST
data that is not present in the \citet{Kelsall1998} model. This diffuse light level could be due to a number of causes:
(a) a remaining HST orbital phase and temperature-dependent TD component, which
may need to include a thermal Earthshine component; (b) a dim (nearly)
spherical component missing in the \citet{Kelsall1998} Zodiacal Light model;
(c) a spherical {\it diffuse} EBL component {\citep{Sano2020}}; or (d) some combination of these
possibilities. The \citet{Wright1998} model leaves little or no room for
diffuse light after subtracting the Thermal Dark signal and DGL in
Fig.~\ref{fig:f125_sa90}--\ref{fig:f160_sa90} and Table~\ref{tab:sa90_vals}. In this
context, we compare our HST results with the following recent results by other
groups:

\sn (1) \cite{Matsuura2017} analyze CIBER rocket spectra, and find that the
sky-SB of diffuse light at 1.4 \mum\ wavelength is
$\sim$42.7$^{+11.7}_{-10.3}$ \nWsqmsr\ compared to the \citet{Kelsall1998}
model. After subtraction of a 1.4 \mum\ iEBL+eEBL signal of $\sim$11.8 
\nWsqmsr\ (\S\ \ref{sec:results_eebl}), this would correspond to a net diffuse light
signal of $\sim$31 \nWsqmsr. They find no significant excess in diffuse light
compared to the \citet{Wright1998} model. They suggest that compared to the
\citet{Kelsall1998} model their results may require ``a new diffuse light
component, such as an additional foreground or an excess EBL with a redder
spectrum than that of the ZL.'' \citet{Korngut2022} use subsequent CIBER
spectra to estimate the Equivalent Width (EW) of the Ca triplet around 8542
\AA, and suggest a simple modification to the \citet{Kelsall1998} model that
adds a constant (spherical) component of 46$\pm$19 \nWsqmsr\ to best fit their
inferred Zodiacal level at 1.25 \mum. The \citet{Korngut2022} CIBER experiment
directly estimates the depth of the Ca triplet Fraunhofer lines in the Zodiacal
spectrum, so it is plausible that much of this excess diffuse light is of
Zodiacal origin. Within the errors, our 1.25--1.4 \mum\ HST--Kelsall
differences of \cle 29--40 \nWsqmsr\ are consistent with the diffuse light
signal suggested by both \cite{Matsuura2017} and \citet{Korngut2022}. Given
that the lower envelopes of our 1.25--1.6 \mum\ HST data--Kelsall model
differences are rather constant at all higher Ecliptic Latitudes, it
is thus possible that a dim, large, and largely spherical component may need to
be added to the \citet{Kelsall1998} model with an amplitude of \cle 29--40
\nWsqmsr\ at 1.25--1.6 \mum\ as seen from Low Earth Orbit. 

\begin{figure*}[!hptb]
	{
		\vspace*{-0.7cm}
		\includegraphics[width=0.950\txw]{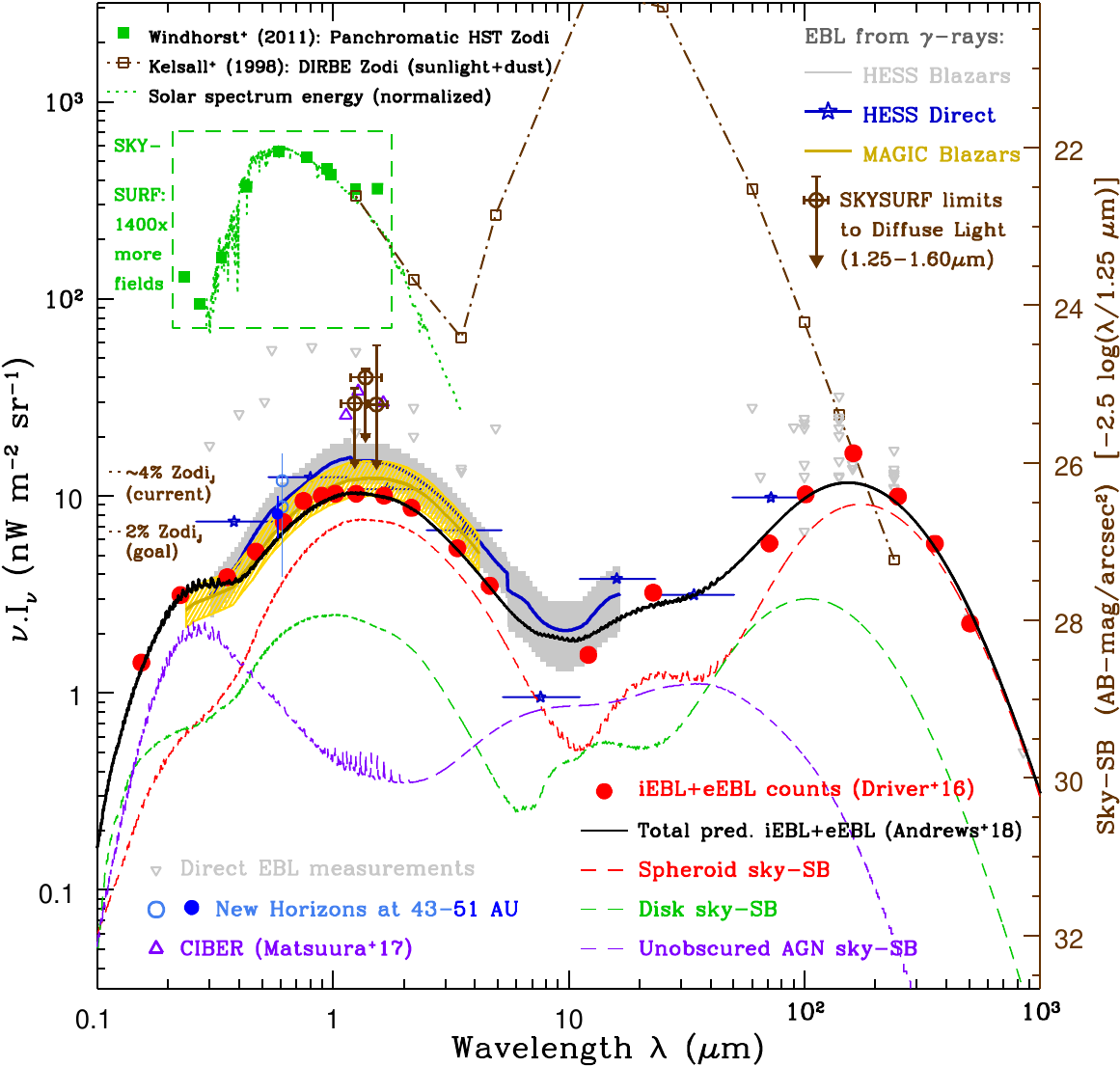}
	}

	\caption{
		Summary of astrophysical foreground and background energy relevant to SKYSURF, along with the first SKYSURF measurements at $1.25$, $1.37$, and $1.53$ microns.
		The left scale indicates the total energy $\nu$.\Inu\ in \nWsqmsr, and the right
		scale the corresponding sky-SB in AB-\magarc\ at 1.25 \mum\ (which can be scaled
		to other wavelengths as indicated). The discrete measurements of D16 from {\it
			integrated and extrapolated galaxy counts (iEBL+eEBL)} (red-filled circles) and
		other published data are shown. Grey triangles indicate
		the {\it total EBL measurements} {\citep{Puget1996, Fixsen1998, Dwek1998a, Hauser1998, Lagache1999, Finkbeiner2000,
			Dole2006, Bernstein2002, Bernstein2007, Cambresy2001, Matsumoto2005,
			Matsuura2011,Matsumoto2011, Tsumura2013, Sano2020}}.
		Also shown are more recent results from the CIBER experiment {(purple triangles; \citealt{Matsuura2017})}, Pioneer (light blue points without errors; \citealt{Matsumoto2018}), and New Horizons (medium and dark-blue points with errors; \citealt{Lauer2021} and \citealt{Lauer2022}, which is offset for clarity) that aim to more accurately subtract the Zodiacal foreground.
		{All of these measurements require accurate modeling of foreground DGL, and, except for Pioneer and New Horizons points, ZL. For a more direct comparison with SKYSURF diffuse-light limits, the iEBL level (taken from the respective references) has been subtracted from the CIBER and New Horizons points.}
		Green squares are panchromatic HST
		sky-SB measurements compared to the Solar spectrum. Brown
		squares indicate the COBE/DIRBE data and the Zodiacal dust model of
		\citet{Kelsall1998}. The grey and orange wedges and blue stars are
		$\gamma$-ray Blazar EBL constraints from the MAGIC and HESS TeV experiments \citep{MAGICCollaboration2008,HESS2013}.
		The black line is the sum of the \citet{Andrews2018} EBL models for
		star-formation in spheroids (red), disks (green), and unobscured AGN (purple).
		The currently achieved calibration+zeropoint accuracy in the 1.25--1.6 \mum\
		HST sky-SB estimates is $\sim$4\% of the Zodiacal sky-SB 
		\citep{Windhorst2022}, as indicated. SKYSURF's goal is a \cle 2\%-accurate
		sky-SB model across the sky at 0.2-1.6 \mum\ wavelengths to address the
		discrepancy between the total EBL and the {\it discrete} iEBL+eEBL. The brown points are the SKYSURF diffuse light limits, {which implicitly subtract discrete iEBL}, identified in Table~\ref{tab:sech_vals}. These limits are lower than previous analyses, but still leave room for an isotropic diffuse light component, whether in the Solar System or at cosmological distances.}
	\label{fig:eblfig1}
\end{figure*}

\citet{Korngut2022} discuss that the heliocentric isotropic IPD distribution in
the inner Solar System at 10--25 AU may be supplied by debris from long-period
Oort Cloud Comets (OCC; \citealp{Oort1950}, see also, \eg\
\citealp{Nesvorny2010} and \citealp{Poppe2016}), and suggest that such a
component may need to be added to the \citet{Kelsall1998} model with possibly a
5\% amplitude. Our upper limits to the 1.25--1.6 \mum\ sky-SB of \cle 29--40
\nWsqmsr\ in Table~\ref{tab:thermal} and Fig.~\ref{fig:eblfig1} suggest that any diffuse light is \cle
10\% of the Zodiacal sky-SB at these wavelengths. Hence, if most of this light
were due to a missing component in the \citet{Kelsall1998} ZL model, such a
component must be dim and extend to high Ecliptic latitudes. Future work is needed to add such a
component to the \citet{Kelsall1998} model and match it to the SKYSURF
observations. Revised models may need to include (a) collisional processes in
the Solar System that can make the Zodiacal dust smaller over time, and (b)
Solar radiation pressure that may drive these smaller dust particles further
out into the Solar System, perhaps forming a tenuous ellipsoidal or more
spherical cloud of dust around the Sun compared to the known Zodiacal IPD
cloud. 

The \citet{Kelsall1998} model includes IPD model uncertainties and lists
possible changes that could improve the IPD modeling. Quoting their paper, one
of their suggested improvements is: ``7. Permit a variation of the albedo for
the shorter wavelength bands to accommodate the clues in the observations that
point to a variation with (Ecliptic) latitude, which may well result from the
differences in the dust contributed by comets as compared to that coming from
asteroids.'' Table 2 of \citet{Kelsall1998} adopt an albedo at 1.25 \mum\
wavelength for their Zodiacal components of $a$=0.204$\pm$0.0013. Recent thermal
IR observations of TNOs imply geometric albedos of \cle 20--30\%, while some
have albedos as large as $\sim$60\% \citep[\eg][]{Duffard2014, Kovalenko2017,
Vilenius2012, Vilenius2014, Vilenius2018}, possibly indicating a more icy
surface for some TNOs. The four small satellites of Pluto have albedos
ranging from 55\%--85\% \citep{Weaver2016}. While the nature of any OCC dust
component at higher Ecliptic latitudes may be substantially different from that
of TNOs and their collision or scattering products, these results suggest that
albedos higher than the $a$$\simeq$0.2 value adopted by \citet{Kelsall1998} are
possible. Future improvements of Zodiacal IPD models may therefore need to 
consider a different albedo distribution for any additional OCC dust component
at higher Ecliptic latitudes, including albedos as appropriate for a larger
fraction of dust particles with icy surfaces. {For example, \citet{Sano2020}
analyze DIRBE results, and find that the observed
Sun Angle dependence of the mid-IR and near-IR background is consistent with an
additional diffuse isotropic component with an amplitude of $\sim5\%$ of the
\cite{Kelsall1998} IPD cloud.}

\begin{figure*}[!hptb]
	{
		\includegraphics[width=1.000\txw]{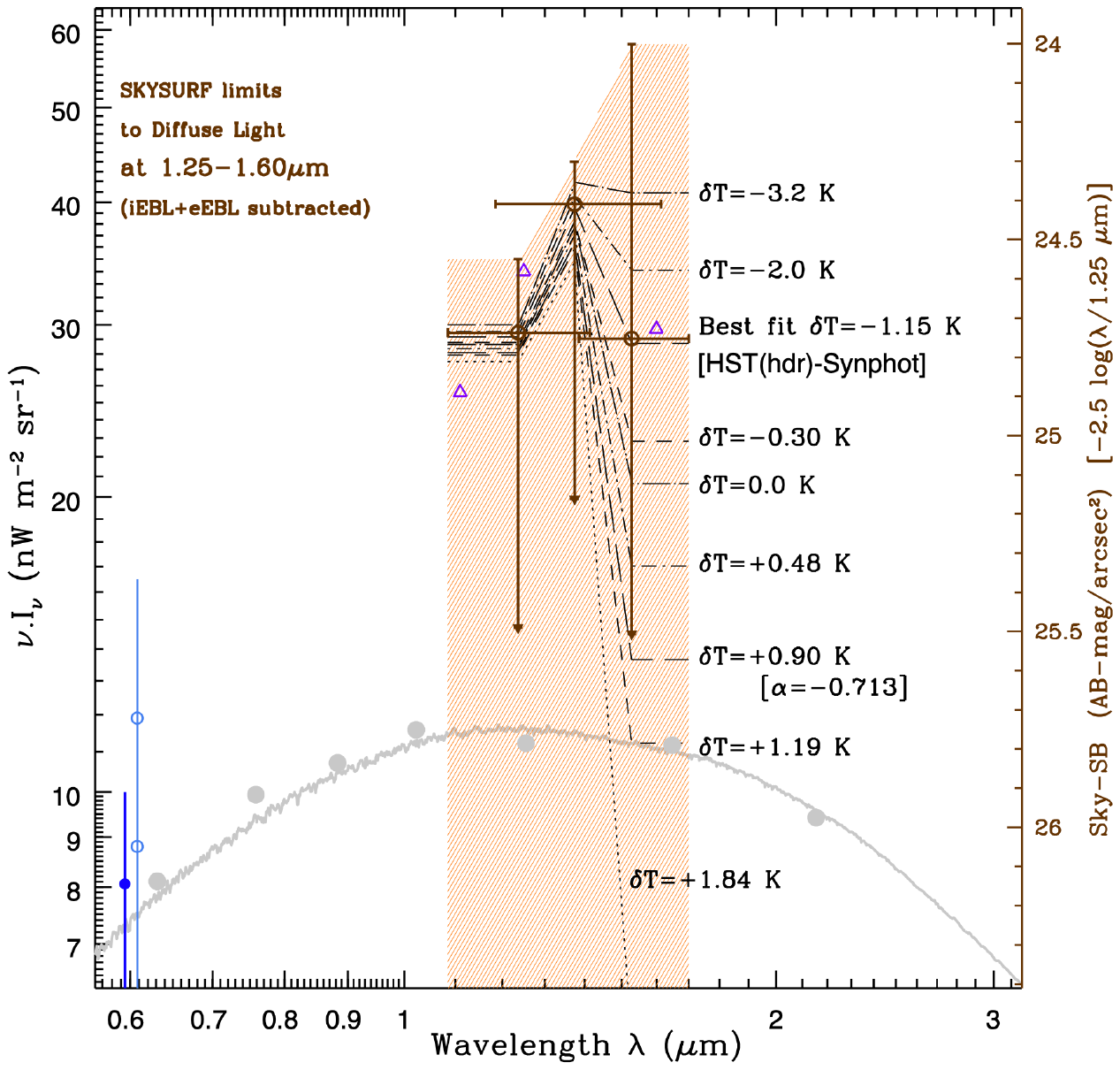}
	}

	\caption{
		{SKYSURF on the level of \emph{diffuse} light compared with limits on diffuse light. As in Fig.~\ref{fig:eblfig1}, purple triangles and blue open and filled circles are from \cite{Matsuura2017}, \cite{Lauer2021}, and \cite{Lauer2022} respectively. Also shown for context are iEBL estimates from \cite{Koushan2021} as the grey points, and  the \cite{Driver2016b} model fit to those measurements as the grey line.} The
		large open orange circles with their error ranges are the SKYSURF diffuse 
		light upper limits in the WFC3/IR filters F125W, F140W, and F160W, as discussed
		in \S\ \ref{sec:hstkelsallresults}--\ref{sec:results_final}, Figs.~\ref{fig:f125_skydiff}--\ref{fig:f160_sa90} and
		Table~\ref{tab:sech_vals}--\ref{tab:sa90_vals}, from which the iEBL+eEBL has already
		been subtracted. The shaded orange wedge indicates our diffuse light limits
		given the current best knowledge of the HST temperature range and its resulting
		WFC3/IR Thermal Dark signal, and the subtracted {\it discrete} iEBL+eEBL signal
		(\S\ \ref{sec:results}). The $synphot$ HST component temperature predictions can vary
		by $\Delta$T$\simeq$$\pm$2 K from those in the HST engineering data and FITS
		headers. The best \chisq-fit occurs for an average ($pysynphot$--HST-header)
		difference of $\Delta$T=--1.15 K. The $\Delta$T value that is most consistent
		with the 1--2 \mum\ Kelsall slope of the Zodiacal spectrum ($\alpha$=0.713; see
		\S\ \ref{sec:results_diffuselim}) is also indicated, and occurs for (HST-header--$synphot$)
		temperature difference of $\Delta$T=+0.90 K. Therefore, while our limits to the
		F125W and F140W diffuse light are a factor $\sim$3 above the best available
		discrete EBL counts of \citet{Koushan2021}, any diffuse signal in the F160W
		filter could be fainter still, as indicated by the orange wedge. The upper limits of the orange wedge
		correspond to the limits derived from the SA90 method. Details are
		given in \S\ \ref{sec:results}--\ref{sec:discussion}. 
		At 1.25 and 1.4 microns, the detection of diffuse light is not very
		dependent on the assumed thermal dark signal. While our fiducial F160W
		results are consistent with the F125W and F140W limits, the F160W
		limit depends much more strongly on the assumed thermal dark signal.}
	\label{fig:eblfig_zoom}
\end{figure*}

\sn (2) \citet{Lauer2021} present 0.6 \mum\ object counts from New Horizons
images of 7 fields taken around Pluto's distance, where the Zodiacal sky-SB is
substantially lower than in LEO. They suggest a possible excess diffuse signal
of unknown origin with an amplitude in the range (8.8--11.9)$\pm$4.6 \nWsqmsr\
at 0.6 \mum. These data are plotted with their two quoted error ranges as the
blue points in Fig.~\ref{fig:eblfig1} and Fig.~\ref{fig:eblfig_zoom}. \citet{Lauer2022}
add a single new NH field with lower DGL contribution, which has a similar 0.6
\mum\ excess diffuse signal with a smaller error bar: 8.1 $\pm$ 1.9 \nWsqmsr (shown as the dark blue point in Fig~\ref{fig:eblfig1}-\ref{fig:eblfig_zoom}).
While their number of NH fields is limited, their images do provide a 0.6 \mum\
diffuse light sky-SB estimate in the very dark sky environment at a distance of
43--51 AU from the Sun. Fig.~\ref{fig:eblfig1} and \ref{fig:eblfig_zoom} suggest their
0.6 \mum\ upper value at 43--51 AU is about 8--10 \nWsqmsr\ above the
integrated and extrapolated {\it discrete} EBL of \citet{Driver2016b} and
\citet{Koushan2021}, respectively, while our HST WFC3/IR 1.25--1.6 \mum\ upper
limits in Fig.~\ref{fig:eblfig_zoom} are about 29--40 \nWsqmsr\ above the {\it
discrete} EBL values at 1.25--1.6 \mum.

The possible origin of 8--10 \nWsqmsr\ of cosmological diffuse light remains an open question.
For example, \citet{Conselice2016} and
\citet{Lauer2021} suggested that some missing light could be caused
by the galaxy counts rapidly steepening at V\cge 24 mag, because existing
surveys are missing a substantial population of low-SB objects. Given the decreasing abundance of low-surface brightness objects with AB$\sim$24 and large sizes \citep[\eg][]{Greene2022,Zaritsky2022}, as well as recent limits on the abundance of low-surface brightness galaxies \citep[\eg][]{Jones2018}, it is hard to imagine that a factor of two
or more in sky-SB comes from faint, undetected low-SB objects at V\cge 24 mag.
Accounting of this diffuse light from even fainter galaxies becomes more difficult because they would have to be even more abundant to account for their corresponding faintness (\eg\ Fig. 2 of \citetalias{Windhorst2022}). Further investigation of this possibility, as well as a more detailed analysis of the impact of surface brightness and confusion-based completeness on EBL estimations will be conducted with future SKYSURF analyses (e.g.~Kramer \etal\ 2022, in preparation).

Any missing diffuse EBL would then also need to be present in our HST--Kelsall
comparison, which allows for \cle 29--40 \nWsqmsr\ of diffuse light at 1.25--1.6
\mum. If for instance \cle 10 \nWsqmsr\ of our HST--Kelsall difference were due
to truly diffuse EBL of cosmological origin (\ie\ very faint, low-SB objects), then the \citet{Kelsall1998} model would only need
$\sim$20 \nWsqmsr\ of additional uniform Zodiacal component. However, our HST
data--Wright model comparison does not require this, and, in fact, leaves
little or no room for any additional diffuse light components, neither an
unrecognized HST Thermal Dark signal component, nor an additional Zodiacal
component, nor a diffuse EBL component.

In conclusion, the darkest $\sim$1\% of our 34,000 HST WFC3/IR 1.25--1.6 \mum\
images closely follow the shape of the \citet{Kelsall1998} model, and suggest
that the \citet{Kelsall1998} model may need an additional (nearly spherical)
component of \cle 29--40 \nWsqmsr, while HST shows no such excess over the
\citet{Wright1998} model. A possible explanation is that the Kelsall model may
be missing $\lesssim$29--40 \nWsqmsr\ of high albedo OCC dust as seen from 1 AU, which
\citet{Wright1998} included by default, because of his assumed strong
no-Zodiacal condition at 25 \mum\ wavelength.

Through the ``Sungrazer''
project \citep{Sekanina2013}~\footnote{see also
	\url{https://sungrazer.nrl.navy.mil/}}, orbiting Solar observatories like SOHO
and STEREO have found thousands of comets since 1995 that are getting in close
proximity of the Sun. \citet{Silsbee2016} have modeled the nearly isotropic
comet population that are expected to show up in very large numbers --- also at
larger distances from the Sun --- with the Rubin
Telescope~\footnote{\url{https://www.lsst.org/}}.
Hence, updated Zodiacal IPD models may be able to include a more spherical
component from such cometary dust left behind in the inner solar system. 

HST studies of KBO's at $\sim$10--100 AU show remarkably blue colors in the
WFC3/IR medium-band filters F139M--F153M \citep[\eg][]{Fraser2012,
Fraser2015}, which have similar central wavelengths but are narrower than our
F140W and F160W filters. While it remains to be seen that OCC dust in the outer
Solar System has similar blue near-IR colors and high reflectance, scattering
models of icy particles (including amorphous and crystalline H$_2$O ice) do
suggest that high albedos with a near-IR wavelength dependence are possible. ZL
model refinements may need to include such considerations in more detail to
better match the LFS envelopes of the HST data at all Ecliptic latitudes. We
consider this in future papers when the full panchromatic SKYSURF database has
been processed including all the UV--optical and remaining near-IR filters.
Once Zodiacal Light models have been updated to fully match the panchromatic
SKYSURF data, this may result in firmer limits to, or estimates of, the amount
of diffuse light that can come from beyond our Solar System, including diffuse
EBL.


\n \section{Summary and Conclusions} \label{sec:conclusions} 

\mn In this paper, we present the first results from the Hubble Space Telescope Archival project
``SKYSURF'', first outlined in \citepalias{Windhorst2022}. Sky surface brightness measurements conducted on HST data are confirmed to be stable and precise, in line with the $\sim2-4\%$ errors estimated in \citetalias{Windhorst2022}.
By comparing measured HST sky-surface brightness measurements with predictions describing Zodiacal and Galactic foregrounds, we place competitive limits on the presence of an isotropic diffuse light component, either within the Solar System or at cosmological distances.

\sn \bul (1) Without having reprocessed the entire HST imaging Archive for
SKYSURF as yet, we illustrate our methods and first results from 34,412 images
in the HST Wide Field Camera 3 IR filters F125W, F140W, and F160W. Compared to
the COBE/DIRBE 1.25 \mum\ and K-band Zodiacal sky-SB predictions of
\citet{Kelsall1998}, our darkest WFC3 F125W, F140W, and F160W sky-SB
measurements appear to be on average $\sim$15--55\% higher (or 
$\sim$0.0145$\pm$0.008, 0.025$\pm$0.009, and 0.048$\pm$0.009 MJy/sr,
respectively) than the \citet{Kelsall1998} model predictions. With both taken at
face value, this places an upper limit of \cle 29--40 \nWsqmsr\ on any
1.25--1.6 \mum\ {\it diffuse light} in excess of the \citet{Kelsall1998} ZL
model components. 

\sn \bul (2) The largest uncertainty in our darkest HST WFC3/IR sky-SB
measurements comes from the WFC3/IR Thermal Dark signal subtraction at 1.6
\mum. From multi-wavelength WFC3/IR images, we assess and subtract the WFC3/IR
Thermal Dark {\it signal} for a range of HST orbital temperatures (--3\cle
$\Delta$T\cle 3 K). In the F160W filter, the Thermal Dark signal {\it may} be
as large as the value of our upper limit to any remaining diffuse light {\it
if} HST were to run hotter than nominal by \cge +2 K, leaving in that case
little room for a significant diffuse light component. However, for the
best-fit $\Delta$T=--1.15 K below the HST reference temperature,
the nominal F160W Thermal Dark signal is $\sim$0.077
e$^-$/pix/sec (or $\sim$0.031 MJy/sr), resulting in a fairly consistent net
diffuse light signal in all three HST filters less than 29--40 \nWsqmsr\ over
the \citet{Kelsall1998} model predictions that were made for the {\it same} sky
pointings and filters, observing day of the year, {\it and} Sun Angle. 

\sn \bul (3) Compared to the \citet{Wright1998} ZL model, HST appears to detect
{\it no} significant 1.25--1.6 \mum\ {\it diffuse light} within the current
uncertainties. The lower envelope of the HST data--Wright model values suggests
no remaining signal in the F125W and F140W filters or perhaps a slightly
negative one. In the F160W filter, the HST data--Wright model values suggest a
remaining diffuse signal of at most $\sim$0.0077 MJy/sr (15 \nWsqmsr). Hence, if
the \citet{Wright1998} model best represents the ZL, this model would leave
little or no room for additional diffuse light components.

\sn \bul In conclusion, given
our Lowest Fitted Sky-SB measurements in the HST WFC3/IR F125W, F140W, and F160W
filters, an update of the \citet{Kelsall1998} ZL model may be needed to better
understand and constrain any additional diffuse light components that may come
from the outer Solar System. Once those are modeled in more detail and over a
wider range of wavelengths, better constraints may also be obtained on any
remaining diffuse light component, including diffuse EBL. This will be
addressed in future SKYSURF papers. 

\acknowledgments
We thank Annalisa Calamida, Phil Korngut, and Tod Lauer for helpful discussions.
Additionally, we thank John Mather for his helpful comments regarding his suggestion of a spherical distribution of Sun approaching comets from SOHO/STEREO and his reference to the Sungrazer project.
We thank HST Archive staff at STScI for their
expert advice on HST component temperatures. All of the data presented in this
paper were obtained from the Mikulski Archive for Space Telescopes (MAST). This
project is based on observations made with the NASA/ESA Hubble Space Telescope
and obtained from the Hubble Legacy Archive, which is a collaboration between
the Space Telescope Science Institute (STScI/NASA), the Space Telescope European
Coordinating Facility (ST-ECF/ESA), and the Canadian Astronomy Data Centre
(CADC/NRC/CSA).

We thank Ms. Desiree Crawl, Prof. Thomas Sharp, and the NASA Space Grant
Consortium in Arizona for consistent support of our many undergraduate SKYSURF
researchers at ASU during the pandemic. We acknowledge support for HST programs
AR-09955 and AR-15810 provided by NASA through grants from the Space Telescope
Science Institute, which is operated by the Association of Universities for
Research in Astronomy, Incorporated, under NASA contract NAS5-26555. 
Work by RGA was supported by NASA under award number 80GSFC21M0002.

We are grateful to the anonymous referee, whose suggestions greatly improved this paper.

We also acknowledge the indigenous peoples of Arizona, including the Akimel
O'odham (Pima) and Pee Posh (Maricopa) Indian Communities, whose care and
keeping of the land has enabled us to be at ASU's Tempe campus in the Salt
River Valley, where this work was conducted.

\sn \software{
Astropy: \url{http://www.astropy.org} \citep{Robitaille2013, Astropy2018};\ 
IDL Astronomy Library: \url{https://idlastro.gsfc.nasa.gov} \citep{Landsman1993};\ 
Photutils: \url{https://photutils.readthedocs.io/en/stable/} \citep{Bradley20};\
\ProFound: \url{https://github.com/asgr/ProFound} \citep{Robotham2017};\ 
\ProFit: \url{https://github.com/ICRAR/ProFit} \citep{Robotham2018};\ 
SourceExtractor: \url{https://www.astromatic.net/software/sextractor/} or
\url{https://sextractor.readthedocs.io/en/latest/} \citep{Bertin1996}. 
}

\facilities{
Hubble Space Telescope Mikulski Archive \url{https://archive.stsci.edu};\ 
Hubble Legacy Archive (HLA) \url{https://hla.stsci.edu};\ Hubble Legacy
Catalog (HLC) \url{https://archive.stsci.edu/hst/hsc/}
}


\bibliographystyle{aasjournal}

\bibliography{skysurf_paper2}

\ve 

\appendix

\n \section{Thermal Behavior of HST}\ \label{sec:thermalappendix} 

\n Temperatures of the HST components are monitored through various thermal
sensors throughout the telescope and WFC3.
The HST component temperatures utilized as reference temperatures in \S \ref{sec:results_thermal} were taken
from tables in \url{https://www.stsci.edu/hst/instrumentation/reference-data-for-calibration-and-tools/synphot-throughput-tables}, and are summarized below in Table~\ref{tab:temps}.
Here, we also reproduce some representative values of relevant telescope components directly from the STScI HST telescope group (private communication), which
are not all directly available through the HST image FITS headers, $ima$- or 
or \jit-files. The $T_{\rm ref}$ values are generally consistent with temperatures measured by sensors on the telescope to a few degrees C.

\begin{deluxetable}{| c | c |}[htb!]
	\tablecolumns{2}
	\tablewidth{1.0\linewidth}
	\tablecaption{$T_{\rm ref}$ of HST optical components 
		\label{tab:temps}}
	\tablehead{
		\colhead{Component}               $\vert$ & 
		\colhead{Temperature (\degree C) }                                       
	}
	\startdata
	Primary Mirror & 15.15\\
	Mirror Pads & 15.15\\
	Secondary Mirror & 17.15\\
	Pick of Mirror & 14.75\\
	IR Chanel Select Mechanism & 0.15\\
	Fold Mirror & 0.15\\
	WFC3IR Mirror 1 & 0.15\\
	WFC3IR Mirror 2 & 0.15\\
	WFC3IR Refractive Corrector Plate  & -35.85\\
	WFC3IR Filter & -35.85\\
	\enddata
	
	\vspace*{-4em}
\end{deluxetable}

HST's thermal variations across each orbit will matter the most for WFC3/IR, as
the IR detector is most sensitive for measurable thermal variations at
wavelengths $\lambda$\cge 1.4 \mum. (To avoid excessive Thermal Dark signal in
the H-band, the WFC3/IR detector was therefore designed to cut-out all 
wavelengths $\lambda$\cge 1.73 \mum, so that the WFC3/IR F160W filter 
therefore is really a ``short H-band'' filter with \leff$\simeq$1.53 \mum). 
During 2020, typical temperatures measured were (working backwards from the IR
detector; all in units of degrees C): IR Detector = --127.8 C; IR Shield =
--100.2 C (the black inner housing surrounding the detector); Outer IR Detector
Housing = --48.75 C; Housing of the IR Filter Select Mechanism (FSM) = --55 C;
Refractive Corrector Plate (RCP) = --33.62 C; and the WFC3/IR Cold Enclosure
(CE) = --33.6 C. These are the temperatures of the components seen by the
WFC3/IR detector directly, and can vary by a couple degrees C.

When the (aluminum) Blank is selected for Dark Current measurements in the
WFC3/IR FSM, it blocks the detector's view of the WFC3/IR RCP, so only indirect
illumination from the CE is possible. This Blank has {\it higher} emissivity
than the WFC3/IR filters, so a measured Dark Current frame looking at the
aluminum Blank will contain additional Thermal Dark signal and have a somewhat
higher amplitude than the dark frame that is applicable to most filters. WFC3
does not have a temperature sensor on the FSM filters and the Blank, but their
temperatures are likely somewhere between the FSM Housing at --55 C and the RCP
at --33.6 C. All of this thermal emission comes from the
entire passband of the IR detector (\ie\ 0.6--1.73 \mum). Therefore, when a
WFC3/IR dark frame is taken to form a Dark Current calibration file to be
subtracted in the WFC3 pipeline, only the thermal sources between the Blank and
the detector listed above plus the actual detector generated Dark Current are
measured. 

The WFC3/IR filters have very high transmission and will also transmit some
Thermal Dark signal from the camera and the telescope, which are at different
temperatures but come from much smaller solid angles. The {\it Dark 
	Current} calibration required thus does depend upon {\it which} filter was used
for the science observation. With a WFC3/IR filter in place, we see the RCP, four mirrors within the WFC3 optical bench cold-enclosure, and the
WFC3 Pick-Off Mirror (POM) in the OTA Hub Area. Temperatures of these are
somewhat less precisely known due to the lack of close temperature sensors. 
The POM picks up more thermal radiation from the Earth during occultation,
which subsequently cools off when the observations start during the next darker
part of an orbit. The four mirrors inside the WFC3 enclosure are at
temperatures of about 0--4 C. They are all silver coated and thus have
fairly low emissivity within the IR filter passbands, slightly lower than that
of gold. 

The WFC3 optical bench and all of its associated baffles provide an environment
kept colder than +4 C, but with higher emissivity. The cold mask at the
location of the RCP has about the same temperature as the RCP and should block
all direct views of these high emissivity surfaces, \ie\ the WFC3 detector only
has a direct view of HST's mirrors. The WFC3 POM consists of a MgF$_2$ flat
substrate overcoated with aluminum to ensure excellent near-UV performance.
Its temperature is less certain, because of the lack of nearby temperature
sensors, but the arm to which it is attached is at +12.5 C, and the ``snout''
leading into the WFC3 optical bench is at +7.9$\pm$0.7 C. The POM sees the
illuminated Earth during most orbits and therefore fluctuates in temperature.
It is unknown by exactly how much, as this depends on the Earth scenes
transiting during bright time, but it is likely that the
POM varies between 10--15 degrees C. The OTA primary and secondary mirrors, and
their associated baffles are also typically at 10--17 C, although the baffles
will not matter much for Thermal Dark signal estimates. 

All these surfaces with their measured or estimated temperatures and
approximate geometries have been modeled using simple black-body approximations
in the $pysynphot$ 
tool and the appropriate emissivities and solid angles as seen from the detector.
Given the incomplete knowledge of exact temperatures and their ranges, as well
as of all the precise geometries inside WFC3 and HST, these $pysynphot$
predictions of the WFC3/IR Thermal Dark signal will have their limitations.
Hence, in \S\ \ref{sec:thermal} we present the best available
$pysynphot$ estimates of the WFC3/IR Thermal Dark signal based on these {\it
	average} temperatures in order to analyze our WFC3/IR sky-SB measurements,
including a plausible temperature range of HST's main components as modeled in
$pysynphot$. We refer to Carleton \etal\ (2022b, in preparation) for a detailed
analysis of the most likely Thermal Dark signal for each WFC3/IR exposure in
the SKYSURF database.

\ve 

\section{Acronyms Used in SKYSURF} \label{sec:acronyms}
{
	\mn 
}{
	\begin{tabular}{ll}
		\hline
		\hline
		\n Acronym   &Explanation                              \\[-2pt]
		\hline
		AB-mag       &--2.5 log (Object-Flux / Zeropoint-Flux) \\  
		ACS          &Advanced Camera for Surveys \\  
		AGN          &Active Galactic Nucleus \\  
		APT          &Astronomers Proposal Tool \\
		ASU          &Arizona State University \\  
		AWS          &Amazon Web Services \\
		CCD          &Charged Coupled Device \\  
		CDM          &Cold Dark Matter \\  
		CERES        &Clouds and the Earth's Radiant Energy System \\
		CIB          &Cosmic Infrared Background \\
		COB          &Cosmic Optical Background \\
		COBE         &Cosmic Background Explorer \\
		COS          &HST's Cosmic Origins Spectrograph \\
		CR           &Cosmic Ray \\
		CTE          &Charge Transfer Efficiency \\ 
		CV           &Cosmic Variance \\
		CVZ          &Continuous Viewing Zone \\
		DC           &(Electronic) Dark Current \\
		DGL          &Diffuse Galactic Light \\
		DIRBE        &Diffuse Infra-Red Background Experiment \\ 
		EBL          &Extragalactic Background Light \\ 
		dEBL         &diffuse Extragalactic Background Light \\ 
		eEBL         &extrapolated Extragalactic Background Light \\ 
		iEBL         &integrated Extragalactic Background Light \\ 
		ERS          &(HST WFC3) Early Release Science program \\
		FOC          &HST's Faint Object Camera \\  
		FOS          &HST's Faint Object Spectrograph \\  
		FOV          &Field of View \\  
		FWHM         &Full-Width Half-Maximum \\  
		GDC          &Geometrical Distortion Corrections \\
		GOODS        &Great Orbiting Observatories Deep Survey \\  
		\HAB         &H-band (1.6 \mum) AB-mag \\  
		HDF          &Hubble Deep Field \\  
		HLA          &Hubble Legacy Archive \\
		HLC          &Hubble Legacy Catalog \\
		HST          &Hubble Space Telescope \\  
		HUDF         &Hubble UltraDeep Field \\  
		HWHM         &Half Width Half Maximum (=0.5$\times$FWHM) \\  
		ICL          &Intra-Cluster Light \\
		IEF          &Illuminated Earth Fraction \\ 
		\hline
	\end{tabular}
} \label{tab:acronyms}

\ve

{
	\cl { ACRONYMS USED IN SKYSURF (continued) }
	\vspace*{-.5em}
	\mn 
}{
	\begin{tabular}{ll}
		\hline
		\hline
		\n Acronym  &Explanation                               \\[-2pt]
		\hline
		IGL          &Intra-Group Light \\
		IPD          &InterPlanetary Dust \\
		IRAF         &Image Reduction and Analysis Facility \\  
		ISM          &Interstellar Medium \\
		\JAB         &J-band (1.25 \mum) AB-mag \\  
		Jy           &Jansky or Flux Density unit (=10$^{-26}$\WsqmHz) \\
		KBOs         &Kuiper Belt Objects \\
		LA           &Earth's Limb Angle \\
		LEO          &Low Earth Orbit \\
		LES          &Lowest Estimated Sky-SB \\ 
		LFS          &Lowest Fitted Sky-SB \\
		MA           &Moon Angle \\
		MAST         &Mikulski Archive for Space Telescopes \\
		NED          &NASA Extragalactic Database \\ 
		NEP          &North Ecliptic Pole \\  
		NICMOS       &Near-Infrared Camera and Multi-Object Spectrograph \\  
		OCC          &Oort Cloud Comets \\
		OTA          &Optical Telescope Assembly \\
		PAM          &Pixel Area Map \\
		PSF          &Point Spread Function \\  
		QSOs         &Quasi Stellar Objects \\
		RA           &HST Roll Angle \\
		R.A.         &Right Ascension \\
		RC3          &Third Reference Catalog of Bright Galaxies \\  
		SAA          &South Atlantic Anomaly \\
		SA           &Sun Angle \\
		SB           &Surface Brightness \\  
		SDSS         &Sloan Digital Sky Survey \\  
		SED          &Spectral Energy Distribution \\ 
		SEP          &South Ecliptic Pole \\
		SFR          &Star-Formation Rate \\ 
		SF           &Star-Forming \\
		SM           &Servicing Mission \\
		STIS         &Space Telescope Imaging Spectrograph \\
		STScI        &Space Telescope Science Institute \\  
		TD           &Thermal Dark {\it signal} \\
		TNOs         &Trans-Neptunian Objects \\
		UVIS         &WFC3 UV--Visual channel \\  
		UV           &Ultraviolet ($\sim$0.1--0.3 \mum) \\  
		WFC3         &HST's Wide Field Camera 3 \\  
		WFPC2        &HST's Wide Field Planetary Camera 2 \\  
		WF/PC        &HST's Wide Field/Planetary Camera \\
		ZL           &Zodiacal Light \\
		\hline
	\end{tabular}
} \label{tab:acronyms2}


\end{document}